\documentclass[%
aip,
amsmath,amssymb,
reprint,%
]{revtex4-1}

\usepackage{graphicx}
\usepackage{dcolumn}
\usepackage{bm}
\usepackage{subfig}
\usepackage[utf8]{inputenc}
\usepackage[T1]{fontenc}
\usepackage{mathptmx}
\usepackage{caption}
\usepackage{color}
\usepackage[most]{tcolorbox}
\usepackage{xcolor}
\usepackage{xfrac}
\usepackage{float}
\usepackage{bigints}
\usepackage[export]{adjustbox}
\usepackage{tabularx}
\usepackage{epstopdf, epsfig}

\usepackage{newtxmath}
\usepackage{hyperref}
\usepackage{url}
\usepackage{graphicx}
\usepackage{amsmath}
\usepackage{empheq}
\usepackage{bm}
\usepackage{accents}
\usepackage{booktabs}
\usepackage{array}

\makeatletter
\def\@email#1#2{%
	\endgroup
	\patchcmd{\titleblock@produce}
	{\frontmatter@RRAPformat}
	{\frontmatter@RRAPformat{\produce@RRAP{*#1\href{mailto:#2}{#2}}}\frontmatter@RRAPformat}
	{}{}
}%
\makeatother

\newcolumntype{P}[1]{>{\centering\arraybackslash}p{#1}}

\newcolumntype{P}[1]{>{\centering\arraybackslash}p{#1}}
\newcommand*{\dt}[1]{%
	\accentset{\mbox{\large\bfseries .}}{#1}}
\newcommand*{\ddt}[1]{%
	\accentset{\mbox{\large\bfseries ..}}{#1}}
\newcommand{\IN}{^{\mathcal{I}}}
\newcommand{\OUT}{^{\mathcal{O}}}

\begin{document}

\preprint{AIP/123-QED}

\title[Sessile Bubble]{Waves and jets on a sessile, incompressible bubble}

\title{Standing waves and jets on a sessile, incompressible bubble}

\author{Yashika Dhote}
\author{Anil Kumar}
\author{Lohit Kayal}
\author{Partha Sarathi Goswami}
\author{Ratul Dasgupta}
\email{dasgupta.ratul@iitb.ac.in}
\affiliation{
 Department of Chemical Engineering, IIT Bombay,
 Powai, Mumbai, India - 400 076
}%


\date{\today}

\begin{abstract}
We show numerically that large amplitude, \textit{shape deformations}, imposed on a spherical-cap, incompressible, sessile gas bubble pinned on a rigid wall can produce a sharp, wall-directed jet. For such a bubble filled with a permanent gas, the temporal spectrum for surface-tension driven, linearised perturbations has been studied recently in \citet{ding2022oscillations} in the potential flow limit. We reformulate this as an initial-value problem analogous in spirit to classical derivations in the inviscid limit by \citet{kelvin1890oscillations}, \citet{rayleigh1878instability} or by \citet{prosperetti1976viscous,prosperetti1981motion} for the viscous case. The first test of linear theory is reported here by distorting the shape of the pinned, spherical cap employing eigenmodes obtained theoretically, as the initial perturbation for our numerical simulations. It is seen that linearised predictions show good agreement with nonlinear simulations at small distortion amplitude producing standing waves. Beyond the linear regime as the shape distortions are made sufficiently large, we observe the formation of a dimple followed by a slender, wall-directed jet analogous to similar jets observed in other geometries from collapsing wave troughs\cite{farsoiya2017axisymmetric,kayal2022dimples}. This jet can eject with an instantaneous velocity exceeding nearly twenty times that predicted by linear theory. By projecting the shape of the bubble surface around the time instant of jet ejection, into the linearised eigenspectrum we show that the jet ejection coincides with the nonlinear spreading of energy into a large number of eigenmodes. We demonstrate that the velocity-field associated with the dimple plays a crucial role in evolving it into a jet and without which, the jet does not form. It is further shown that evolving the bubble shape containing a dimple and zero initial velocity-field via linear theory, does not lead to the formation of the jet. These conclusions accompanied by first principles analysis, provide insight into the experimental observations of \citet{prabowo2011surface} where similar jets were reported earlier. Our inferences also complement well-known results of \citet{naude1961mechanism} and \citet{plesset1971collapse} demonstrating that wall-directed jets can be generated from \textit{volume preserving}, shape deformation of a pinned bubble.
\end{abstract}

\maketitle

\newcommand{\mj}{{\mathrm{J}}}

\section{Introduction}\label{sec:intro} 		
Imagery of oscillating spherical interfaces bring to our mind pictures of wobbling rain-drops \cite{beard1989natural}, oscillating vapour bubbles \cite{prosperetti2017vapor}, shape distortions of gas bubbles in a glass of champagne wine \citep{liger2002effervescence} or a foamy ocean patch with its distinct acoustic signature due to oscillating, entrained, air bubbles \citep{deane2002scale}. Oscillations about spherical shapes arise not only at these daily-life length scales, but also at astrophysical scales where perturbations about the near spherical form of stars or planets have been of keen interest \citep{chandrasekhar1981hydrodynamic}. Amongst the early quantitative studies was the one by \citet{kelvin1890oscillations} who obtained the frequency for small amplitude, shape oscillations of a sphere (globe) or that by Dirichlet of finite amplitude oscillations of ellipsoids (section $382$ in \citet{lamb1993hydrodynamics}), the restoring force being self-gravitation in both (also see \citet{lamb1881oscillations}). Kelvin's analysis demonstrated that the disturbance velocity potential in the sphere, satisfies the simple harmonic oscillator equation (eqn. $98$ in \citet{kelvin1890oscillations}, page $385$). Some years prior to Kelvin's study, \citet{rayleigh1879capillary} had studied surface tension driven, \textit{shape oscillations} of a freely suspended, spherical, liquid droplet obtaining the corresponding dispersion relation in the small amplitude limit. The extension into three dimensions was done by \citet{lamb1993hydrodynamics} in 1932 employing the spherical harmonic $\mathbb{Y}_k^{(m)}(s,\psi) \equiv P^{(m)}_k(\cos(s))\exp\left[Im\psi\right]$ ($|m| \leq k$), $P^{(m)}_k(s)$ the associated Legendre function, $s$ being the polar angle and $\psi$ the azimuthal angle. Lamb's analysis also accounted for density of fluids on either sides of the interface with outer fluid density $\rho\OUT$, inner fluid density $\rho\IN$, surface tension $T$ and equilibrium radius $\hat{R}_0$ leading to the dispersion relation in eqn. \ref{eqn1}. One may easily obtain the natural frequencies for shape oscillations of a freely suspended droplet ($\rho\OUT << \rho\IN$) or a bubble ($\rho\OUT >> \rho\IN$) from this Rayleigh-Lamb relation (RL hereafter) 
\begin{eqnarray}
	\omega^{(k,m)}_{\text{RL}} =\left(\frac{T}{\hat{R}_0^3}\frac{k\left(k+1\right)\left(k-1\right)\left(k+2\right)}{\left(k+1\right)\rho\IN +\; k\rho\OUT}\right)^{1/2}, k=1,2,3\ldots \label{eqn1}.
\end{eqnarray}

In contradistinction to shape oscillations the study of volume oscillations characterised by the isotropic spherical harmonic $\mathbb{Y}_0^{(0)}$, of relevance to compressible bubbles with gas or vapour inside, also originated from \citet{rayleigh1917viii}. His study of the collapse of a spherical cavity led to the now well-known Rayleigh-Plesset equation \citep{plesset1977bubble}. Radial, volumetric oscillations can occur about the equilibrium, (stable) solution(s) to this equation and the frequency of these in the small amplitude limit was independently obtained by \citet{xvi1933musical} assuming adiabatic behaviour of the gas (Ch. $4$, section G in \citet{leal2007advanced} and \citet{chapman1971thermal}). Since these early pioneering theoretical studies, a large and fascinating scientific literature on the physics of bubbles (both gas and vapour) has developed \cite{yang2003transient,prosperetti1982generalization,prosperetti1978linear,hao1999dynamics,ory2000growth}. Important advances have been reviewed periodically over several articles e.g. see \citet{plesset1977bubble,blake1987cavitation,feng1997nonlinear,lauterborn2010physics,prosperetti2017vapor}.

Our aim here is to understand the linear and nonlinear, surface tension driven, \textit{shape deformations} of a pinned, sessile, spherical-cap, air bubble of millimetric dimensions placed on a rigid, flat substrate and surrounded by quiescent water. Due to our focus on the millimetric size range, this being below the capillary length ($\approx 2.7$ mm) as well as the visco-capillary length scale for air-water ($\approx 0.01$ micron \cite{duchemin2002jet}), gravity and viscosity are neglected in a first approximation and we return to a discussion of this at the end of the study. As surface tension is the only restoring force, we approximate the sessile bubble shape as a truncated sphere (spherical-cap). Sessile gas bubbles can appear in several applications spanning length scales: at nano-metric scales, they appear as gas pockets on liquid immersed substrates, see fig. $1$ and $2$ in \citet{lohse2015surface} for AFM images as well as fig. $41$ in the same study, for much larger ones attached to a goblet of water. At micron length scales, such bubbles may undesirably appear on catalyst surfaces and techniques aimed at their removal, such as coalescence have been studied experimentally in \citet{lv2021self}. While the Rayleigh-Lamb spectrum for shape oscillations or the Minnaert frequency for volume oscillations of a free bubble have been known for more than a century, the modification to these when the bubble is attached to a substrate at arbitrary contact angle, have been presented relatively recently. The modification to the Minnaert frequency due to a bubble on a substrate was obtained in \citet{blue1967resonance,maksimov2005volume} for natural contact angle conditions and for pinned as well as natural contact line boundary conditions recently in \citet{ding2022oscillations} (also see references therein for earlier work on hemispherical bubble oscillations). The linear spectrum for a pinned bubble is tested here, employing modal initial conditions obtained from theoretical analysis. In addition, we also foray into the nonlinear regime where a novel observation is that of wall-directed jets which arise from large amplitude modal deformations, reminiscent of such jets obtained in modal deformations in other geometries \cite{farsoiya2017axisymmetric,basak_farsoiya_dasgupta_2021,kayal2022dimples}. 

As we simulate the shape distortion modes only, we provide justification below for the neglect of the volume mode, particularly in the large distortion regime. For this, it is instructive to contrast the shape modes against the volume mode (also sometimes called the `breathing' mode). Consider first a freely suspended, air bubble (hereafter referred to as a free bubble), a rough estimate of whose natural  breathing mode frequency (without considering surface tension) is given by the formula \citep{longuet1989monopole1} $\left(\dfrac{3\gamma \hat{P}_b}{\rho\OUT \hat{R}_0^2}\right)^{1/2}$ where $\gamma$ is the ratio of specific heats ($1.4$) of air and $\rho\OUT = 1$ gm/cm$^3$ is the density of water at STP conditions. For a $1$ mm air bubble in water at atmospheric pressure $\hat{P}_b$, this frequency is $\approx 2\times 10^4$ rad/s. In contrast, the frequency of the lowest shape mode ($k=2$) of the same bubble is given by the RL dispersion relation $\left(\dfrac{T}{\rho\OUT \hat{R}_0^3}(k+1)(k-1)(k+2)\right)^{1/2}\approx 10^3$ rad/s. This large separation in frequency between the lowest shape mode and the volume mode of oscillation for a free bubble, implies that shape oscillations can persist long after the faster volume oscillations have damped out. While in the linear regime, only modes that are excited initially can persist, at larger distortion amplitude (that we also study here) this is not necessarily so. At large amplitude, exciting only the shape modes initially may generate the volume mode through nonlinear interactions. The two-part study by \citet{longuet1989monopole1,longuet1989monopole2} investigated this for a free, air bubble showing that this nonlinear transfer of energy can be very pronounced when twice the natural frequency of any given shape mode, matches the volume oscillation frequency. This is due to second harmonic resonance well-known in the theory of surface waves \citep{wilton1915lxxii}. For a $1$ mm air bubble, in order for this transfer of energy from a shape  to a volume mode to occur efficiently, however requires excitation of a shape mode with a high mode index see table $1$ in \citet{longuet1989monopole1}. Also see \citet{williams1991resonant,longuet1991resonance} for debate on the relevance of this mechanism and the role of viscosity. 

These aforementioned arguments for a free bubble also extend qualitatively to pinned, sessile ones with radius $\hat{R}_0$ and contact angle $\alpha$. The results by \citet{ding2022oscillations} (their fig. $3$, note their definition of contact angle differs from ours by a shift of $\pi$), imply that for sufficiently high equilibrium bubble pressure, a sessile bubble is stable against collapse via the breathing mode ($k=0,m=0$) at all values of contact angle $\alpha$, for pinned contact line conditions. Their non-dimensional pressure $\Pi \equiv \dfrac{3\gamma \hat{P}_b \hat{R}_0\sin(\alpha)}{T}$ may be interpreted as $\sin(\alpha)$ times the ratio of square of the frequency of the volume mode $\left(\dfrac{3\gamma \hat{P}_b}{\rho\OUT \hat{R}_0^2}\right)^{1/2}$ to the square of  the shape mode frequency scale $\left(\dfrac{T}{\rho\OUT \hat{R}_0^3}\right)^{1/2}$. Fig. $6$ in \citet{ding2022oscillations} for $\alpha=110^{\circ}$ indicates that at sufficiently large value of $\Pi$, the ratio of the frequency of the lowest pinned, shape mode and the pinned volume mode do not satisfy the second harmonic resonance condition outlined earlier. Thus if such a bubble is distorted without altering its volume, we expect the energy transfer (to the breathing mode) to be insignificant at short times. 

The physical situation that we simulate in this study, corresponds to the large $\Pi$ regime and thus one may qualitatively extrapolate the above conclusions to contact angles  $< \pi/2$ (representative of simulations presented in this study) and ignore energy transfer to the volume mode in a first  approximation. We also note that in the presence of acoustic forcing, the volume mode is excited at the forcing frequency. In such a case, the shape modes can be parametrically excited by the volume modes and this has close analogy to subharmonic Faraday waves, due to the underlying Mathieu equation which governs both phenomena. Starting from the seminal work by \citet{plesset1954stability,benjamin1958excitation}, this has been studied extensively for free bubbles in \citet{mei1991parametric} with continuing investigations more recently \cite{doinikov2004translational,guedra2018bubble}. 

In contrast to these studies, here our focus is on the \textit{natural} oscillations of a sessile bubble distorted initially and without any externally imposed acoustic forcing. Of relevance here, the well-known results of \citet{naude1961mechanism} demonstrated that when a hemispherical cavity collapses, a wall-directed jet is ejected. In this study, we will demonstrate that such jets can also be produced from large amplitude, shape deformations even when the (sessile) bubble preserves its volume and we will return to a discussion of this interesting point at the end. Our numerical study of shape deformations in the nonlinear regime, may also be viewed as being complementary to the one by \citet{tsamopoulos1983nonlinear} where too, the nonlinear behaviour of shape modes for a free bubble was treated analytically, ignoring the breathing mode. 

\begin{figure}
	\centering
	\subfloat[Jet from an air bubble in water]{\includegraphics[scale=0.13]{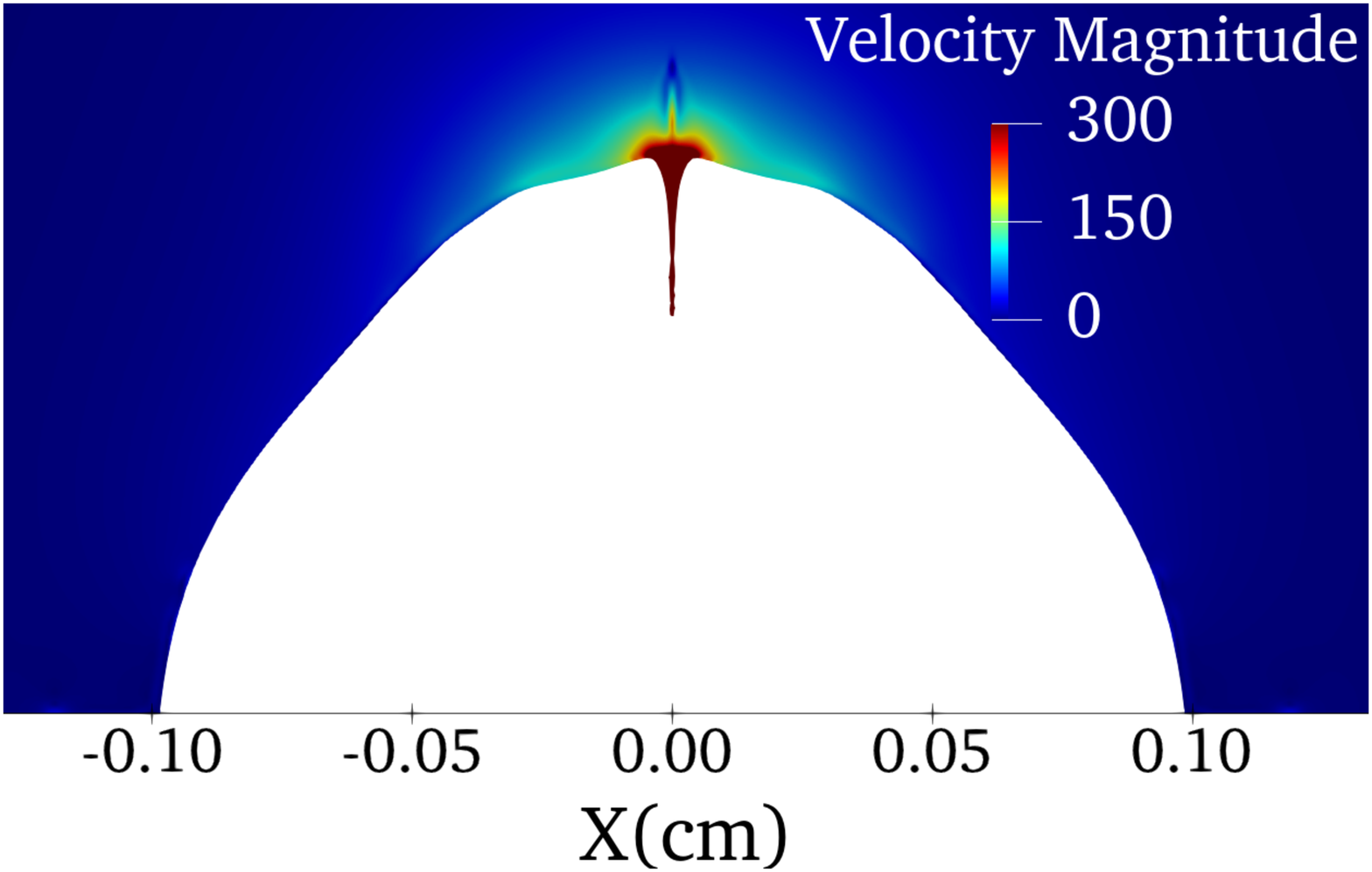}\quad}
	\subfloat[Magnified view of the left panel]{\includegraphics[scale=0.16]{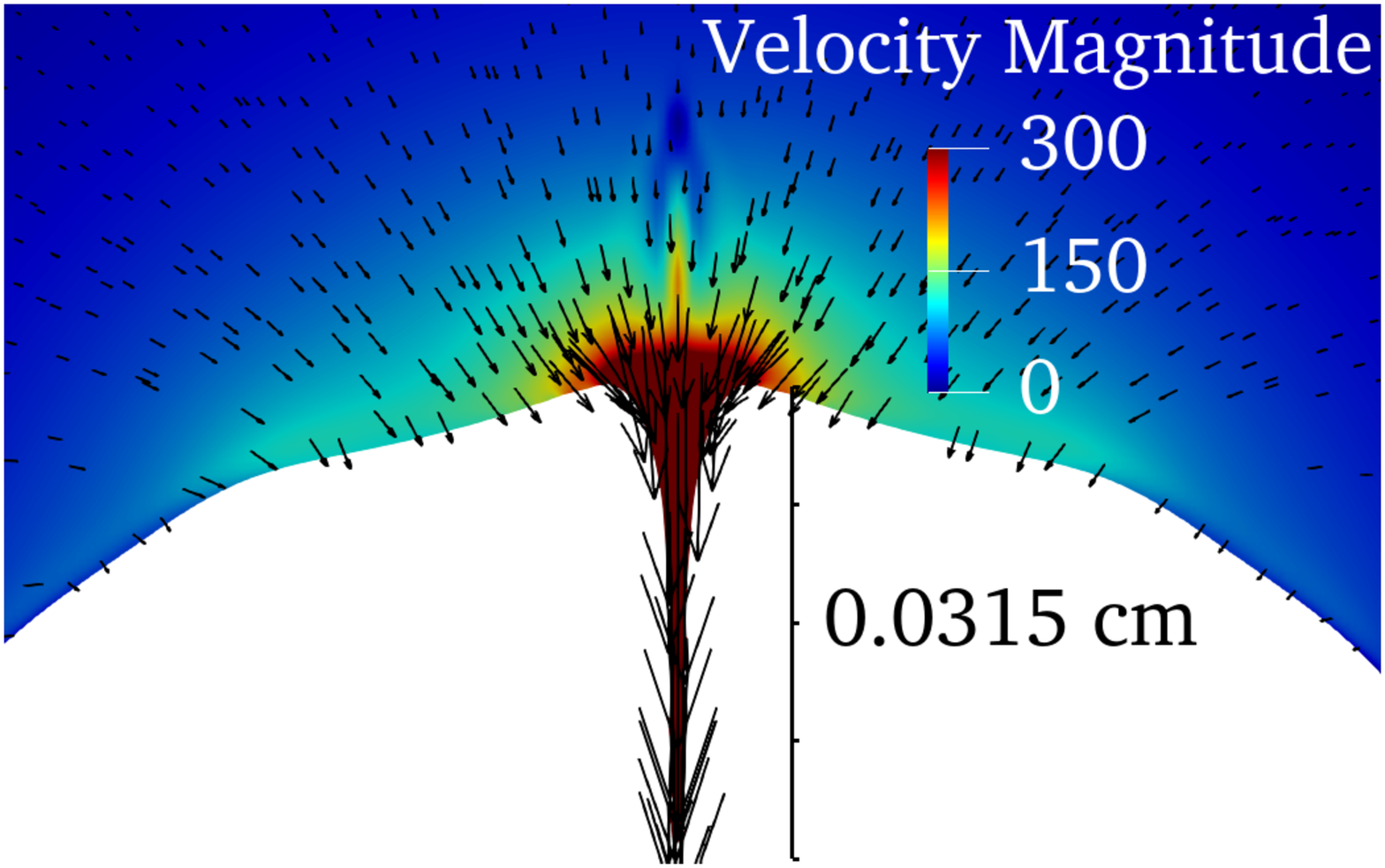}}
	\caption{(a) Jet (in deep brown) with tip drops generated from large amplitude deformation of a spherical-cap, air bubble in water and pinned at the contact line. The contours on the liquid side are for total velocity (in $\text{cm/s}$). Note the formation of an jet directed opposite to the wall, in the liquid. The simulation also solves for air inside the bubble indicated in white, although these velocity contours are not indicated in the figure. (b) Magnified image of the left panel showing the (total) velocity as arrows }
	\label{fig0}
\end{figure}

To our knowledge, our current results represent the first validation of the linear predictions for the pinned shape modes. A novel finding beyond linear predictions is that for large amplitude surface distortions, a slender wall-directed jet can form at the symmetry axis. A preview of this result is shown in fig. \ref{fig0}. Panel (a) depicts a jet obtained from a $1$ mm radius spherical cap. The surface of this cap has been distorted initially at large amplitude using the lowest mode obtained from theory (section II). A thin, slender jet subsequently emerges at the symmetry axis towards the wall. Panel (b) presents a closer view of this. Note the presence of a counter jet (deep blue contour) in both images, directed opposite to the wall and expected from momentum conservation. To place our study in a wider context, we summarise in the next section some of the important studies on jet formation from bubble collapse that have appeared in literature.
\section{Literature survey: jet formation from bubbles}
\begin{figure}
	\centering
	\includegraphics[scale=0.3]{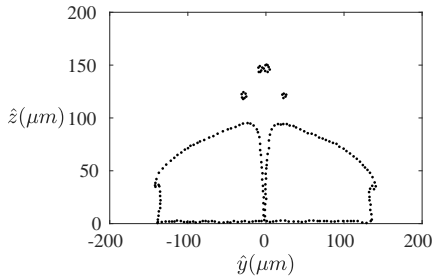}
	\caption{Jet from a collapsing, hemispherical, compressible gas bubble at a wall obtained from numerical simulations reported in \citet{lechner2020jet}. The shape of the bubble is extracted from their fig. $8$, panel (b) at $\hat{t}=114.46\;\mu$s.}
	\label{fig1}
\end{figure}  
Fig. \ref{fig1} represents a jet formed in numerical simulations of \citet{lechner2020jet} (data extracted from their study) from a collapsing, hemi-spherical, \textit{compressible} gas bubble ($\approx 25$ microns in initial diameter) placed near a wall. In these simulations, the liquid and gas are both modelled as being compressible; the gas is ideal following an adiabatic process while the liquid follows the Tait equation of state \citep{lechner2020jet}. This figure represents the late stages of bubble collapse when the jet at the symmetry axis strikes the wall, see their\cite{lechner2020jet} fig. 8 panel (b). In contrast to fig. \ref{fig1} taken from \citet{lechner2020jet}, fig. \ref{fig0} representing our simulations treating the gas inside as well as the liquid outside the bubble as incompressible. Consequently, the bubble volume in these simulations remains invariant with time unlike \citet{lechner2020jet}.

Formation of jets from a collapsing vapour bubble when close to a wall or attached to it has been long known. The experiments of \citet{naude1961mechanism} reported observation of a jet generated from the collapse of a spark generated, cavitating vapour bubble of millimetre size range, close to a solid boundary. They also formulated a second order perturbative theory in the potential flow limit, based on shape distortions of a collapsing hemispherical cavity, which predicted the generation of this wall-directed jet (see fig. $6$ in \citet{naude1961mechanism}). In contrast, the subsequent numerical simulations also in the potential flow limit by \citet{plesset1971collapse} of a collapsing, \textit{spherical} cavity without any imposed shape distortion, demonstrated clearly the formation of this jet reaching speeds upto $170$ m/s. Notably they concluded that compressibility effects inside the cavity (as well as in the liquid outside) during jet formation could be neglected entirely. These computational results received strong experimental support from very high speed ($\sim 900,000$ frames per second) photographic measurements reported in \citet{lauterborn1975experimental} of a laser generated, collapsing water vapour bubble. Fig. $6$ in \citet{lauterborn1975experimental} compares their experimentally observed bubble shape with that predicted numerically by \citet{plesset1971collapse}, demonstrating very good agreement. 

Post these early studies on jet formation, there have been several subsequent investigations on vapour bubble collapse near a solid boundary, focussed particularly on the effect of shear on bubble collapse and the jet, which can be important in engineering scenarios; we refer the reader to the review by \citet{blake1987cavitation} for a summary of literature upto $1987$ and that by \citet{feng1997nonlinear}, third paragraph, page $202$ where a brief summary of the effect of shear on cavitation bubble collapse and jet formation is discussed. The Direct Numerical Simulations (DNS) of \citet{popinet2002bubble} employing a free-surface code, investigated jet formation from a collapsing vapour bubble, for bubbles far smaller compared to earlier studies. Their bubbles were in the micron size range where viscous effects are dominant. These authors demonstrated that for such small bubbles, compressibility (and thermal) effects inside the cavity can be important with jet formation nearly suppressed, when the viscosity of the liquid was increased beyond a threshold (see fig. $11$ in their study). Very recently, the DNS and theoretical investigation by \citet{saini2022dynamics} has solved the mass, momentum and energy equations numerically for bubble collapse at arbitrary contact angles. Note that their definition of contact angle differs from ours ($\alpha$) by a shift of $\pi$ and the authors report observing wall-directed jets only for $\alpha < \pi/2$. In concluding this literature survey, we emphasize the study by \citet{lechner2020jet}, particularly their Appendix C where very interesting physical mechanisms of flow focussing have been discussed as well as the recent experimental work by \citet{rossello2023bubble} where a reasonably extensive bibliography on  bubble collapse is presented. We also refer the reader to fig. $2$ (image on the right between $(7)$ and $(9)$) of the experimental study by \citet{biasiori2023synchrotron} which depicts such a jet. Also, of particular interest is the previous study by \citet{prabowo2011surface} which reported the formation of a jet from a wall-attached bubble (their fig. $5$), when excitation of several shape modes conincides with the generation of a wall-directed jet. We will return to a discussion of this later in the study. 

The remainder of this study is organised as follows: in the next section, we discuss the inviscid, temporal spectrum of linearised, surface tension driven shape oscillations of a sessile bubble at arbitrary contact angle. The methodology is derived from \citet{bostwick2014dynamics} but formulated as an initial value problem as opposed to normal mode analysis. Linearised predictions are tested against numerical  simulations next obtaining very good agreement. At larger distortion amplitude, we observe jets in our simulations and their generation and evolution is discussed in detail. We conclude by summarising the significance of our results and scope of further work.

\section{Linear theory}\label{sec:linTh}
In this section, the linear theory of small-amplitude perturbations of a sessile bubble of radius $\hat{R}_0$ and contact angle $\alpha$ in the base state, pinned at its contact line is discussed in the potential flow limit. The linearised, inviscid spectrum for this has been presented recently in \citet{ding2022oscillations} and we re-formulate this as an initial value problem (IVP hereafter) here. The motivation to do this comes from the IVP approach to study shape oscillations of a free, spherical bubble. This was first explored in a series of seminal studies by \citet{prosperetti1977viscous,prosperetti1980free,prosperetti1980normal,hao1999effect} and this problem has been re-visited recently by \citet{farsoiya2020azimuthal}, in the context of shape modes of a free, \textit{cylindrical} bubble. In these studies, it has been shown that for a free bubble (cylindrical or spherical) with viscous liquid in the bubble exterior treated as being radially unbounded, the spectrum has a discrete and \textit{a continuous part}, the contribution of the latter being relatively easily captured in the IVP approach in contrast to the normal mode one. Initial shape deformations of the bubble can excite the discrete modes as well as the continuous spectrum of vorticity modes (see discussion around page $337$ in \citet{prosperetti1980free}, section $4$ in \citet{prosperetti1980normal} and section $3.2$ in \citet{farsoiya2020azimuthal}). This excitation manifests as a second order, \textit{integro-differential equation} in time governing the amplitude $\hat{a}(\hat{t})$ of the linearised, standing, shape modes on the bubble surface. The memory term in such equations come from the excitation of these aforementioned vorticity modes and represent the damping in the unsteady boundary layer, at the bubble surface. On the other hand, the first order (in time) damping term in these equations arise from dissipation in the bulk of the fluid and this may be estimated by calculating dissipation on the potential flow \citep{lamb1993hydrodynamics,joseph1994potential}.

In the inviscid, potential flow limit (of interest in our current study) the aforementioned vorticity modes reduce to zero frequency modes \cite{Ramana2023} and the damped, integro-differential equations \citep{prosperetti1980free,farsoiya2020azimuthal} reduce to simple-harmonic oscillator equations. In the analysis that follows, we derive this simple harmonic oscillator equation governing shape modes on a sessile bubble analogous to the derivations by \citet{kelvin1890oscillations} or \citet{rayleigh1879capillary} (see eqn. $39$ in Appendix II of \citet{rayleigh1879capillary}) for free drops or liquid cylinders. Apart from the novelty of the analytical derivation which follows, a practical motivation for solving the problem is also to obtain the eigenmodes beyond the first few, whose shapes have been presented in \citet{ding2022oscillations}. It will be seen in our study that for large amplitude distortion, significant energy transfer occurs into these higher modes in the spectrum and knowledge of these are necessary for projecting the instantaneous shape of the interface as well as to resolve the small scale features at the bubble surface which precede the formation of the jet. In addition, it is also expected that the analytical framework presented next will serve as the starting point for future linear (and weakly nonlinear) solutions to the \textit{viscous} IVP for a pinned sessile bubble.
\begin{figure}
	\centering
	\includegraphics[scale=0.35]{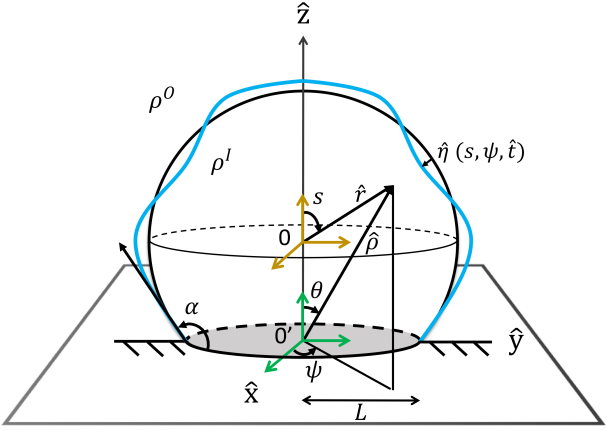}\label{fig2a}
	\caption{Pinned spherical cap of radius $\hat{R}_0$, contact angle $\alpha\in \left[0,\pi\right]$, partitioning immiscible fluids of density $\rho^{I}$ and $\rho^{0}$ and interfacial tension $T$. Following \citet{bostwick2014dynamics} we employ two spherical coordinate systems centred at the sphere (yellow) with origin $O$ and at the centre of the circular footprint (grey) of the spherical gap (green) with origin $O^{'}$. Referred to the axes in yellow, the coordinates of any point in the bubble is ($\hat{r},s,\psi$). The same point has coordinates ($\hat{\rho},\theta,\psi$) referred to the axes in green. The line in blue indicates a (volume preserving) perturbation to the equilibrium bubble shape, respecting pinned conditions at the CL. The variable $\hat{\eta}(s,\psi,\hat{t})$ measures the local shape deviation from sphericity. We assume $\rho\OUT >> \rho\IN$ and neglect the dynamics of the gas inside.}
	\label{fig2}
\end{figure}
\subsection{Initial Value Problem (IVP) formulation}
Our inviscid IVP analysis, is formulated in the potential flow approximation employing the novel, two coordinate systems proposed by \citet{bostwick2014dynamics}, see fig. \ref{fig2}. The inverse operator formalism (Green function) for solving the resultant generalised eigenvalue problem, was first formulated by \citet{strani1984free,strani1988viscous} for analysing natural oscillations of a spherical liquid cap supported on a spherical bowl and suitably reformulated by \citet{bostwick2014dynamics} in their insightful study of the inviscid temporal spectrum of a sessile, spherical cap on a flat substrate. We work at a level of approximation where the motion at the bubble surface and in the liquid outside is accounted for, but pressure or velocity fluctuations in the gas inside are ignored. The justification for this is the large density ratio for air-water bubbles satisfying $\rho\OUT >> \rho\IN$. In addition, we neglect compressibility effects, thus allowing for only volume preserving perturbations.

As seen in figure \ref{fig2}, in the base (equilibrium) state, the bubble shape is a spherical cap of radius $\hat{R}_0$ with contact angle $\alpha$. Gravity is neglected and we self-consistently restrict ourselves to spherical caps of radius smaller than the capillary length for air-water. The spherical cap base state is assumed to entrap a quiescent \& permanent gas of density $\rho\IN$ inside and is surrounded by unbounded, quiescent fluid of density $\rho\OUT$ outside. In the base state (indicated with subscript `b'),
\begin{eqnarray}
	\hat{\phi}_b\OUT =0,\quad \hat{\eta}(s,\psi,\hat{t})=0,\quad \hat{P}\OUT_b = -\frac{2T}{\hat{R}_0}, \label{eqn2_1}
\end{eqnarray}
where the pressure in the gas inside the bubble has been set to be zero at all time and $T$ is surface tension. Following \citet{bostwick2014dynamics}, we non-dimensionalise all perturbation variables as (those with hat are dimensional),
\begin{eqnarray}
	\phi\OUT \equiv \frac{\hat{\phi}\OUT}{\frac{L^2}{\left(\frac{\rho\OUT L^3}{T}\right)^{1/2}}} \; t \equiv \frac{\hat{t}}{\left(\frac{\rho\OUT L^3}{T}\right)^{1/2}},\; \nonumber\\
	\eta \equiv \frac{\hat{\eta}}{L},\; r\equiv \frac{\hat{r}}{L},\rho\equiv \frac{\hat{\rho}}{L},\; \bm{\nabla} \equiv L\hat{\bm{\nabla}}  \label{eqn2_2}
\end{eqnarray}
where from fig. \ref{fig2} $L=\hat{R}_0\sin(\alpha)$. Here $\phi\OUT$ and $\eta$ represent the perturbation velocity potential in the liquid outside and the perturbed free surface of the bubble respectively. Note that $\hat{\rho}$ (or $\rho$) in eqns. \ref{eqn2_2} represents length (see fig. \ref{fig2}), distinct from $\rho\OUT$ which is used for the density of the fluid outside the bubble. 

We follow the mixed notation of \citet{bostwick2014dynamics} where the independent variables for the free-surface deformation $\eta$ are chosen to be $(s,\psi, t)$ while that for $\phi\OUT$ are chosen to be $(\rho,\theta,\psi, t)$. This has the advantage that the implementation of the no-penetration boundary condition becomes easy due to the substrate being at $\theta=\pi/2$, independent of the bubble contact angle $\alpha$. We note the following relations from the geometry of fig. \ref{fig2} allowing us to express $\rho$ and $\theta$ in terms of $r$ and $s$
\begin{eqnarray}
	\rho(r,s) &=& \left[\cot^2\left(\alpha\right) + r^2 - 2r\cot(\alpha)\cos(s)\right]^{1/2},\nonumber \\
	\text{and}\quad \theta(r,s) &=& \cos^{-1}\left[\frac{r\cos(s) - \cot(\alpha)}{\rho(r,s)}\right]. \label{eqn2_3}
\end{eqnarray}
Within the linear approximation, the following equations and boundary conditions govern perturbations to the shape of the spherical cap ($\eta$) and the perturbation velocity potential $\phi$ in the liquid outside (we drop the superscript on $\phi\OUT$ henceforth):
\begin{subequations}\label{eqn2_4}
	\begin{align}
		& \nabla^2\phi = 0, \tag{\theequation a} \\
		&\frac{\partial\eta}{\partial t} = \left(\frac{\partial\phi}{\partial r }\right)_{r=\csc(\alpha)}, \tag{\theequation b}\\
		& \left(\bm{\nabla}\phi\cdot\mathbf{e}_z\right)_{\theta=\pi/2} = 0\quad \text{for} \; \rho \in \left[1,\infty\right),\; \psi \in \left[0,2\pi\right]\tag{\theequation c} \\
		&\left(\frac{\partial\phi}{\partial t} \right)_{r=\csc(\alpha)}= -\sin^2\left(\alpha\right)\left(2 + \nabla_{||}^2\right)\eta, \tag{\theequation d}\\
		& \texttt{Pinned CL:}\quad \eta(s=\alpha,\psi,t) = 0,
		\tag{\theequation e} \\
		&\eta\left(s=0,\psi,t\right) \rightarrow \text{finite}, \tag{\theequation f}\\ 
		&
		\int_0^{\alpha}ds\sin(s)\int_{\psi=0}^{2\pi}d\psi\;\eta(s,\psi,t) = 0, \tag{\theequation g}
	\end{align}
\end{subequations}
where $\nabla^2_{||} \equiv \frac{1}{\sin(s)}\frac{\partial}{\partial s}\left[\sin(s)\frac{\partial}{\partial s}\right] + \frac{1}{\sin^2(s)}\frac{\partial^2}{\partial\psi^2}$. Eqns. \ref{eqn2_4} a is the Laplace equation governing velocity perturbations in the liquid (outside the bubble). Eqns. \ref{eqn2_4} b,c,d are the linearised kinematic boundary condition, the no-penetration condition at the substrate and the pressure jump due to surface tension at the linearised interface respectively. Eqn. \ref{eqn2_4} e is the pinned condition at the contact line (CL).  Eqn. \ref{eqn2_4} g represents volume conserving perturbations in the linearised limit. 

Consistent with the linear approximation, all boundary conditions are applied at the unperturbed spherical cap, expressed in non-dimensional form by $r=\csc\left(\alpha\right)$. Note the usage of mixed notation in eqn. \ref{eqn2_4} b. Using equations \ref{eqn2_3}, we may express the variable $\phi(\rho,\theta,\psi,t)$ as $\phi(\rho(r,s),\theta(r,s),\psi,t)$ and evaluate the derivative $\frac{\partial\phi}{\partial r}|_{r=\csc\left(\alpha\right)}$. Similarly, the dot product in \ref{eqn2_4} c is evaluated in  $\left(\rho,\theta,\psi\right)$ coordinates. Following \citet{prosperetti1976viscous}, we look for standing wave solutions to eqns. \ref{eqn2_4} and set
\begin{subequations}\label{eqn2_5}
	\begin{align}
		&\eta(s,\psi,t) = a(t)y(s)\cos(m\psi), \tag{\theequation a}\\
		&\phi(\rho,\theta,\psi,t) = \dt{a}(t)\;\Phi(\rho,\theta)\cos(m\psi),\;\; m \in \mathbb{Z}^{+}\tag{\theequation b}
	\end{align}
\end{subequations}
where $\dt{a}(t) \equiv \frac{da}{dt}$, the usage of $\dt{a}$ in eqn. \ref{eqn2_5}b being motivated by the kinematic boundary condition eqn. \ref{eqn2_4} b. It is important to remark here that the time dependence assumed in \ref{eqn2_5} is arbitrary and we specifically \textit{do not} assume the time dependence to be exponential apriori, in distinction to normal mode analysis\cite{ding2022oscillations}. As a consequence, we will \textit{derive an equation} governing $a(t)$ which needs to be solved with specific initial conditions. This is to be contrasted against the normal mode approach \cite{ding2022oscillations} where an exponential time dependence is assumed apriori and initial conditions do not appear explicitly in the analysis. 

Due to the mixed notation usage, the Laplace eqn. \ref{eqn2_4}a is to be written with $\rho,\theta,\psi$ as independent variables rather than $r,s,\psi$. Using eqns. \ref{eqn2_5} a,b into eqn. \ref{eqn2_4} b and  \ref{eqn2_4} d, we obtain
\begin{eqnarray}
&\ddt{a}(t)\Phi|_{r=\csc(\alpha)} = -a(t)\sin^2(\alpha)\Bigg(\left(1-x^2\right)\frac{d^2}{dx^2} - 2x\frac{d}{dx}  \nonumber \\
& 
+ \left(2 - \frac{m^2}{1-x^2}\right)\Bigg)\left(\dfrac{\partial\Phi}{\partial r}\right)_{r=\csc(\alpha)}, \label{eqn2_6}
\end{eqnarray}
which may be written as
\begin{eqnarray}
&&\left(\ddt{a}(t)\;\mathbf{M}  +a(t)\sin^2(\alpha)\;\mathbf{K}\right)\cdot \left(\frac{\partial\Phi}{\partial r}\right)_{r=\csc(\alpha)}\hspace{-10mm}(x) = \mathbf{0},    \label{eqn2_7}
\end{eqnarray}
where $\mathbf{M} \equiv \int dr$ and $\mathbf{K} \equiv \left(1-x^2\right)\frac{d^2}{dx^2} - 2x\frac{d}{dx} + \left(2 - \frac{m^2}{1-x^2}\right)$\cite{ding2022oscillations}. It is clear from the structure of eqn. \ref{eqn2_7} that in order for space and time to be separable (necessary for obtaining standing wave solutions as assumed earlier in eqn. \ref{eqn2_5} a,b), the linear operators $\mathbf{M}$ and $\mathbf{K}$ must satisfy the eigenvalue problem
\begin{eqnarray}
\mathbf{K}\cdot\left(\frac{\partial\Phi}{\partial r}\right)_{r=\csc(\alpha)}\hspace{-10mm}(x)\quad  = \quad  \lambda^2_{j,m}\;\mathbf{M}\cdot\left(\frac{\partial\Phi}{\partial r}\right)_{r=\csc(\alpha)}\hspace{-10mm}(x)\;,  \label{eqn2_8}
\end{eqnarray} 
in turn implying that $a(t)$ in eqn. \ref{eqn2_7} satisfies the \textit{simple harmonic oscillator equation}
\begin{eqnarray}
\ddt{a}_{(j,m)}(t) +\lambda_{(j,m)}^2\sin^2(\alpha)a_{(j,m)}(t) = 0. \label{eqn2_9}
\end{eqnarray}
The kinematic boundary condition \ref{eqn2_4} b, allows us to relate the shape of the eigenmodes viz. $y_{(j,m)}(s)$ from $\left(\frac{\partial\Phi}{\partial r}\right)_{r=\csc(\alpha)}^{(j,m)}$ while the eigenvalue $\lambda_{(j,m)}$ are related to the frequency of the standing waves obtained from this analysis. 

We thus accomplish our stated objective of deriving a simple harmonic oscillator equation for $a(t)$ in an IVP framework. Note that we have attached subscripts $(j,m)$ to the eigenvalues in \ref{eqn2_8}. The index $j=1,2,3\ldots$ represents the eigenvalues in ascending order while $m=0,1,2,\ldots$ from eqn. \ref{eqn2_5}. Eqn. \ref{eqn2_8} is  the same eigenvalue problem that was earlier obtained by \citet{ding2022oscillations} albeit via normal mode analysis. As demonstrated by \citet{prosperetti1976viscous},\citet{prosperetti1980free},\citet{farsoiya2017axisymmetric} and \citet{farsoiya2020azimuthal}, the IVP approach is more rigorous, particularly in the viscous regime where due to the radial unboundedness of the domain, one also expects a continuous spectrum. We note that the vorticity modes which are present even in the inviscid problem \cite{Ramana2023}, however do not evolve in time when viscosity is set to zero and thus are not reflected in equation \ref{eqn2_9}.

Note in particular, that eqn. \ref{eqn2_8} is an eigenvalue problem with an integro-differential structure \cite{bostwick2014dynamics,ding2022oscillations} (as $\mathbf{M}$ is an integral operator while $\mathbf{K}$ is a differential operator). It is also worthy of emphasis that this integro-differential structure of the eigenvalue problem in \ref{eqn2_8} governing shape modes, is \textit{not} unique to a sessile bubble but appears also for a free bubble as eqn. \ref{eqn2_8} remains applicable to both cases. This is easily demonstrable analytically as follows: recall that for a free bubble, the second coordinate system in fig. \ref{fig2} is not necessary and the independent variables are $\left(r,s,\psi\right)$ for all dependent variables. Thus $\Phi(r,s)$ in eqn. \ref{eqn2_5} b represents non-diverging solutions to the Laplace equation for $\csc(\alpha) \leq  r < \infty$ which are of the form $\Phi_{(j,m)}(r,s)=r^{-(j+1)}\mathbb{P}_{j}^{(m)}(s)$ for $j=1,2\ldots$ (shape modes only) and $ |m| \leq j$. It is straightforward to substitute this into the generalised eigenvalue problem \ref{eqn2_8} and compare the resultant equation with the associated Legendre differential equation governing $\mathbb{P}_{j}^{(m)}(s)$ \citep{ALP}. From this one may deduce that the eigenvalues for a free bubble are given by the formula $\lambda_{(j,m)}^2 = (j-1)(j+1)(j+2)$. This is simply the non-dimensional version of the RL dispersion relation eqn. \ref{eqn1} in the bubble limit ($\rho\OUT>>\rho\IN$) discussed at the introduction. 

For a sessile bubble on the other hand, the eigenvalue  problem represented by eqn. \ref{eqn2_8} has to be solved subject to the constraints that the no-penetration condition in eqn. \ref{eqn2_4} c, the pinning condition in \ref{eqn2_4}e and the volume conservation condition in eqn. \ref{eqn2_4} g be explicitly satisfied. For a sessile bubble, the independent variables for $\Phi$ are $\Phi(\rho,\theta)$ and the Laplace eqn. \ref{eqn2_4} a with these as independent variables, admits solutions (which do not diverge at $\rho\rightarrow \infty$) of the form $\rho^{-(j+1)}\mathbb{P}_{j}^{(m)}\left(\cos(\theta)\right)$ for $j\in \mathbb{Z}^{+}$ where $\mathbb{P}$. We use these solutions as basis functions with unknown coefficients, employing the same procedure as laid out in \citet{ding2022oscillations}. This is a numerical procedure and explicitly satisfies eqns. \ref{eqn2_4} c, \ref{eqn2_4}e and \ref{eqn2_4} g. In implementation, we choose to solve the inverse operator version of eqn. \ref{eqn2_8} viz.  $\left(\frac{\partial\Phi}{\partial r}\right)_{r=\csc(\alpha)}\hspace{-10mm}(x)\quad  = \quad  \lambda^2_{j,m}\;\mathbf{K}^{-1}\cdot\;\mathbf{M}\cdot\left(\frac{\partial\Phi}{\partial r}\right)_{r=\csc(\alpha)}$\cite{ding2022oscillations} expressed using the Green function of the operator $\mathbf{K}$ \cite{ding2022oscillations}
\begin{eqnarray}
&&\left(\frac{\partial\Phi_{(j,m)}}{\partial r}\right)_{r=\csc(\alpha)}\hspace{-10mm}\left(x\right)= \lambda_{(j,m)}^2\int_{b}^{1}G(x,y;m)\Phi_{(j,m)}(r=\csc(\alpha),y)dy, \nonumber\\
&&\text{where}\quad b \equiv \cos(\alpha), \label{eqn2_11}
\end{eqnarray}
by expansion in the aforementioned basis functions and taking suitable inner products. The analytical expression for the Green function corresponding to $\mathbf{K}^{-1}$ was provided earlier in \citet{ding2022oscillations} and the detailed derivation for this, subject to pinned boundary condition \ref{eqn2_4} f, is provided in Appendix 1 of our study. 

As details of the numerical procedure for solving the eigenvalue problem \ref{eqn2_8} has already been explained in \citet{ding2022oscillations}, we refer the reader to their study for more details. In our current study, we require access to a large number of modes in the spectrum (for calculating energy transfer to other modes in the nonlinear regime, discussed later), we have calculated the derivative $\left(\frac{\partial\Phi}{\partial r}\right)_{r=\csc(\alpha)}$ and the integrals arising from the inner products numerically. While analytical expressions for calculating $\left(\frac{\partial\Phi}{\partial r}\right)_{r=\csc(\alpha)}$ are also available to do this, the expressions for derivatives of $\mathbb{P}_j^{(m)}\left(\cos(\theta)\right)$ become increasingly lengthier for $j > 10$ and thus symbolic packages have been found to become prohibitively slow in evaluating these. In contrast, numerical integration and differentiation are relatively much faster. It has been checked that the eigenmodes obtained numerically versus those obtained using analytical expressions for $\left(\frac{\partial\Phi}{\partial r}\right)_{r=\csc(\alpha)}$, agree well with each other.

\begin{figure}
	\centering
	\subfloat[Non-dimensional frequency $\lambda_{(k,m)}$ versus $\alpha$]{\includegraphics[scale=0.21]{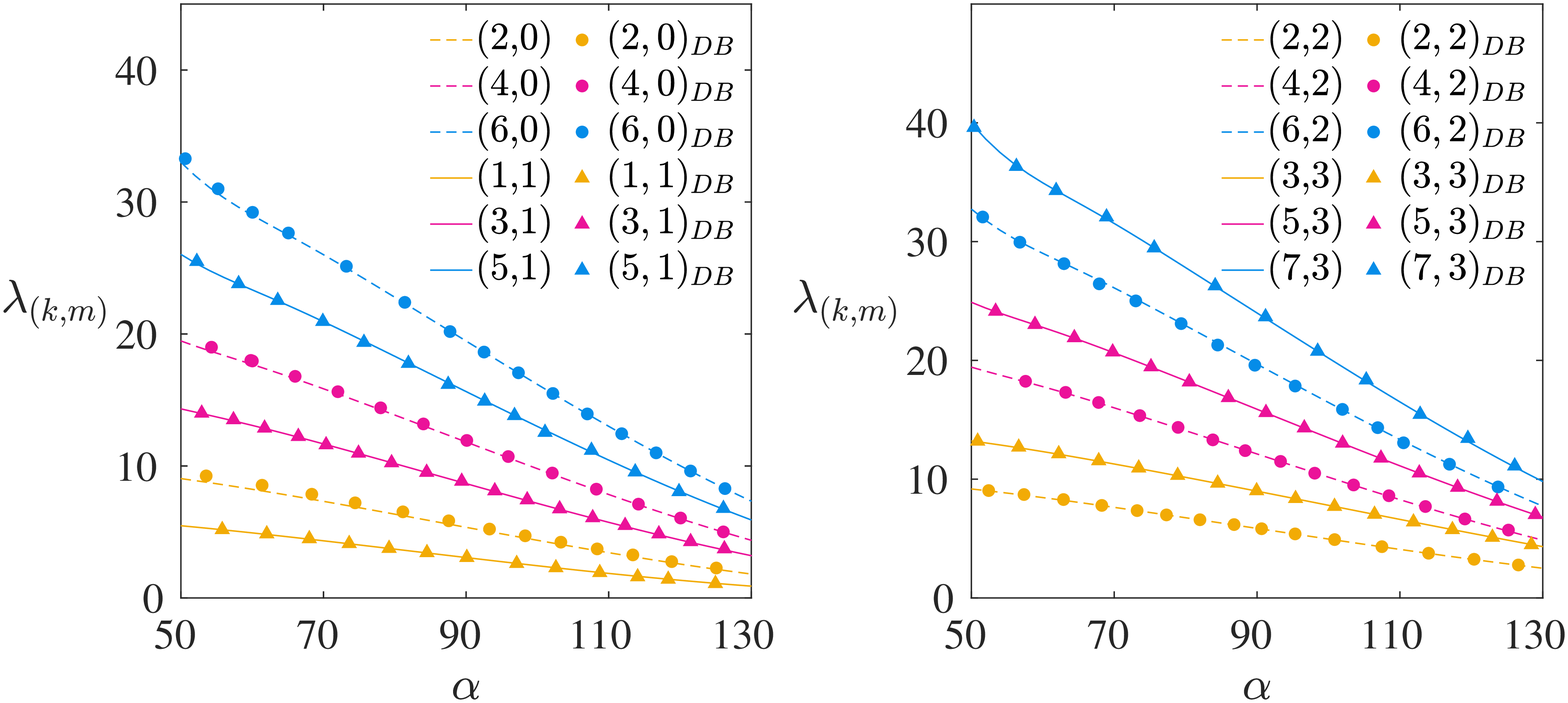}}\\
	\subfloat[]{\includegraphics[scale=0.4]{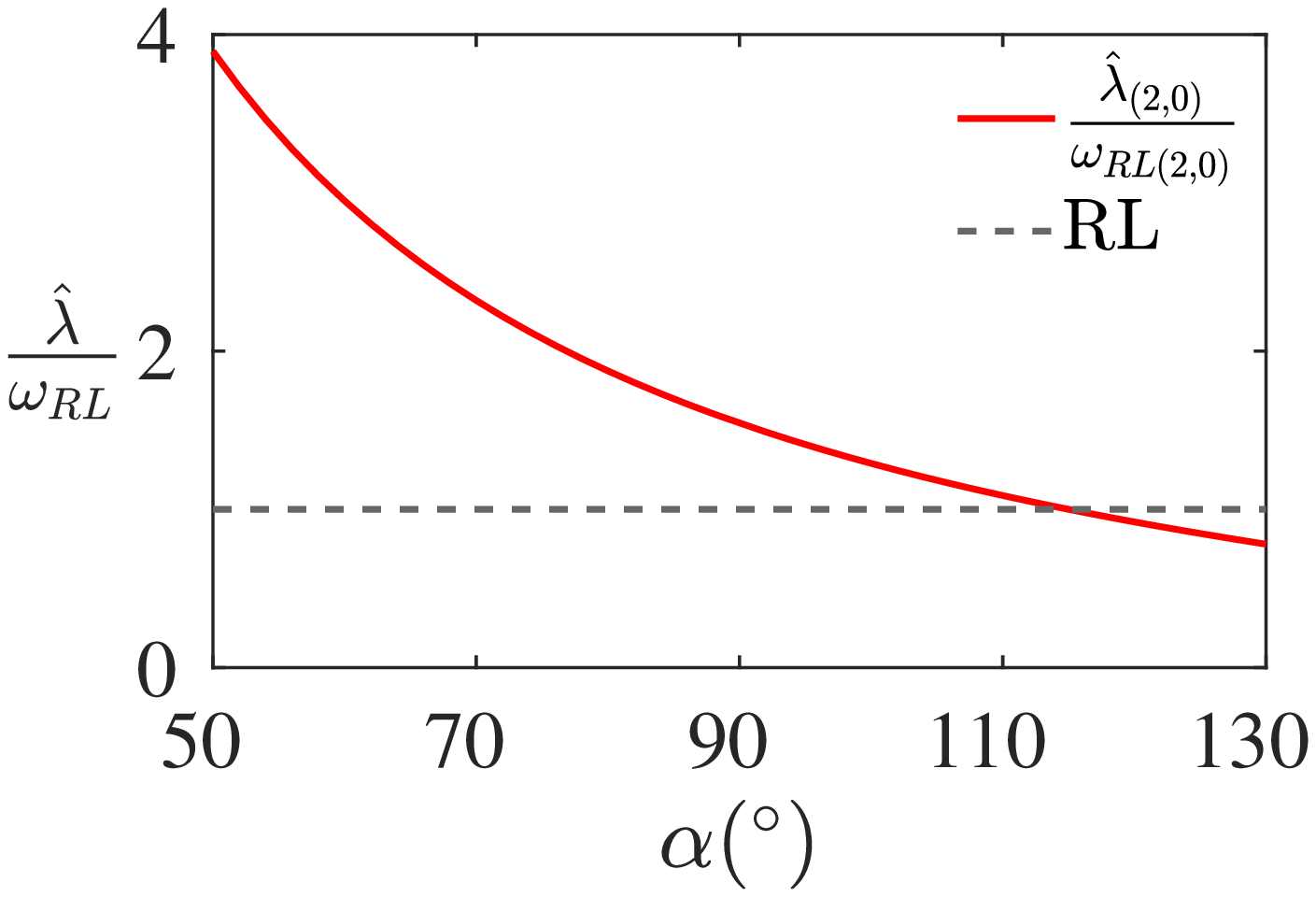}}
	\caption{Panel (a) The eigenvalues of pinned axisymmetric and three dimensional modes as function of contact angle $\alpha$ (in degrees) from solving the eigenvalue problem in eqn. \ref{eqn2_8}. Our numerical results have been matched against those by \citet{ding2022oscillations} referred to as DB in the figure captions. (b) The pinned axisymmetric mode frequency $\hat{\lambda}_{(2,0)}$ normalised by the Rayleigh-Lamb frequency $\omega_{\text{RL}}$ obtained from taking the bubble limit of expression \ref{eqn1} for $k=2$, as function of contact angle $\alpha$ of the spherical cap in the base-state. Note that both RL bubble frequency and the spherical cap bubble frequency have been calculated using the \textit{same} radius $\hat{R}_0$. This implies that the spherical cap volume is changing with $\alpha$.}	
	\label{fig4}
\end{figure}

\begin{figure}
	\centering
	\includegraphics[scale=0.23]{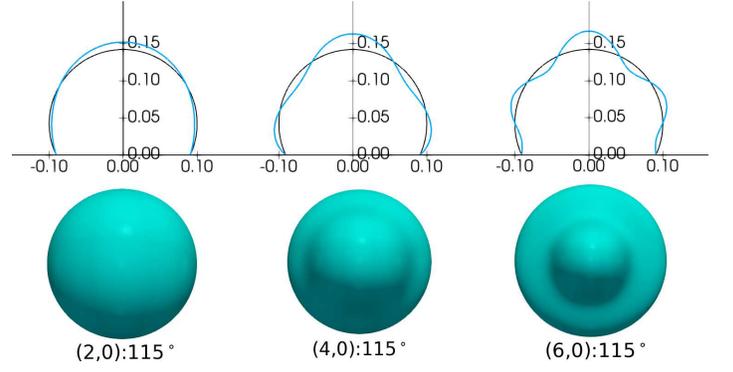}
	\caption{From left to right: Axisymmetric pinned eigenmodes $y_{(k,m)}(s)$ for $(k=2,m=0)$, $(k=4,m=0)$ and $(k=6,m=0)$ with $\alpha=115^{\circ}$. Upper panel: cross-sectional view, lower panel: top-view}
	\label{fig5}
\end{figure}

These eigenvalues and eigenmodes are depicted in figs. \ref{fig4} and \ref{fig5}. Note that in these figures, we relabel the modes as $(k,m)$ instead of $(j,m)$ using the relation $k \equiv 2j + \left(1-\delta_{m0}\right)\left(m - 2\right),\; m=0,1,2,\ldots$ and $\delta_{m0}$ being the Kronecker delta. The index $k$ has the advantage that it's value indicates the (maximum) number of intersections of the eigenmode (nodal circles) with the base state \cite{bostwick2014dynamics} and also respects the condition $m \leq k$.


\section{Numerical simulations}\label{sec:sim}
\begin{figure}
	\centering
	\subfloat{\includegraphics[scale=0.3]{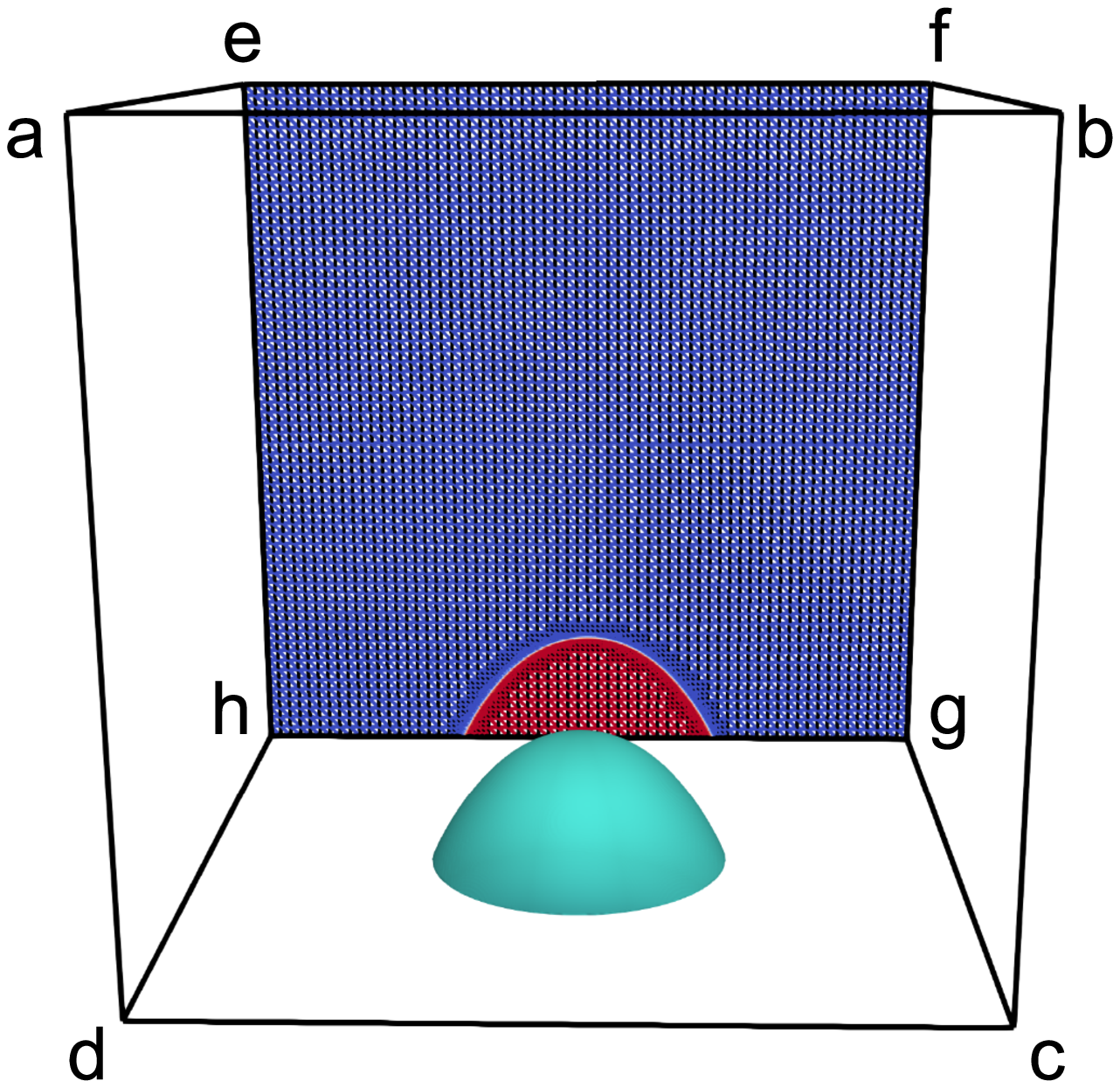}	\label{fig6a}}
	\subfloat{\includegraphics[scale=0.18]{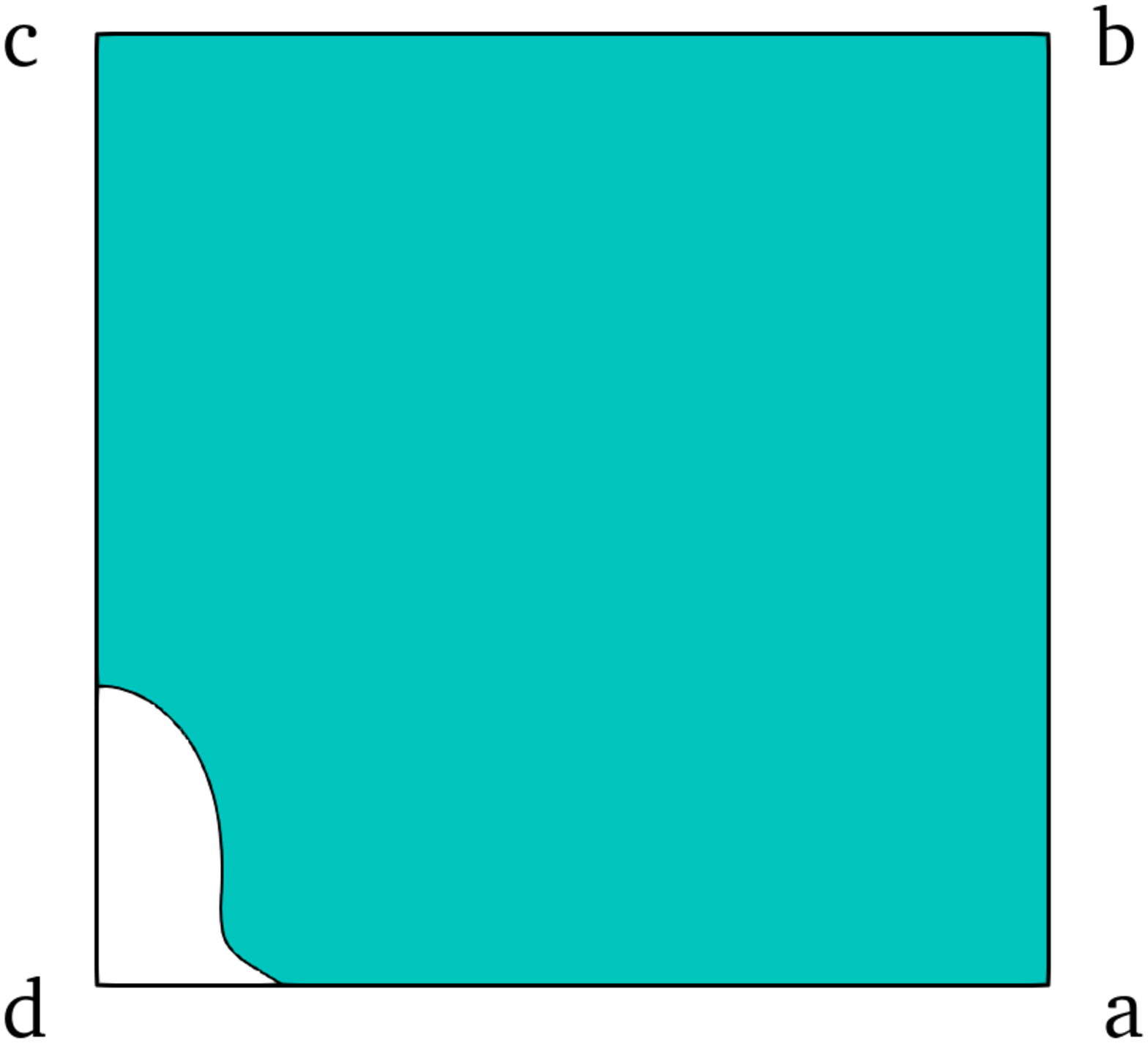}	\label{fig6b}}
	\caption{(Left panel) Computational geometry in three dimensional and (Right panel) axisymmetric case. For visualisation purpose, the grid in the left panel imposed at the mid-plane of the bubble is depicted in the background.}
	\label{fig6}
\end{figure}
The  numerical simulations presented here have been carried out using the open source code Gerris \citep{popinet2009accurate}. Both axisymmetric and three dimensional simulations are reported here. As the linear theory assumes a radially unbounded domain, in order to minimise  boundary effects in our simulations, these have been placed at a distance approximately five times, the equilibrium droplet radius as shown in figs. \ref{fig6a} and \ref{fig6b}. We include surface tension in the simulation but neglect gravity and viscosity. Thus these results correspond to the zero Bond number and zero Ohnesorge number limit of a bubble. Gerris solves the incompressible, Navier-Stokes equations (although viscosity is zero for most of our simulations):
\begin{eqnarray}\label{eqn3_1}
	&\bm{\nabla}\cdot\textbf{u} = 0 \nonumber\\ &\rho\left(\frac{\partial\textbf{u}}{\partial t} +\bm{\nabla}\cdot\left(\textbf{u}\textbf{u}\right)\right) = -\bm{\nabla }p + \bm{\nabla}\cdot(2\mu \mathbf{D}) + T\kappa\delta_s\textbf{n}  \nonumber
\end{eqnarray}
where $\rho$, $\mu$, $\textbf{u}$, $p$
are density, viscosity, velocity and pressure respectively. $\mathbf{D}$ is the strain rate tensor (in  our simulations viscosity is set to zero), $T$ is surface tension coefficient, $\kappa$ represents local curvature and the unit normal of the interface $\mathbf{n}$, with the surface delta function $\delta_s$. Gerris uses the one fluid model where the gas and liquid phases are distinguished by the color function $f$. Here $f = 1$ for the gas (air) inside the bubble and $f = 0$ for the outer fluid (water) and $0 < f < 1$ for the interface between fluids. Density is determined by a weighted combination of the density $\rho\IN$  (gas density inside the bubble) and $\rho\IN$ and $\rho\OUT$ (liquid density outside the bubble),
\begin{align}
	\rho &= f\rho\IN + (1-f)\rho\OUT 
\end{align}
The temporal evolution of volume fraction $f$ is governed by,
\begin{equation}
	\frac{\partial f}{\partial t} + \bm{\nabla}\cdot\left(\textbf{u}f\right) = 0
\end{equation}
and Gerris reconstructs the interface between  the phases using a piecewise linear, volume of fluid based reconstruction algorithm. We have extensively used the adaptive mesh refinement. feature of Gerris. For a typical large amplitude axisymmetric simulation, there are about $150$ grid points along the height of the drop and about $115$ grid points along the wall (level $12$ of refinement at the maximum level). For enforcing the pinned boundary condition  at the contact line,  we impose for $\hat{r} < L =\hat{R}_0\sin(\alpha), \; f = 1$ and for $\hat{r} > L,\;f = 0$. Due to this, the interface stays at $\hat{r}=L$ at the CL upto a precision which equals the width of a single computational cell at the wall.  Table \ref{tab_bc} shows the  boundary conditions at the various computational boundaries with respect to the simulation geometry presented in figs. \ref{fig6a}  and \ref{fig6b} and table \ref{tab:params} presents the parameters.
\begin{table}
	\footnotesize
	\centering
	\begin{tabular}{c c c c c c}
		\multicolumn{6}{c}{\textbf{Axi-symmetric}}\\
		\hline
		Face & \multicolumn{2}{c}{$\textbf{u}$} & & $p$ & \multicolumn{1}{c}{$f$} \\
		\multicolumn{1}{c}{} & ${u}_n$ & \multicolumn{3}{l}{${u}_t$} &\\
		\hline
		c-d  & D & N & & N &  \multicolumn{1}{c}{N} \\
		d-a  & D & N & & N &  \multicolumn{1}{c}{\text{refer to text}} \\
		other faces  & N & N & & D & \\
		\hline
		\multicolumn{6}{c}{\textbf{3D}}\\
		\hline
		Face & \multicolumn{3}{c}{$\textbf{u}$} & $p$ & $f$ \\        
		\multicolumn{1}{c}{} & ${u}_n$ & ${u}_{t1}$ & ${u}_{t2}$ & \multicolumn{1}{c}{} & \multicolumn{1}{c}{} \\ 
		\hline
		\multicolumn{1}{c}{} c-d-h-g  & D & N  & N  & N &  \text{refer to text}\\
		\text{other faces}  & N & N  & N  & D &  N\\
		\hline
		\multicolumn{2}{c}{D: Dirichlet} & \multicolumn{2}{c}{N: Neumann}\\
	\end{tabular}
	\normalsize
	\caption{Boundary conditions on various faces of the computational domain, please see fig. \ref{fig6}}
	\label{tab_bc}
\end{table}

\begin{table}
	\begin{center}
		\def~{\hphantom{0}}
		\begin{tabular}{cccccccc}
			$\hat{R}_0$  & $T$ & $\rho\IN$ &  $\rho\OUT$ & $\mu\IN$ & $\mu\OUT$ &  $\textcolor{black}{g}$\\[3pt]
			\hline
			0.1 & 72 & 0.001 & \textcolor{black}{1} & \textcolor{black}{0}& \textcolor{black}{0-0.01} & \textcolor{black}{0} \\
		\end{tabular}
		\caption{Simulations parameters (CGS units) with air (inside) - water (outside) parameters conducted in Gerris \citep{popinet2009accurate}. $T$ being surface tension, $\rho\IN$, $\rho\OUT$ density $\mu\IN$, $\mu\OUT$ being the dynamic viscosity and $g$ is the acceleration due to gravity.}
		\label{tab:params}
	\end{center}
\end{table}

\subsection{Benchmarking: free-bubble shape oscillations}
\begin{figure}
	\centering
	\subfloat[Free air bubble in water  ]{\includegraphics[scale=0.3]{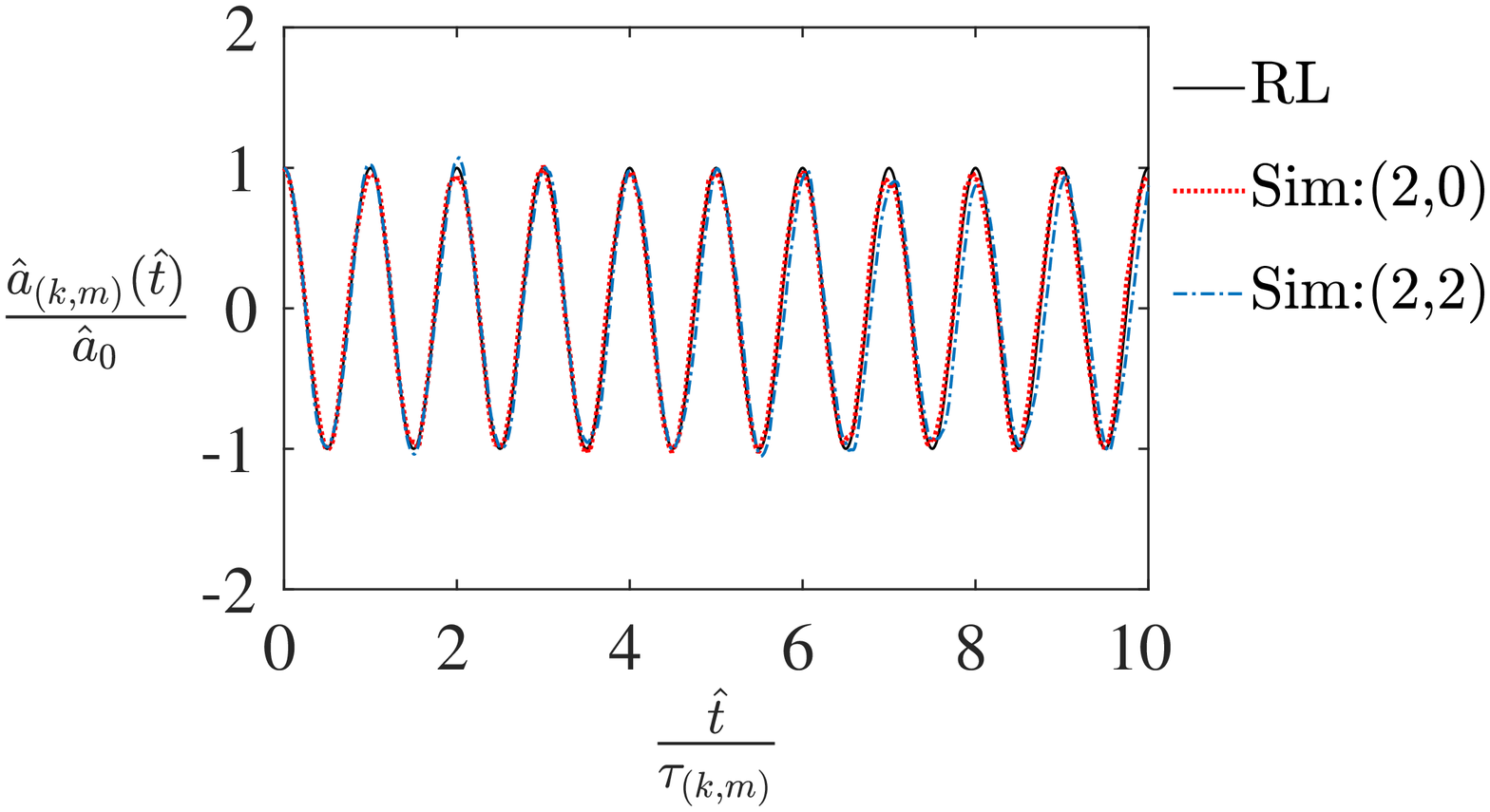}}
	\subfloat[Mode shapes]{\includegraphics[scale=0.2]{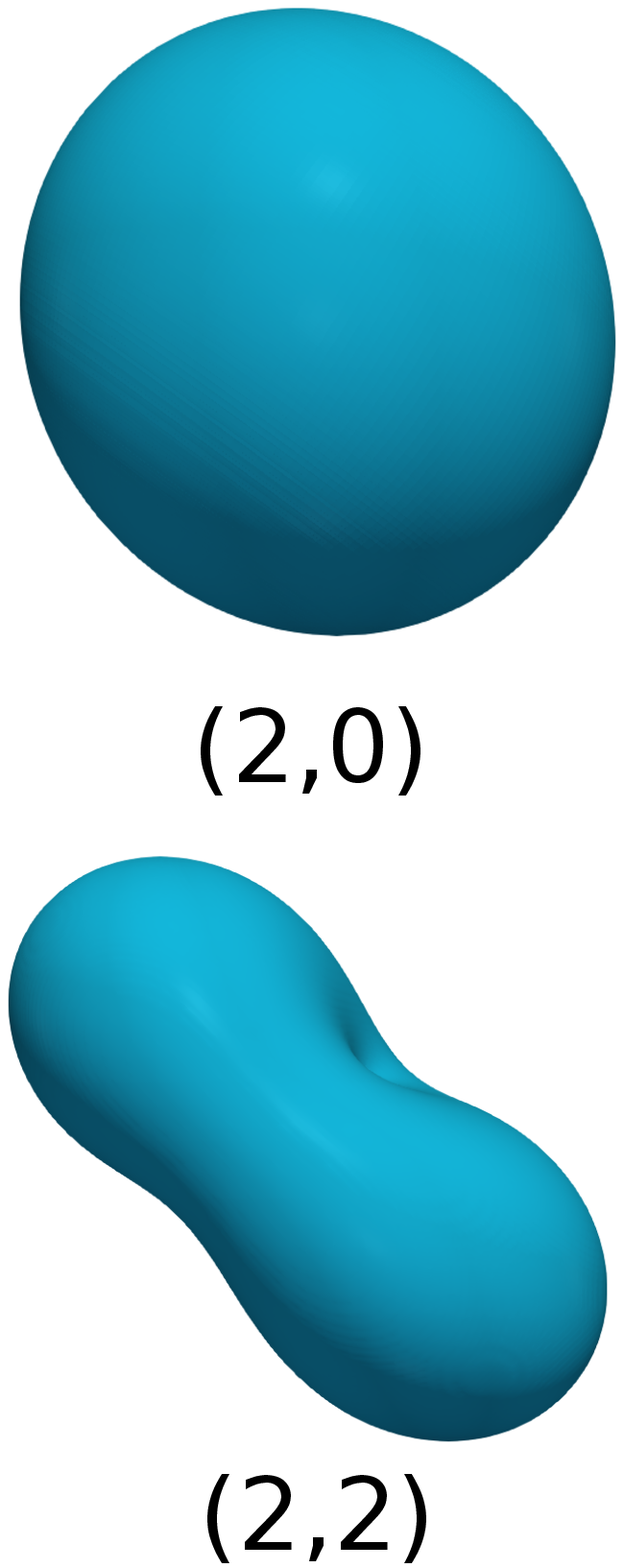}}
	\caption{Comparison of predictions from the Rayleigh-Lamb (RL) spectrum from expression \ref{eqn1} for shape modes of a free air bubble against inviscid, numerical simulations (Sim) with surface tension and no gravity \citep{popinet2009accurate}. The legends identify the chosen spherical harmonic for exciting the spherical interface at $\hat{t}=0$ viz. $\hat{r}_s(s,\psi,0)=\hat{R}_0 + \hat{a}_0\mathbb{Y}_k^{(m)}(s,\psi)$. For the mode $(k=2,m=0)$, in CGS units $\hat{a}_0=0.005$ while for the $(k=2,m=2)$ mode, $\hat{a}_0=0.0005$. The time signals are obtained from tracking the bubble surface in the simulation at $s=\pi/2$. For axisymmetric simulations, the grid is $1024^2$. For the three dimensional mode $(2,2)$, we use adaptive grid with maximum level $9$ in the simulations. The variable $\tau_{(k,m)} \equiv \frac{2\pi}{\omega^{(k,m)}}$ corresponds to the time period of one oscillation.}
	\label{fig7}
\end{figure}

Gerris  has been extensively used in several two-phase simulations in literature. As benchmarking of the inviscid version of the  code, we deform the surface of a freely suspended air-bubble in water (without gravity or viscosity), in the shape of an eigenmode. For an incompressible, free bubble this is the spherical harmonic $\mathbb{Y}_{k}^{(m)}(s,\psi)$, $k>0,\; 0 \leq m \leq k$. The bubble surface at $t=0$ is initialised as $\hat{r}(s,\psi,0)=\hat{R}_0 + \hat{a}_0\mathbb{Y}_k^{(m)}(s,\psi)$ with zero velocity everywhere in the gas inside as well as in the liquid outside. We test the code for two modes (separately) viz. an axisymmetric mode $(k=2,m=0)$ and a three dimensional mode $(k=2,m=2)$ are individually excited with small $\hat{a}_0$ ($<< \hat{R}_0$) and the time signals are depicted in fig. \ref{fig7}. Good agreement is apparent for the first few oscillations. Note that the upper index $m$ of the spherical harmonic does not appear in the dispersion relation in eqn. \ref{eqn1} implying that the $(k=2,m=0)$ mode and the $(k=2,m=2)$ mode have the same frequency, although the mode shapes are distinct. This is the well-known degeneracy of modes for free bubbles and drops \citep{bostwick2014dynamics} which is no longer present, when the bubble is pinned.
\section{Results}\label{sec:res}
In this section, we present results from numerical simulations, first in the linear regime where the linear theory presented in figs. \ref{fig4} is tested for small amplitude, modal perturbations and later in the nonlinear regime at larger amplitudes. 
\subsection{Sessile bubble: linear regime}
We compare our simulations with the linear theory predictions (modal shapes and frequencies) obtained from solving the eigenvalue problem \ref{eqn2_8}. The sessile bubble surface is distorted initially at $\hat{t}=0$ using the lowest, axisymmetric shape mode i.e. $\hat{r}_s(s,\psi,0) = \hat{R}_0 + \hat{a}_0y_{(k,m)}(s)\cos(m\psi)$ with $\hat{a}_0 << \hat{R}_0$, so as to remain close to the linear regime. Note that $\hat{a}_0 \equiv \hat{a}_{(k,m)}(0)$ in the simple harmonic oscillator eqn. \ref{eqn2_9} and as we start the simulations with a distorted bubble surface and zero velocity everywhere in the gas inside and the liquid outside the bubble, this implies $\dt{\hat{a}}_0(0)=0$. For all numerical simulations presented here, $y_{(k,m)}(s)$ was obtained by solving the eigenvalue problem \ref{eqn2_8} and this mode shape $y_{(k,m)}(s)\cos(m\psi)$ was subsequently used to distort the bubble shape and this was then provided as an initial condition to the simulation. 

\begin{figure}
	\centering
	\includegraphics[scale=0.35]{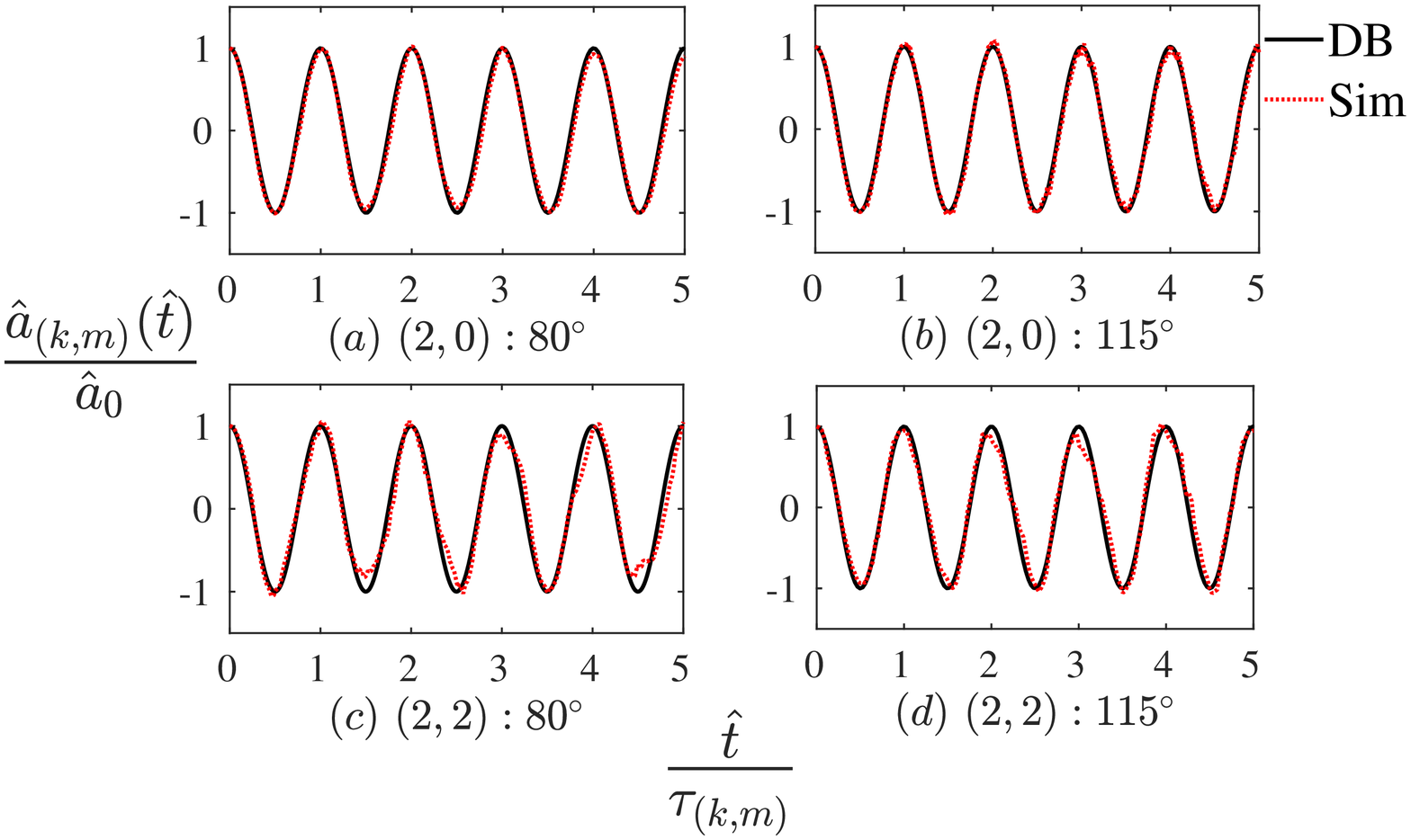}
	\caption{Time signals obtained by exciting pinned CL modes on the spherical base cap. The initial distortion of the cap is set as $\hat{r}_s(s,\psi,0) = \hat{R}_0 + \hat{a}_0y_{(k,m)}(s)\cos(m\psi)$ where $y_{(k,m)}(s)$ is extracted from the numerical solution to eqn. \ref{eqn2_8}. The time signal is extracted from the numerical simulations by tracking the interface at $s\approx 18.4^{\circ}$. The panels are to be read from left to right. Panel (a) $k=2,m=0,\alpha=80^{\circ},\hat{a}_0\approx 0.002$ cm. Panel (b) $k=2,m=0,\alpha=115^{\circ},\hat{a}_0\approx 0.001$ cm. Panel (c) $k=2,m=2,\alpha=80^{\circ},\hat{a}_0\approx 2\times 10^{-3}$ cm. Panel (d) $k=2,m=2,\alpha=115^{\circ},\hat{a}_0\approx 1\times 10^{-3}$ cm. The acronym DB in the figure captions refer to the natural frequency of the mode predicted by \citet{ding2022oscillations} and also provided by us in fig. \ref{fig4}. $\tau_{(k,m)}$ represents the (dimensional) time period of the $(k,m)$ mode. We use adaptive grid of maximum level $11$ (powers of $2$) and minimum $9$ in these simulations near and far from the interface respectively.}
	\label{fig8}
\end{figure}

Fig. \ref{fig8} shows the time signal extracted from numerical simulations by tracking  the interface at $s\approx 18.4^{\circ}$ for various $(k,m)$. As our simulations start with an eigenmode as initial condition, linearised standing waves develop and are visualised at the bubble surface. The time-period of the same viz. $\tau_{(k,m)}$ is known from the eigenvalue $\lambda_{(k,m)}$ and is used to non-dimensionalise time $\hat{t}$ in all the panels of this figure \ref{fig8}. The first two panels (a), (b) are for axisymmetric modes while panels (c), (d) are for three dimensional modes. All the panels show reasonably good agreement with theory (labelled as DB representing \citet{ding2022oscillations}), although small discrepancies are observed for three dimensional modes in the lower panels of the figure. For the three dimensional mode $k=2,m=2$ (panel (c) and (d)) we observe the onset of nonlinear effects relatively early compared to other simulations.
\begin{figure}
	\centering
	\includegraphics[scale=0.18]{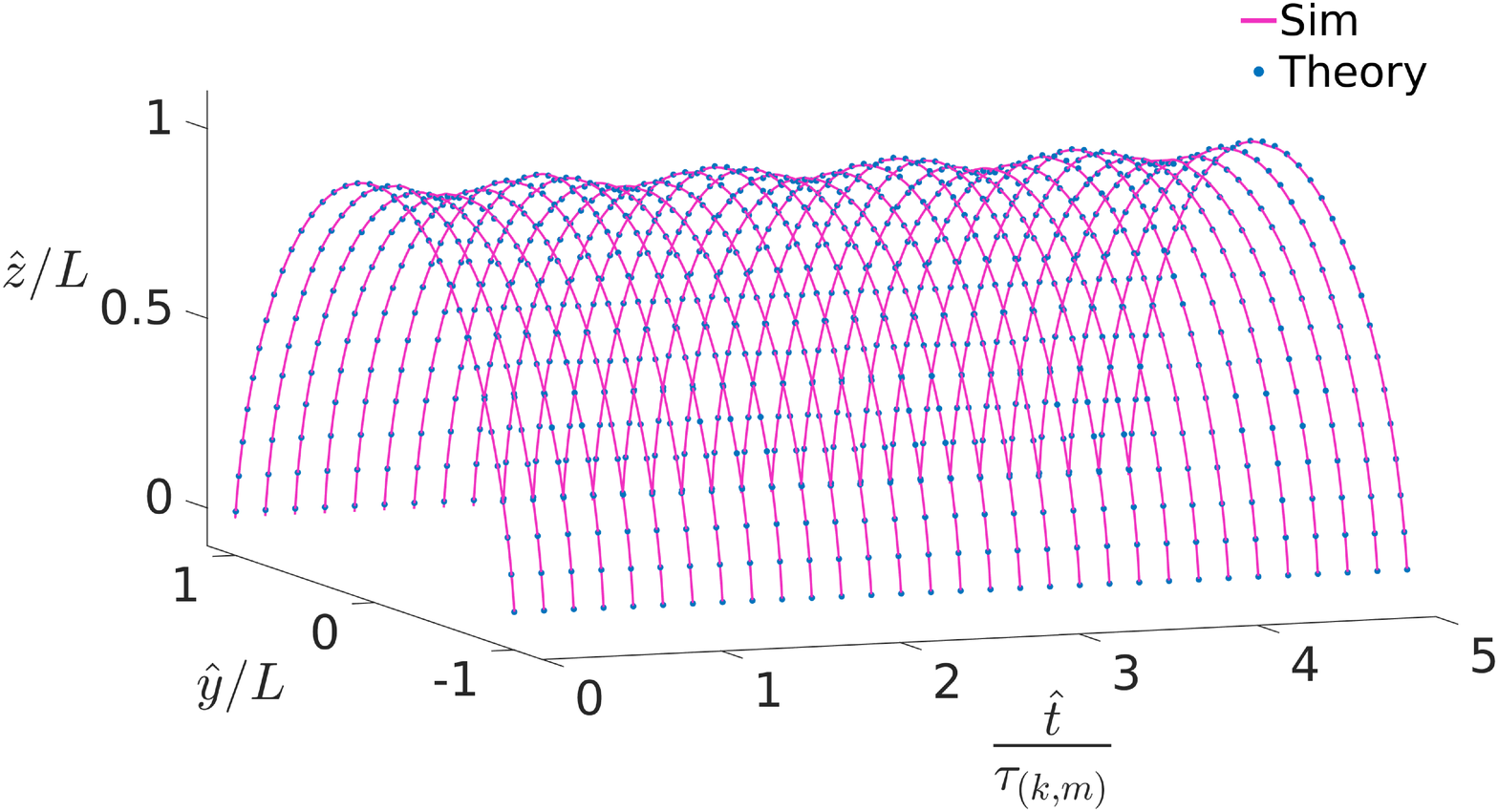}
	\caption{Time evolution of the entire interface for $k=2,m=0,\alpha=80^{\circ}$ with pinned CL. This is the same simulation as reported in fig. \ref{fig8}, panel (a). Note the good agreement between the simulation and linear theory.}	
	\label{fig10}
\end{figure}  
Fig. \ref{fig10} shows the time evolution of the entire bubble surface for the $k=2,m=0$ (axisymmetric) mode. This corresponds to the same simulation whose time signal at a particular point on the interface is reported in fig. \ref{fig8}, panel (a) earlier. It is seen that the entire bubble shape shows good agreement with simulations, this being depicted in the figure up to five (linear) time periods of oscillation. The instantaneous pressure field and the streamlines for this mode $k=2,m=0$ are also presented in fig. \ref{fig11}.
\begin{figure}
	\centering
	\subfloat[$\alpha = 80^{\circ}$]{\includegraphics[scale=0.12]{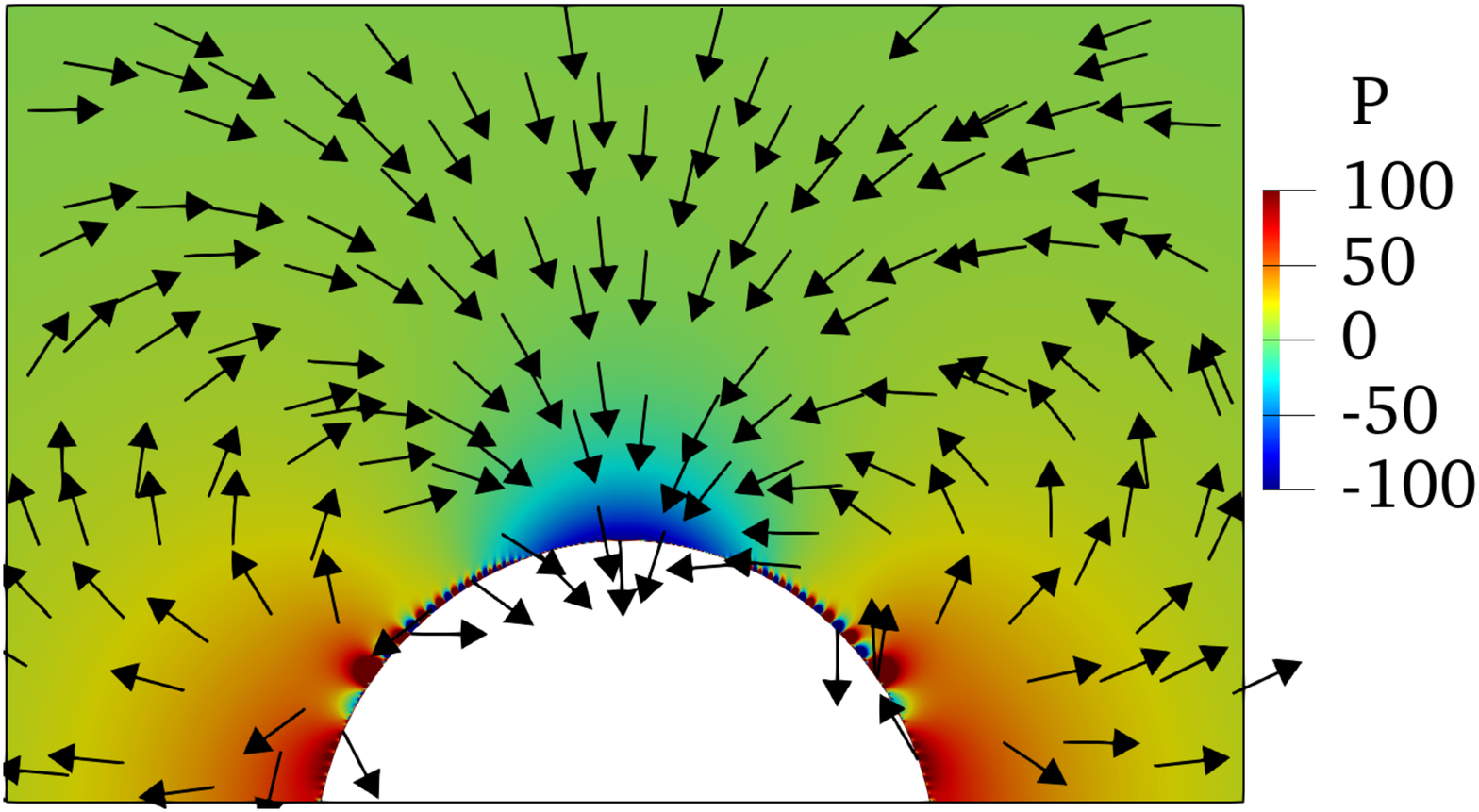}\hspace{2mm}}
	\subfloat[$\alpha = 115^{\circ}$]{\includegraphics[scale=0.12]{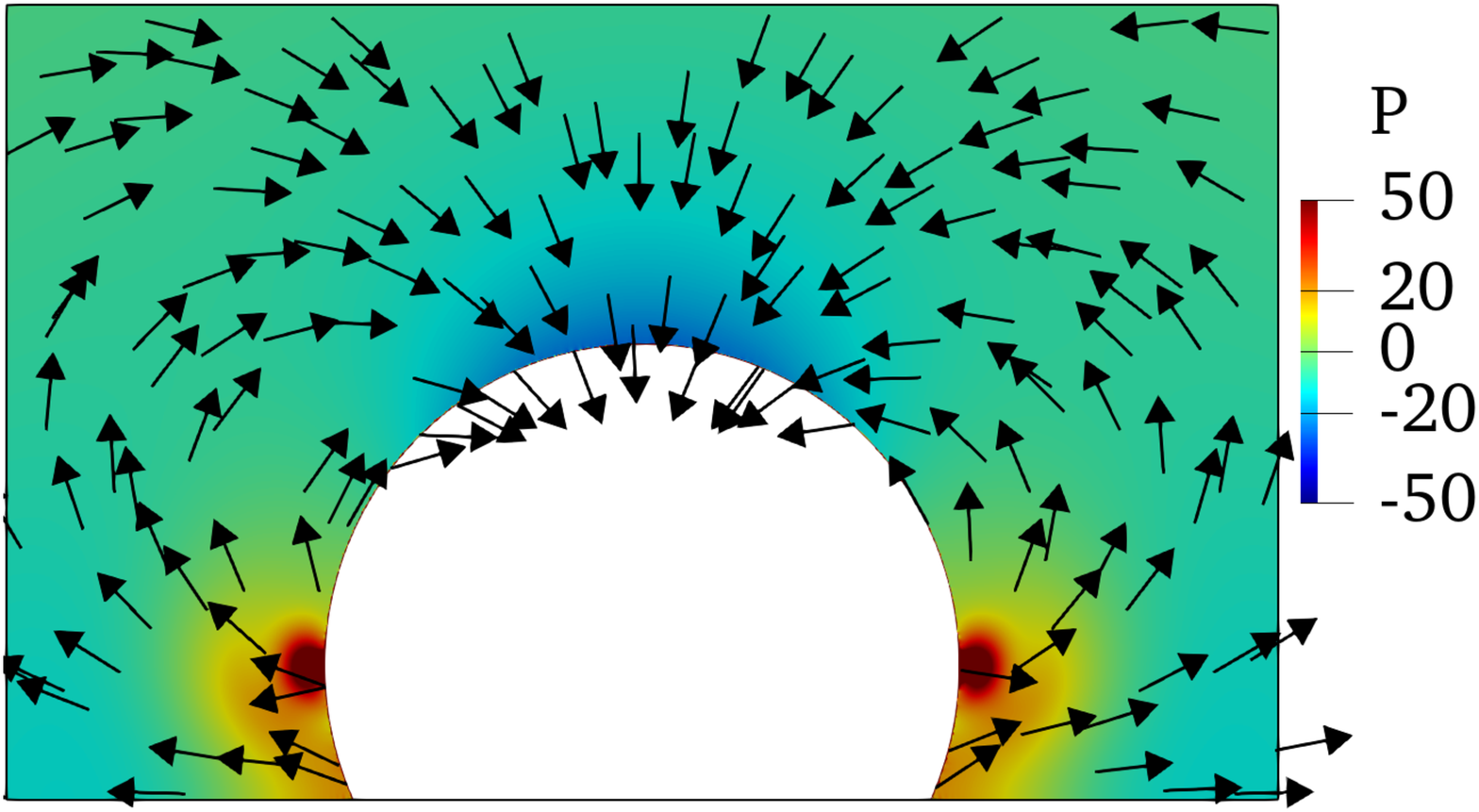}}
	\caption{The instantaneous streamlines and pressure contours as obtained from our simulations for the pinned mode $k=2,m=0$.}	
	\label{fig11}	
\end{figure}

\subsection{Onset of the nonlinear regime}
\begin{figure}
	\centering
	\subfloat[]{\includegraphics[scale=0.32]{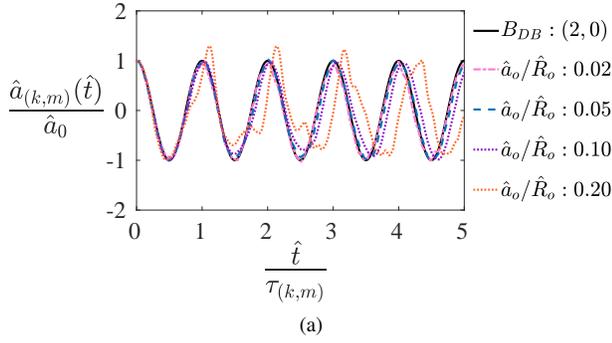}}\\
	\subfloat[]{\includegraphics[scale=0.18]{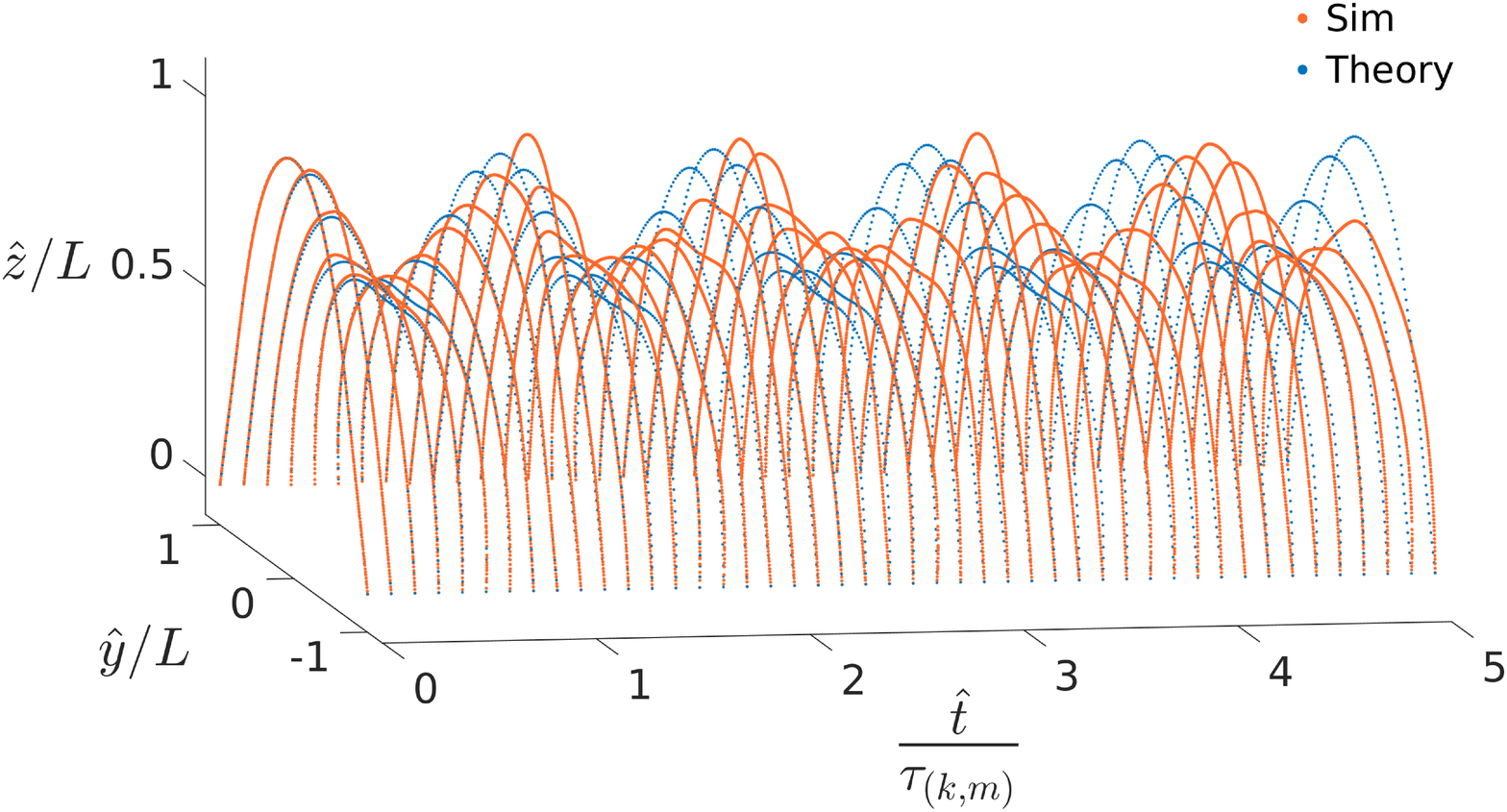}}
	\caption{Panel a) The effect of change of $\hat{a}_0$ on the time signal. For $\hat{a}_0/\hat{R}_0 \geq 0.05$, a reduction in the frequency of the signal is seen  compared to when $\hat{a}_0/\hat{R}_0\rightarrow 0$, where a good agreement with linear theory is observed. At large values of $\hat{a}_0/\hat{R}_0$, other modes are nonlinearly generated and the time signal resembles a superposition of several sinusoids. Here in the base state $\alpha=80^{\circ}$. Panel b) Time evolution of the bubble shape for $\frac{\hat{a}_0}{\hat{R}_0}=0.2$ for which the time signal is shown in the upper panel.}	
	\label{fig12}	
\end{figure}

\begin{figure}
	\centering
	\subfloat[Pinned CL]{\includegraphics[scale=0.25]{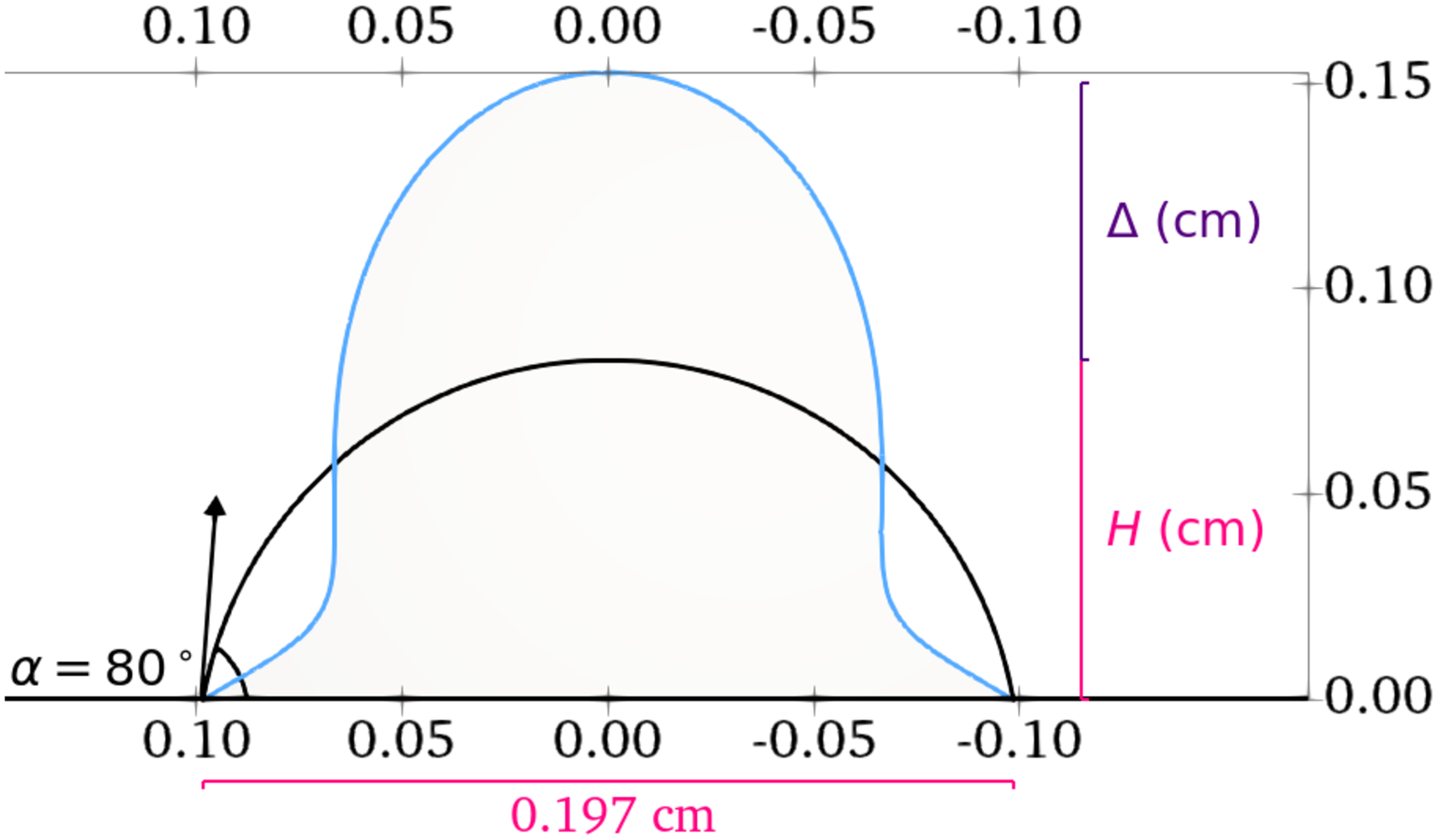}}
	\caption{Large amplitude shape deformation regime. Such initial deformation at $\hat{t}=0$ is generated by choosing $\hat{a}_0$ in linear theory to be sufficiently large (see text). For experimental verification of our large amplitude numerical predictions, we provide the scales in the figure. Here $\Delta=0.07,\; H = 0.0826$.}	
	\label{fig13}
\end{figure}

In this sub-section, we move beyond the linear regime studied so far.  Fig. \ref{fig12} depicts the effect of increasing the modal amplitude $\hat{a}_0$ holding $\hat{R}_0$ fixed. It is clearly seen that as $\hat{a}_0/\hat{R}_0$ is increased, the frequency of the standing wave \textit{decreases}, see the near sinusoidal signal indicated with purple dots which has a slightly higher time period than the linearised prediction. This observation is consistent with the weakly nonlinear calculation by \citet{tsamopoulos1983nonlinear} who employed the Lindstedt-Poincare technique to obtain finite amplitude, time-periodic deformations of a free bubble. \citet{tsamopoulos1983nonlinear}  analytically predicted a quadratic reduction in natural oscillation frequency with increasing amplitude of the shape modes, see expressions $59, 60$ and $61$ in \citet{tsamopoulos1983nonlinear}. An important difference between the results by these authors and our present observations is that, as $\hat{a}_0$ in increased systematically, nonlinear corrections become necessary in order to have finite-amplitude time-periodic initial conditions; these corrections are absent for our initial condition and consequently periodic behaviour is expected only for $\hat{a}_0/\hat{R}_0 \rightarrow 0$ in fig. \ref{fig12}. Note that in contrast the finite amplitude calculation by \citet{tsamopoulos1983nonlinear} by analytical construction, is time-periodic.
Increasing the value of $\hat{a}_0$ even further, we generate larger amplitude, shape deformations for the bubble. In figure \ref{fig13},  a large amplitude deformation of the bubble surface is depicted. This has been obtained by increasing the modal amplitude $\hat{a}_0$ to be sufficiently large for the $(k=2,m=0)$ eigenmode with $\alpha=80^{\circ}$. For generating this shape, we used $\hat{R}_0=0.1$ cm and $\hat{a}_0=0.07$ cm. Before we describe the dynamics resulting from such large amplitude deformations, it is necessary to discuss volume conservation of the gas inside the spherical cap below.

We recall that our simulations are in the incompressible limit, both for the gas inside and the liquid outside the bubble. Eqn. \ref{eqn4_1} equates the (non-dimensional) volume of a distorted bubble to that of a spherical cap, both pinned at the same radial location, 
\begin{eqnarray}
		&\bigint_{s=0}^{\alpha} \; ds\sin(s)\bigint_{\psi=0}^{2\pi}\; d\psi \bigint_{r=0}^{\csc(\alpha) + \eta(s,\psi,t)}\; dr\;r^2 \nonumber \\
		&- \dfrac{\pi}{3}\cot(\alpha) = \dfrac{\pi}{3}\left[\dfrac{2-3\cos(\alpha) + \cos^3(\alpha)}{\sin^3(\alpha)}\right]. \label{eqn4_1}
\end{eqnarray}
Eqn. \ref{eqn2_4} g, presented earlier representing volume conservation of the gas inside of a spherical cap is an approximate expression correct only up to linear order in $\eta$. This equation has been obtained by neglecting quadratic and cubic terms in $\eta(s,\psi,t)$ in the exact equation \ref{eqn4_1} and is a good approximation only when $\hat{a}_0 << \hat{R}_0$. Physically this means that for small modal amplitude ($\hat{a}_0$), the volume of the distorted bubble and the base state spherical cap are nearly the same. However, for large amplitude deformations where $\hat{a}_0 \sim \hat{R}_0$ (see fig. \ref{fig13} and the values of $\hat{a}_0$ and $\hat{R}_0$ mentioned in the previous paragraph), neglecting terms with $\eta^2$ and $\eta^3$ compared to $\eta$ in eqn. \ref{eqn4_1} is no longer tenable. At such large deformation the volume of the perturbed bubble is substantially different from that of the base-state spherical cap. This difference arises at non-linear order due to the eigenmodes to the linearised problem, not satisfying the volume conservation constraint exactly. This feature is not specific to a sessile bubble but also appears for free bubbles or droplets e.g. see eqn. $7$ in \citet{tsamopoulos1983nonlinear} where volume conservation is imposed and this condition is distinct from its linearised version $\int_{-1}^{1}\;\eta(x,t)dx=0$. At these large deformations, the ratio $\hat{a}_0/\hat{R}_0$ is no longer a meaningful measure of nonlinearity as the volume of the perturbed bubble is significantly different from a spherical cap of radius $\hat{R}_0$ and contact angle $\alpha$. We consequently use the non-dimensional ratio $\Delta/H$ (see fig. \ref{fig13}) to characterise such large amplitude deformations. 
\subsection{Large distortion regime: jet formation at $\Delta/H \sim \mathcal{O}(1)$}
\begin{figure}
	\centering
	\includegraphics[scale=0.45]{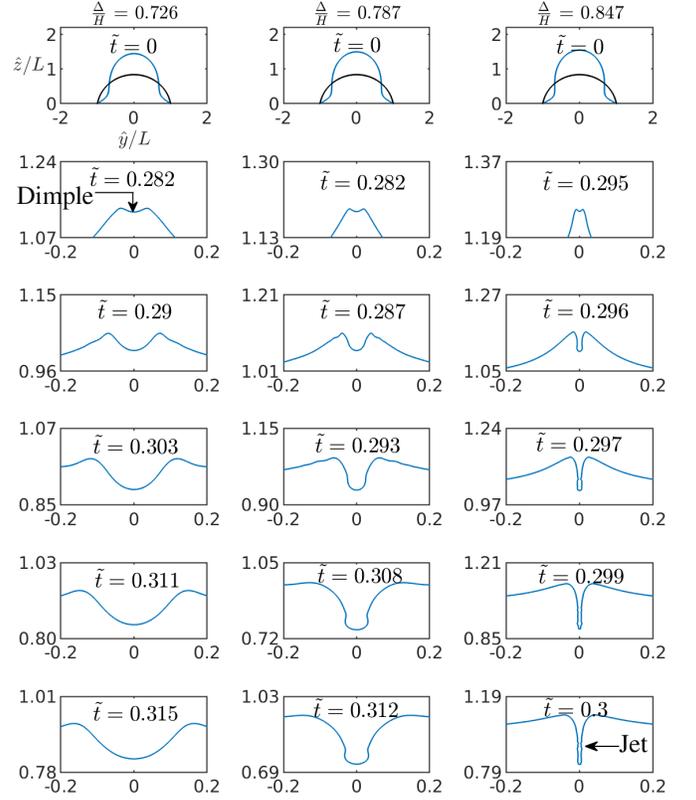}
	\caption{Time evolution of the bubble shape due to large amplitude shape deformation of a pinned bubble. The collage is read column wise and each column shows the time evolution of the bubble shape with non-dimensional time $\tilde{t}\equiv \frac{\hat{t}}{\tau_{(k,m)}}$ for a given value of $\Delta/H$. As illustrated in fig. \ref{fig13}, $\Delta/H$ is a measure of the bubble deformation. For first the left column, $\Delta/H=0.726$, for the middle column $\Delta/H=0.787$ and for the right column, $\Delta/H=0.847$.}	
	\label{fig14}
\end{figure}

Figs. \ref{fig14} depict qualitatively new features (compared to linear theory) which present themselves as the bubble surface is distorted far beyond the linear regime. Each column in this figure is for a specific value of $\Delta/H$ and the columns should be read from top to bottom, to follow the temporal evolution of the bubble surface around the symmetry axis. The initial condition for each column is generated from the modal formula of the bubble shape $\hat{y}_s = \hat{R}_0 + \hat{a}_0y_{(2,0)}(s)$ by systematically increasing $\hat{a}_0$. The base state spherical cap is provided in the first row of each column (black curve) with parameters $\left(\alpha,\hat{R}_0\right)$. Note that each column in fig. \ref{fig14} has a different initial condition ($\Delta/H$) and thus represents a bubble of different volume, although this volume is preserved throughout the course of each individual simulation. Our motivation to increase modal amplitude arises from recent work on nonlinear surface waves where it has been shown theoretically and computationally, that one can obtain jets from by exciting a single eigenmode at large amplitude \cite{farsoiya2017axisymmetric,basak_farsoiya_dasgupta_2021,kayal2022dimples}. It was further shown in \citet{kayal2022dimples} that the radial inward motion of zero crossings of the perturbed state with the interface in the base state, at time $\hat{t}=0$, which behave as nodes in a linearised description, serve as a visual indication of wave focussing preceding jet formation. It will be seen below a similar inward radial motion of zero crossing also occurs for a spherical cap bubble.

For $\Delta/H=0.726$ (first column in fig. \ref{fig14}), we observe the formation of a small structure at the symmetry axis (see the figure at $\tilde{t}=0.282$), which we label as a dimple here, following similar terminology followed in bubble bursting at a free surface \citep{lai2018bubble,basak_farsoiya_dasgupta_2021}. This dimple subsequently broadens without transforming into a thin jet. When $\Delta/H=0.787$ (middle column of fig. \ref{fig14}), the dimple that forms is now sharper than before and the formation of a somewhat slender jet like structure is seen at $\tilde{t}=0.293$. A clear jet like structure is seen to emerge only in the third column of this figure, where $\Delta/H=0.847$ (see $\tilde{t}=0.3$). A natural question then is, when should we call the emerging structure a jet in fig. \ref{fig14}? The answer to this will become clear when we present the velocity of the interface at the symmetry axis, as a function of time in figs. \ref{fig16}. Appendix $2$ provides an approximate formulae for the initial bubble distortion, which can be used to reproduce results shown in fig. \ref{fig14}.

\begin{figure}
	\centering
		\includegraphics[scale=0.23]{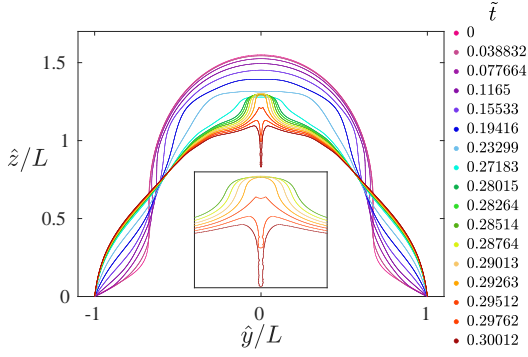}\label{fig15a}
	\caption{Focussing of capillary waves leading to jets. Superposition of bubble surface shapes at various time instants. This simulation corresponds to the third column of fig. \ref{fig14} with $\Delta/H=0.847$. Note the formation of a cylindrical cavity shaped structure at the symmetry axis around $\tilde{t} = 0.282$ from which a sharp dimple and eventually a slender jet emerges. Compare this with fig. $8$ in \citet{duchemin2002jet} or fig. 1$b$, lower centre panel in \citet{ganan2021physics} where a similar cylindrical shaped cavity is seen.}
	\label{fig15}
\end{figure}

\begin{figure}
	\centering
	\includegraphics[scale=0.3]{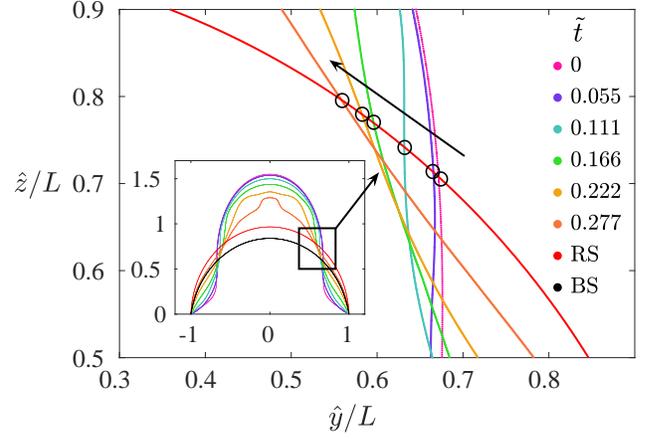}
	\caption{The inward movement of zero-crossings are depicted in black hollow circles. Note that this is a simulation for large amplitude (same as fig. \ref{fig15}. In order to depict the radial inward motion of zero crossings, we construct a spherical cap (labelled as reference state with the acronym RS) which has the same volume as the perturbed bubble shape. This spherical cap (parameters $\tilde{R}_0,\tilde{\alpha}$) is shown in red in the inset and should be contrasted with the spherical cap indicating the base-state (BS) depicted in black with parameters $\left(\hat{R}_0,\alpha\right)$. The arrow indicates that the motion is directed radially inwards with time. Note that the reference state spherical cap has parameters $\tilde{\alpha}=88.05^{\circ},\tilde{R}_0=0.0985$ cm compared to the base-state $\alpha=80^{\circ},\hat{R}_0=0.1$ cm.}
	\label{fig15-0}
\end{figure}

To visually illustrate the focussing of capillary waves which leads to the formation of these jets better, we  superimpose the interface shapes at various instants of time in fig. \ref{fig15}. Note that the time-step spacing in this figure is non-uniform as the jet ejection process is fast compared to other time scales in the problem. In this figure, wave focussing, formation of a cylindrical cavity followed by ejection of a dimple  (see inset) and a jet are clearly visible. To highlight the wave focussing better, we present in fig. \ref{fig15-0}, the inward motion of zero crossings of the perturbed bubble shape, towards the axis of symmetry. Here these zero crossings of the perturbed bubble shape are depicted on a spherical cap with the same volume as the perturbed bubble. We label this as a reference state (RS) spherical cap with parameters $(\tilde{R}_0,\tilde{\alpha})$ in contrast to the base-state (BS) spherical cap, with parameters $(\hat{R}_0,\alpha)$. A clear radial inward motion of the zero-crossing is seen in the figure and is reminiscent of similar inward motion of zero crossings seen in nonlinear capillary-waves in other geometries (see fig. $2$ in \citet{kayal2022dimples}). It is important to recall here, that the inward motion of zero-crossings is a nonlinear effect. In linear theory, the zero crossings of a standing wave constitute nodes which do not get displaced in time.

\begin{figure}
	\centering
	\includegraphics[scale=0.25]{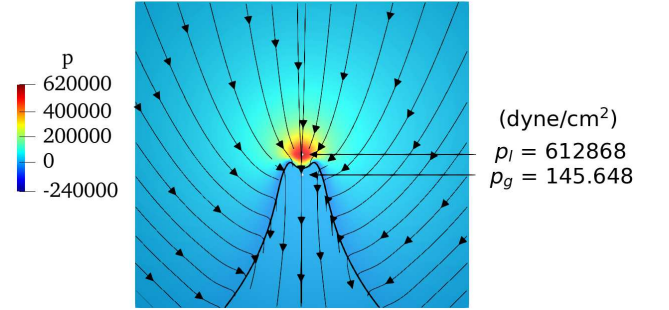}
	\caption{Large pressure gradient along the symmetry axis across the interface (in white) at the instant when the dimple emerges. This simulation is the same as fig. \ref{fig15}. The contours represents instantaneous streamlines. Note that the pressure in the region indicated in red estimated from surface tension induced pressure-jump is (in dyn/cm$^2$) $p_l = \frac{T}{R_d}  + p_g \approx 84,260 + 145.648 \approx 84,405$ dyn/cm$^2$ where $R_d$ is the radius of the small dimple seen in the figure and $p_l$ and $p_g$ are the pressure in the liquid and gas respectively. This value is substantially lower than the actual pressure $p_g=6,12,868$ dyn/cm$^2$ that is measured in the simulation. This difference is due to the high stagnation pressure in the red region, produced from the radial convergence of fluid towards the axis of symmetry.} 
	\label{fig15-1}
\end{figure}

\begin{figure}
	\centering
	\includegraphics[scale=0.28]{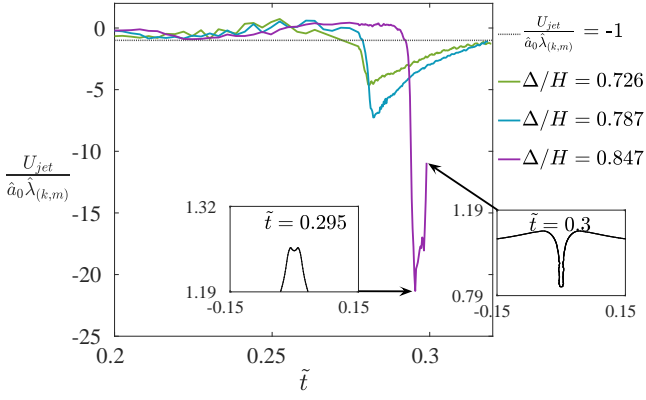}
	\caption{Velocity vs. time at the symmetry axis. The sharp transition into a dimple followed by a jet as $\Delta/H$ is increased as reflected in the vertical velocity of the bubble surface at the symmetry axis. The vertical velocity $U_{\text{jet}}$ is non-dimensionalised by the linear velocity scale $\hat{a}_0\hat{\lambda}_{(k,m)}$ and in the linear regime, this non-dimensional velocity is expected to lie within the range $\left[-1,1\right]$.} 
	\label{fig16}
\end{figure}

Fig. \ref{fig15-1} depicts the pressure field and instantaneous streamlines around the instant when the dimple emerges in fig. \ref{fig15}. Note the very large vertical pressure gradient which manifests as a large vertical velocity leading to formation of a jet subsequently. As explained in the caption to this figure, the large pressure that is seen above the dimple in this figure, cannot be explained purely from surface-tension driven pressure jump and contains a significant contribution from stagnation pressure developed due to radial focussing of fluid towards the symmetry axis (note the instantaneous streamlines on the liquid side in this figure, all of which indicate a significant radial component of velocity towards the symmetry axis.)

Fig. \ref{fig16} depicts the vertical velocity of the bubble surface at $\hat{r}=0$ (symmetry axis) in a time window around the emergence of the dimple. Note that the velocity $U_{\text{jet}}$ has been non-dimensionalised using the linear velocity estimate $\hat{a}_0\hat{\lambda}_{(k,m)}$ obtained from the formula $\hat{\eta} = \hat{a}_0\cos(\hat{\lambda}_{(k,m)}\hat{t}\;)$. If we hypothesize that even at these large amplitude deformation, the bubble surface continues to behave as a linear standing wave indicated by this formula, the velocity at the symmetry axis ($\hat{r}=0$) would then vary between plus and minus one on this scale. Figure \ref{fig16} shows  that as $\Delta/H$ increases from $0.726$ to $0.847$, there is a  sharp increase (i.e. very high acceleration) in the vertical velocity reaching a peak value of close to twenty times the linear estimate. The insets which depicts the shape of the interface around the symmetry axis, show that the peak velocity (for $\Delta/H = 0.847$) coincides with the emergence of the dimple and reduces as the jet emerges around $\tilde{t} \approx 0.3$. Note that this behaviour becomes qualitatively different at a slightly lower value of $\Delta/H=0.726$. Here the acceleration of the dimple is much smaller and the peak velocity is only about five times the linear estimate. We can thus use figure \ref{fig16} to obtain an objective criteria to decide when should we label the emerging structure, as being a `jet'. It is clear that for $\Delta/H=0.847$ around $\tilde{t}\approx 0.3$, the dimple that emerges may be called the precursor to a jet, as it is born with very high acceleration (note the slope of the purple curve) and consequently velocity. In concluding this sub-section, we note the interesting observations made in \citet{gordillo2019capillary} who observe that for an air bubble in water at sufficiently low Bond number and of radius $\sim 0.3$ mm, the jet that is experimentally observed when the bubble bursts has a velocity of $\sim 10$ m/s whereas the linear estimate from the capillary velocity scale based on the bubble radius yields a velocity estimate of only $0.5$ m/s in qualitative agreement with fig. \ref{fig16} of our study.

\subsection{Modal analysis: nonlinear transfer of energy}
Having discussed the emergence of the jet at $\Delta/H = 0.847$, it is useful to enquire about the nonlinear transfer of energy into other modes in the spectrum. That there must be such transfer is clear from the appearance of the dimples in fig. \ref{fig14}, whose width is far smaller than a characteristic length scale of the initial modal excitation at $\tilde{t}=0$. In this section, we further analyse this transfer of surface energy into modes in the spectrum, which are absent initially. For this the instantaneous shape of the (axisymmetric) bubble $\hat{r}_s(s,\hat{t})$ at any time instant $\hat{t}_0$ is projected into the (axisymmetric) eigenmodes $y_i(s)$ (we arrange the axisymmetric shape modes $y_{(k,0)}(s),k=2,4,6\ldots$ in increasing order labelling them as $y_1(s),y_2(s),y_3(s)\ldots$) as
\begin{eqnarray}
	\hat{r}_s(s,\hat{t}_0) = \hat{R}_0 + \hat{\eta}(s,\hat{t}_0) = \hat{R}_0 + \sum_{i=1}^{N}C_i(\hat{t}_0)y_{i}(s). \label{eqn5_1}
\end{eqnarray}
We determine the $C_i(\hat{t}_0)$ in eqn. \ref{eqn5_1} using the usual inner product definition i.e.
\begin{eqnarray}
	&&\int_{b}^{1}\hat{\eta}(x,\hat{t}_0)y_j(x)dx = \sum_{i=1}^{N}\left(\int_{b}^{1}y_j(x)y_i(x)dx \right)C_i(\hat{t}_0),\nonumber \\
	&&j=1,2\ldots N. \label{eqn5_2}
\end{eqnarray}
The integrals in eqn. \ref{eqn5_2} are performed numerically and this leads to a linear set of equations in $C_1(\hat{t}_0),C_2(\hat{t}_0)...C_N(\hat{t}_0)$ which are then solved in Python \cite{python2021python} to determine their values. These are then substituted in eqn. \ref{eqn5_1} to reconstruct the bubble shape. 

\begin{figure}
	\centering
	\subfloat[]{\includegraphics[scale=0.28]{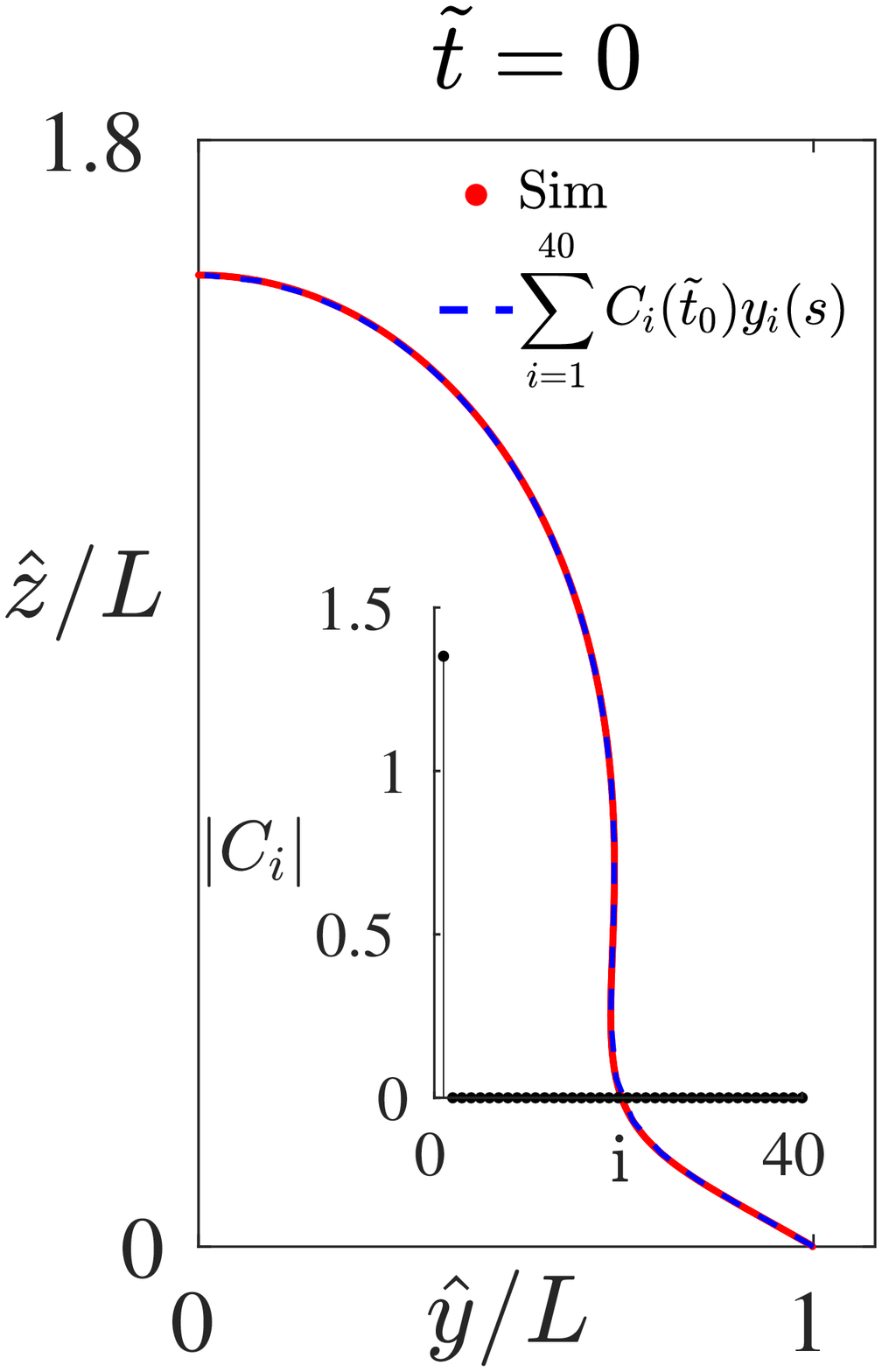}}
	\subfloat[]{\includegraphics[scale=0.28]{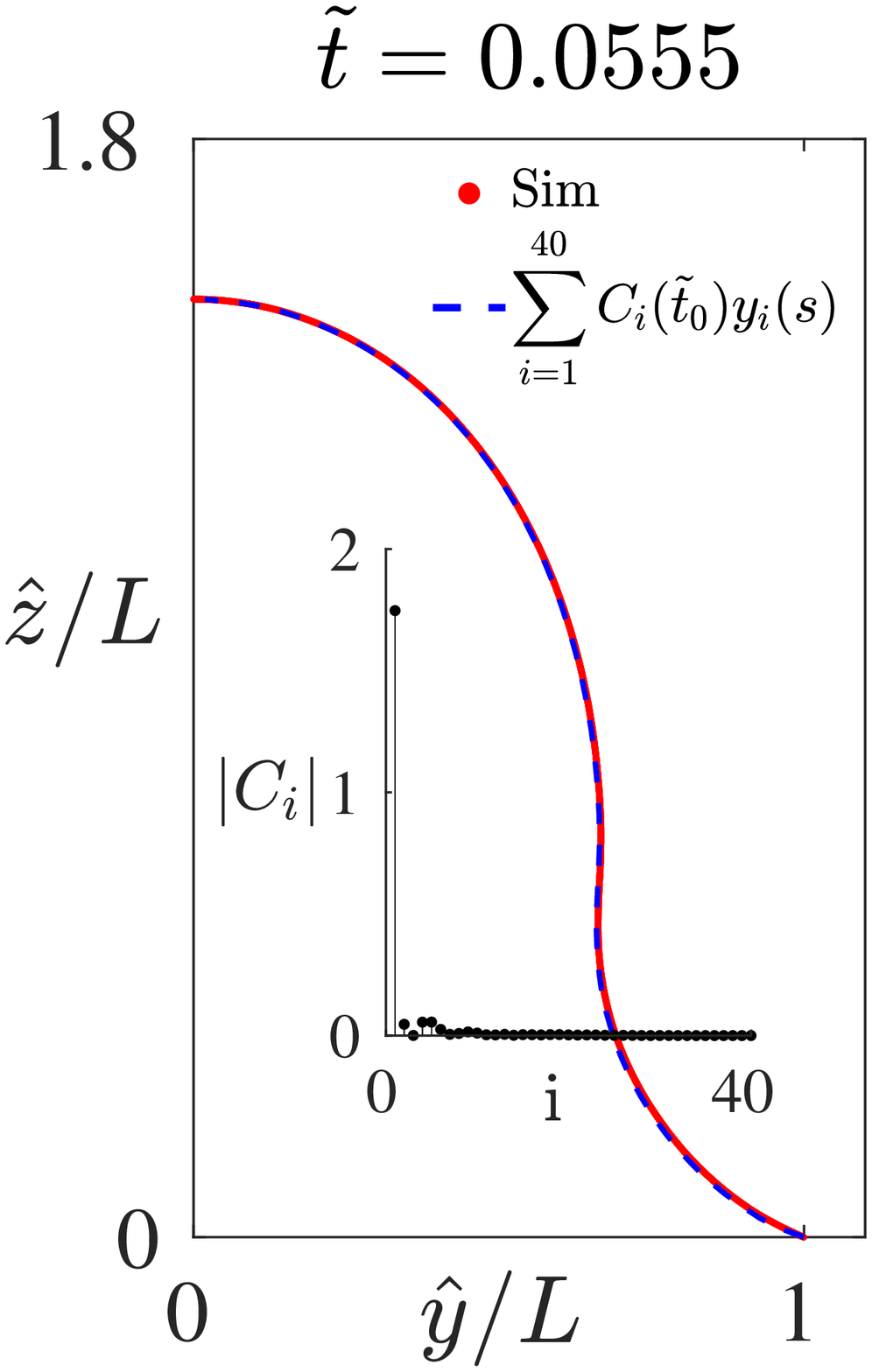}}\\
	\subfloat[]{\includegraphics[scale=0.28]{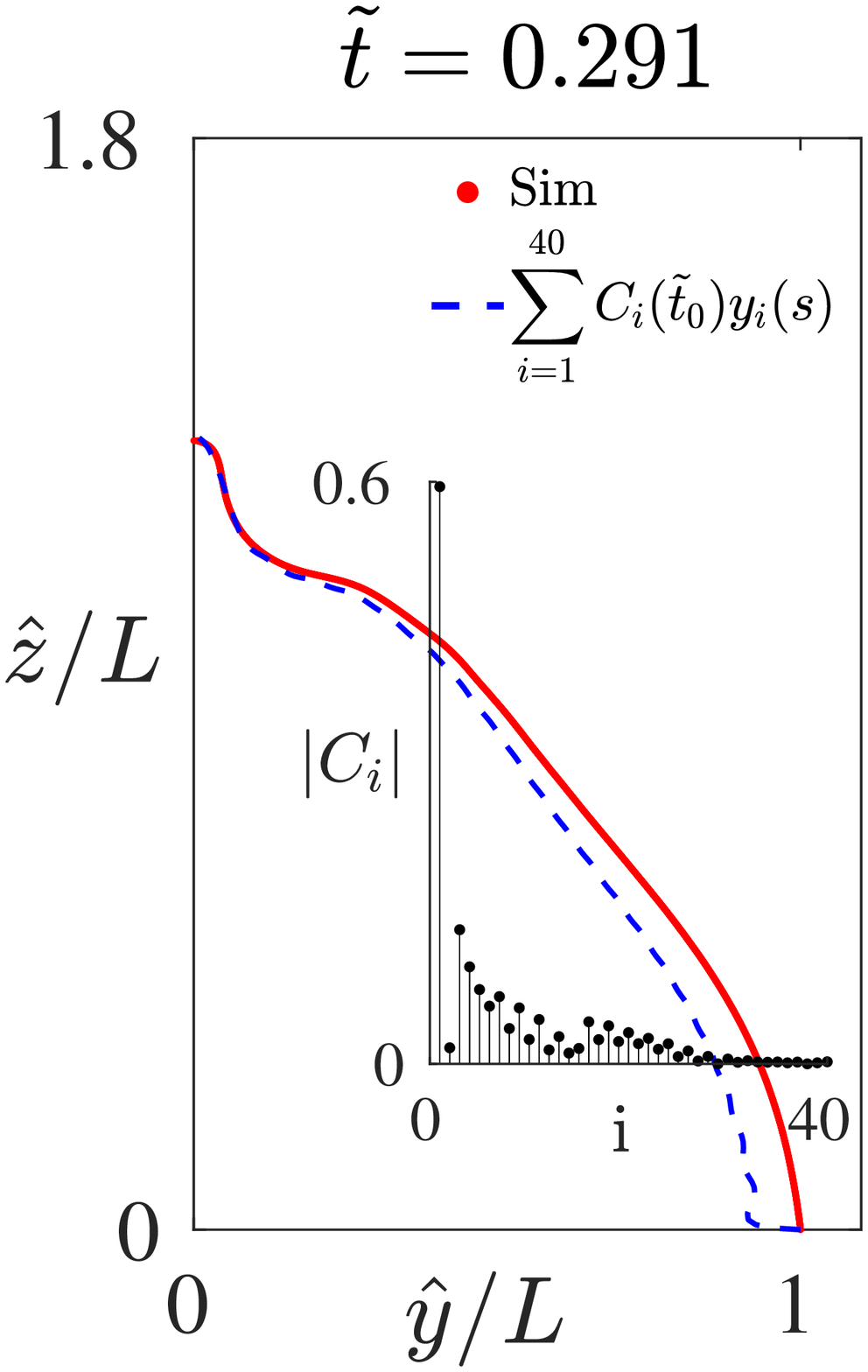}}
	\subfloat[]{\includegraphics[scale=0.28]{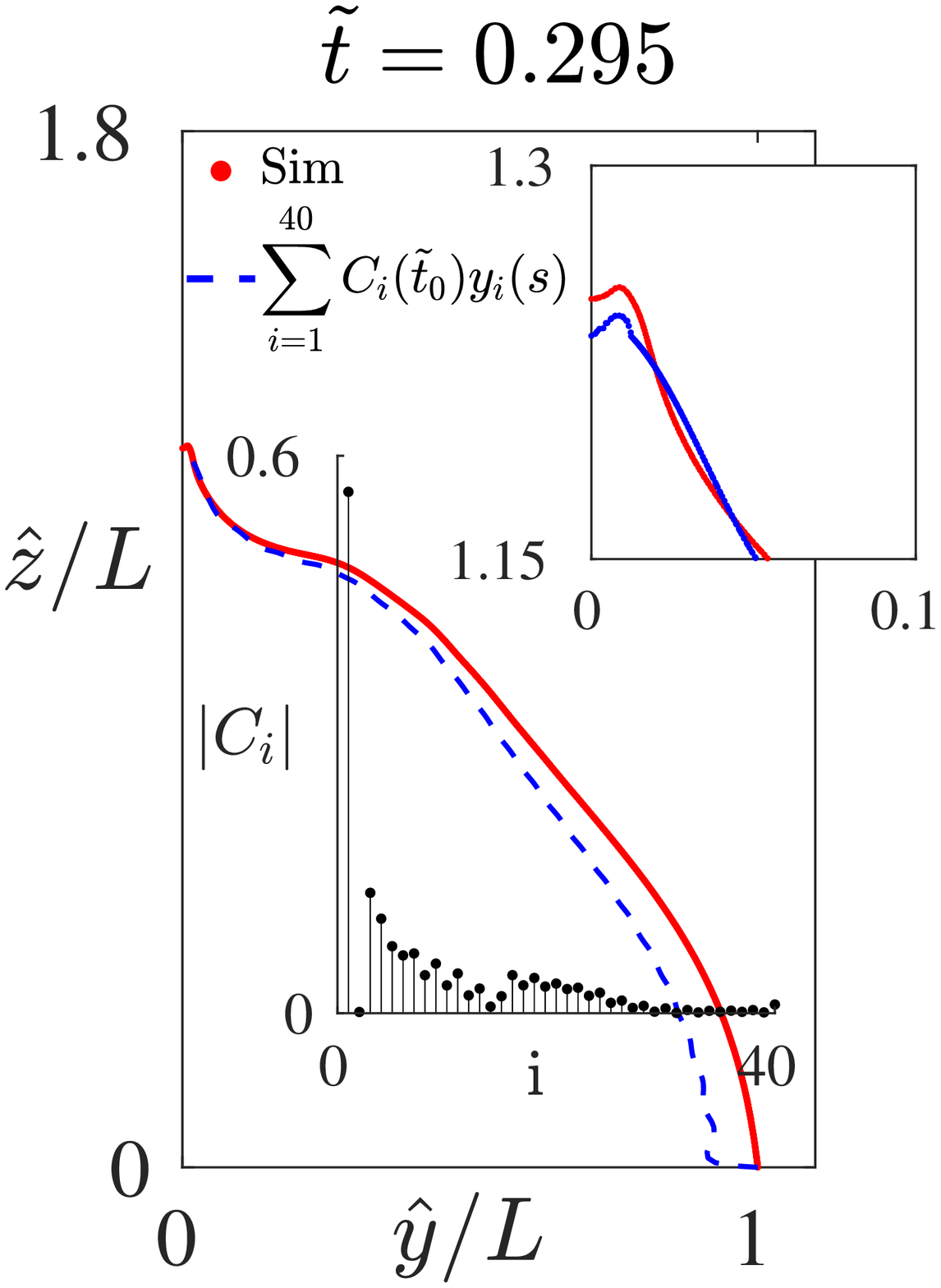}}
	\caption{The instantaneous shape of the interface from simulation (Sim) is decomposed into the shape modes of the bubble at different instants of time using eqn. \ref{eqn5_1} and the bubble shape is reconstructed using this. Panels in the inset depict the absolute value of the modal coefficients viz. $|C_i(\tilde{t}_0)|$ versus $i$ at various instants of time $\tilde{t}_0$. Panel (a) $\tilde{t}_0=0$, only one mode is excited initially viz. $y_{(2,0)}(s)$ and thus $C_1 \neq 0$ (b) $\tilde{t}_0=0.255$ (c) $\tilde{t}_0=0.291$ (d) $\tilde{t}_0=0.295$. Note the presence of energy is modes significantly beyond $C_1$. The inset to panel (d) shows a magnified image of the dimple around the symmetry axis in the main figure.}
	\label{fig18}
\end{figure}

\begin{figure}
	\centering
	\subfloat[]{\includegraphics[scale=0.28]{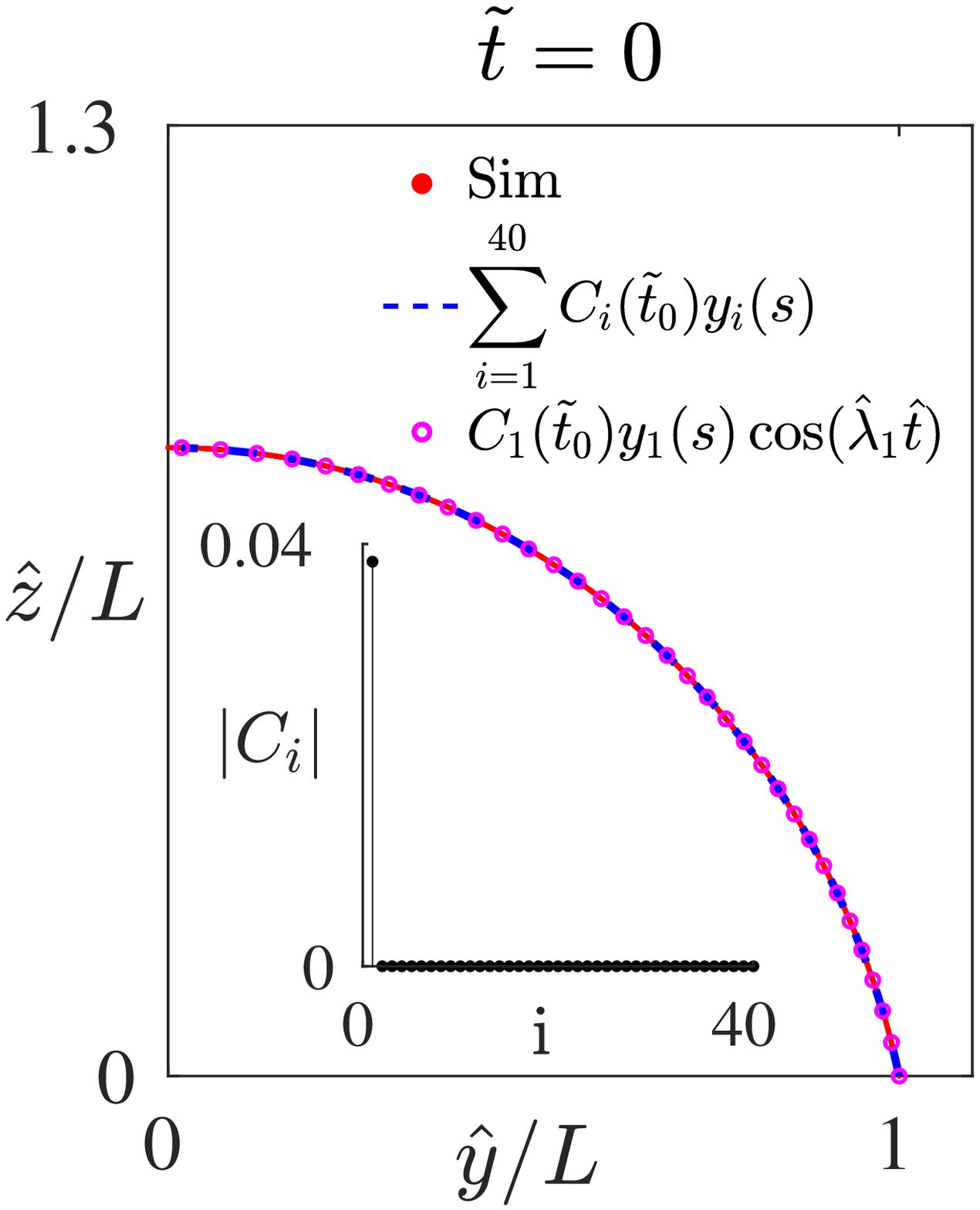}}
	\subfloat[]{\includegraphics[scale=0.28]{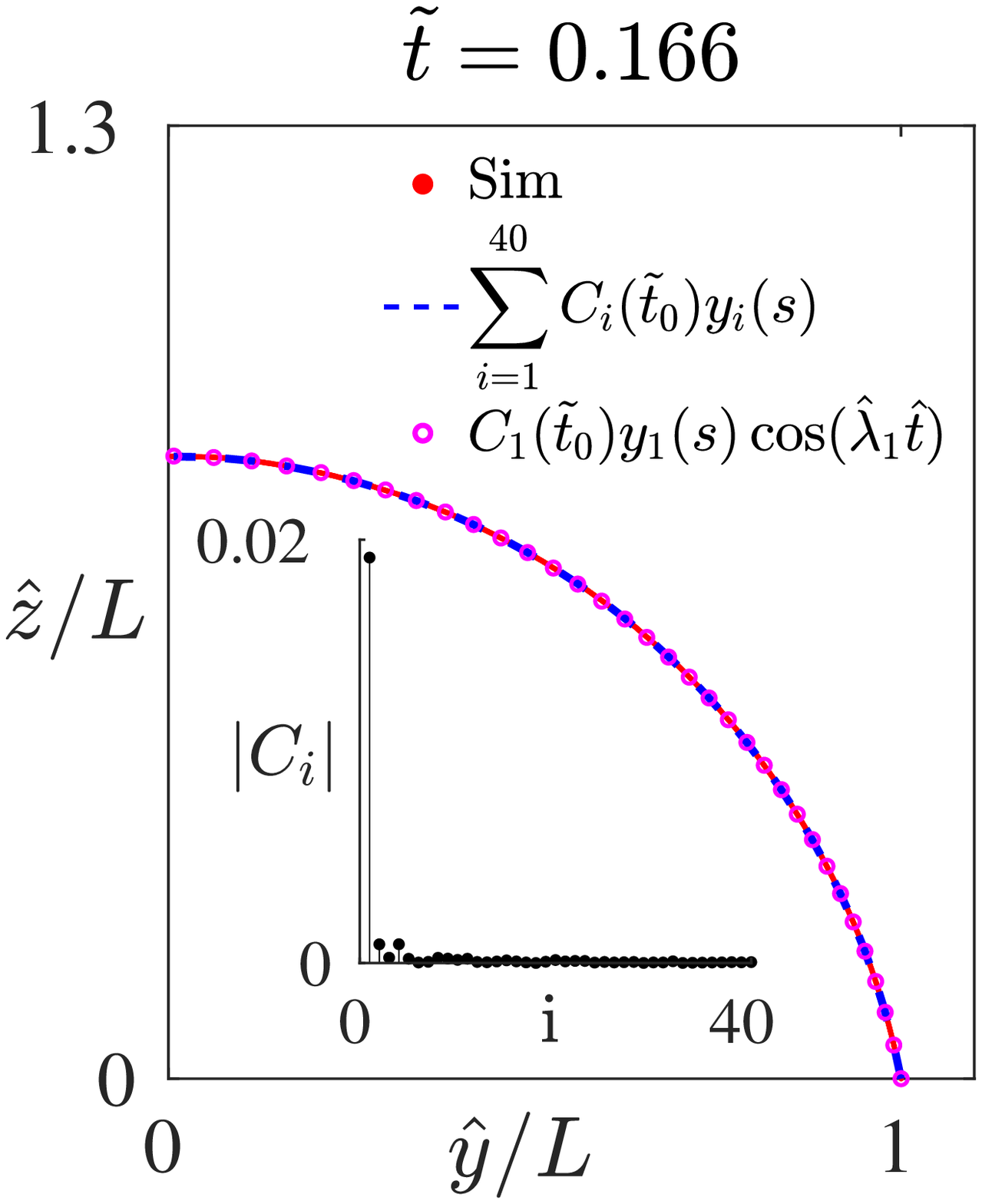}}\\
	\subfloat[]{\includegraphics[scale=0.28]{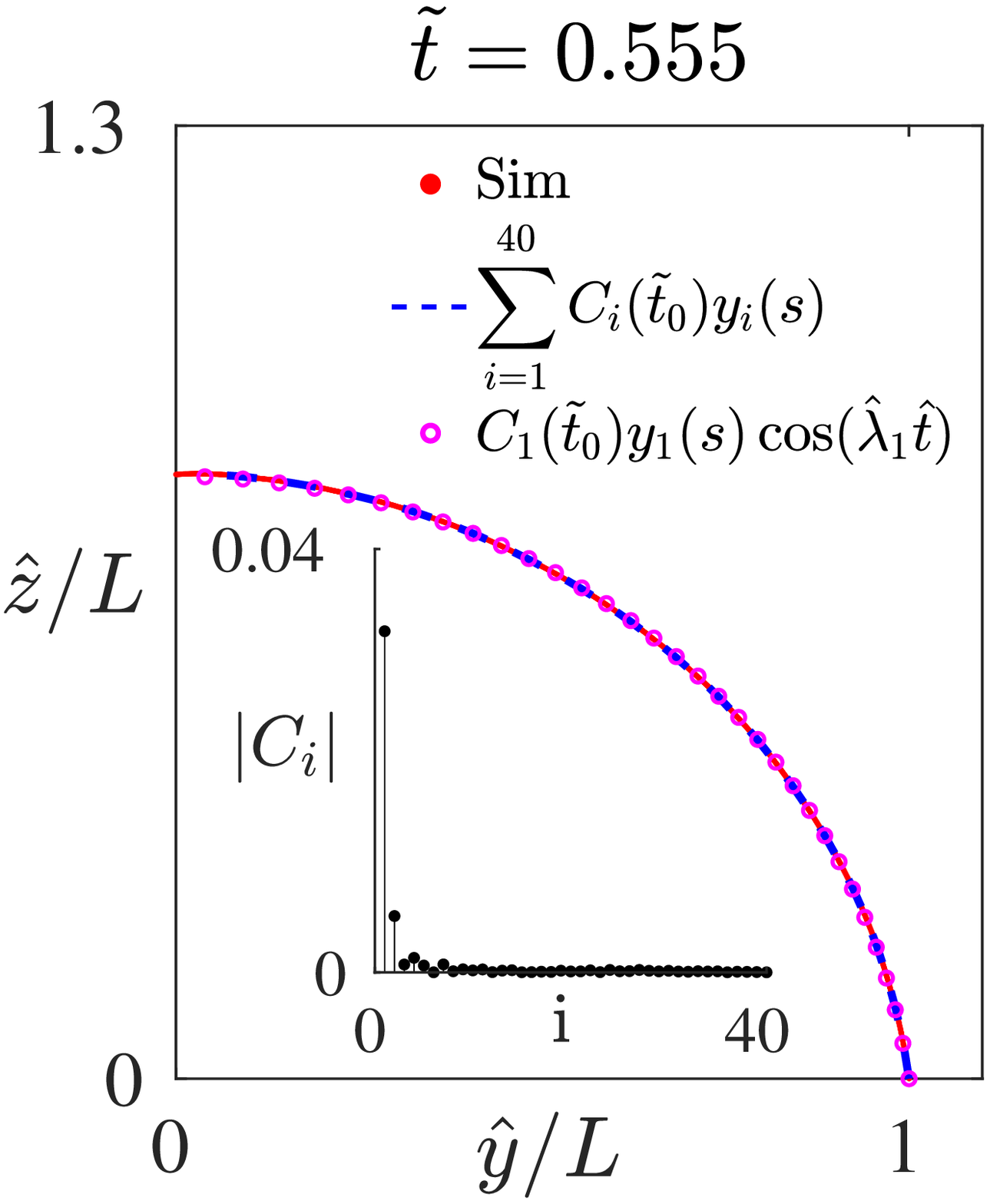}}
	\subfloat[]{\includegraphics[scale=0.28]{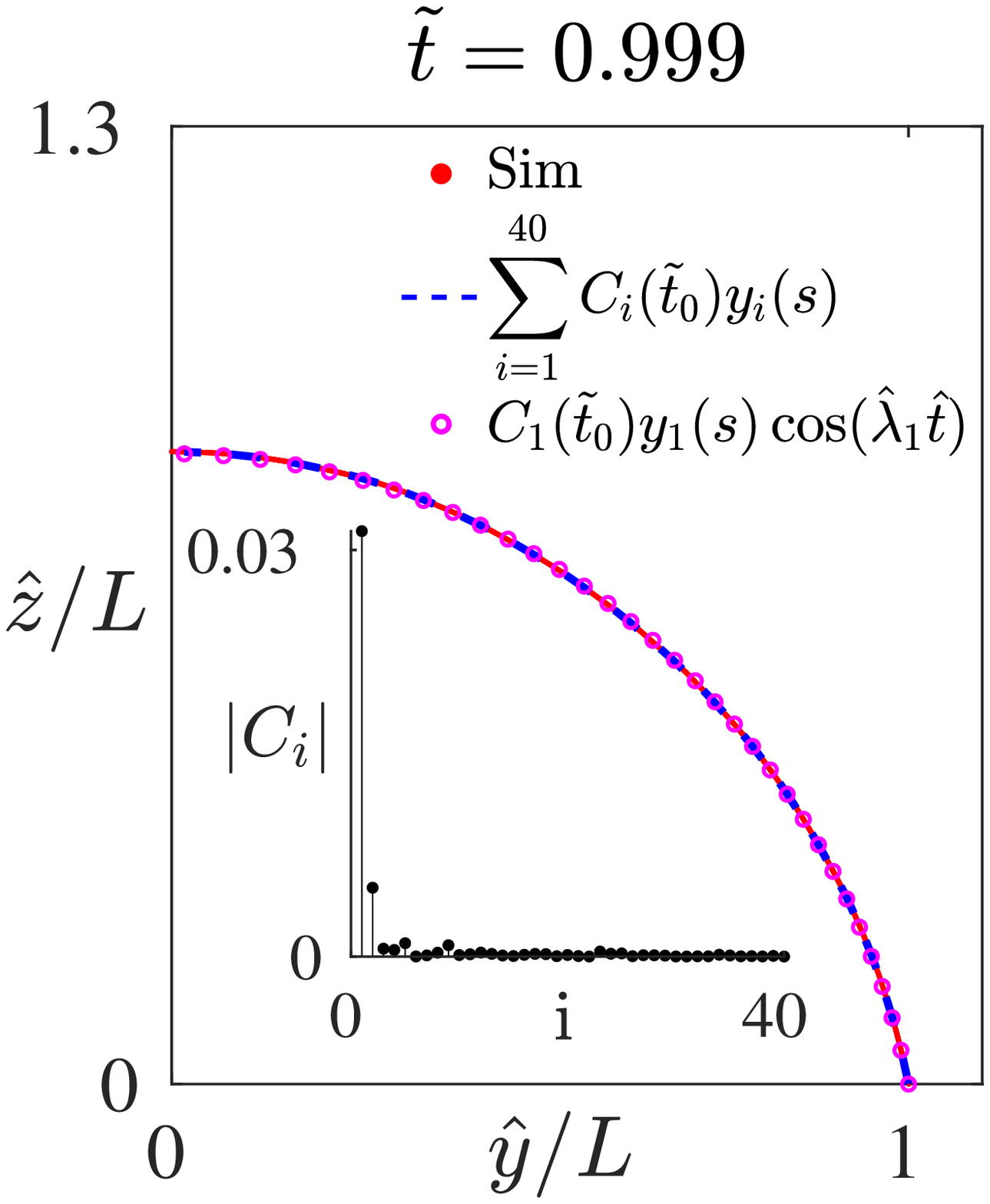}}
	\caption{Modal decomposition and reconstruction at different time instants for panel (d) in fig. \ref{fig8}, panel (a). To a very good approximation, the mode which is excited initially is the only mode which persists at later time. This explains the good agreement of the interface shape with linear theory seen in the figure represented by magenta dots.}
	\label{fig18b}
\end{figure}

Fig. \ref{fig18} depicts the reconstructed bubble shape in this manner for the third colum of figure \ref{fig14} with $\Delta/H=0.847$. In the insets of fig. \ref{fig18}, we depict this modal decomposition as a function of time i.e. the absolute value of the modal coefficients $|C_i(\hat{t}_0)|$ as a function of $i$ (positive integer values only) at different time instants $\hat{t}_0$ upto the generation of the jet. Also depicted in these figures, is the reconstructed interface (in blue) from these modal coefficients. It is seen from the panels that although we start with only a single mode (panel (a) has a single peak for $C_1$), the (surface) energy rapidly re-distributes into several modes as seen from the insets of the other panels. Our numerical eigenvalue calculation, allows us access to forty ($N=40$) modes and it is seen from panels (c) and (d) in fig. \ref{fig18} that while the reconstructed interface (blue curve in each of the panels) matches the bubble shape reasonably accurately close to the symmetry axis, there are differences in the far-field near the pinning location. Fig. \ref{fig18b} represents the same modal decomposition and reconstruction of the bubble shape, but now in the linear regime where we had seen good agreement earlier for fig. \ref{fig8}. Consistent with this, it is observed that the dominant mode at later time remains the mode, which was excited initially. This behaviour in the linear regime in fig. \ref{fig18b} is to be contrasted against the nonlinear behaviour observed near dimple inception  in fig. \ref{fig18}. The observation of excitation of several modes in fig. \ref{fig18} at the instant of dimple formation and subsequent jet ejection, also supports the earlier experimental observations of \citet{prabowo2011surface}. These authors excited a $160$ micron spherical cap bubble at a contact angle of $120^{\circ}$ with a transducer and observed the generation of a wall-directed jet.  Interestingly, these authors show that when the jet ejects from the bubble, there are several shape modes (in addition to the volume mode) which have been excited, roughly consistent with our observations in fig. \ref{fig18} where only shape modes are seen to be \textit{enough} in the generation of this jet.  

\begin{figure}
	\centering
	\includegraphics[scale=0.45]{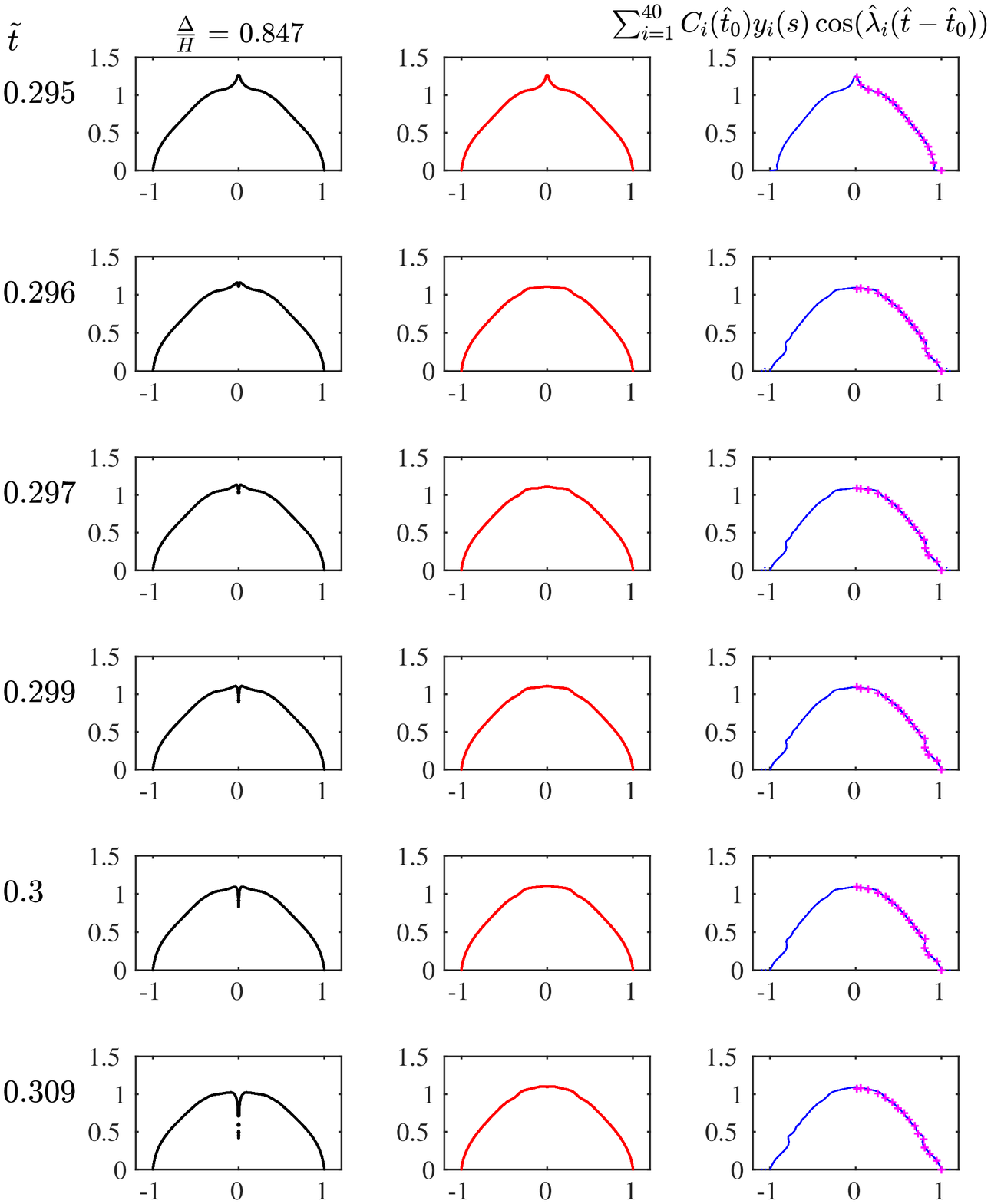}
	\caption{(Left column) Same as column three of fig. \ref{fig14}. (Middle column) Time evolution of the bubble shape indicated in red  in panel (d) of fig. \ref{fig18} in numerical simulation, by setting the velocity field to zero everywhere in the domain. (Right column) Time evolution of the bubble shape in blue in panel (d) of fig. \ref{fig18} in numerical simulation, by setting the velocity field to zero everywhere in the domain. The magenta `+' symbols indicate the bubble shape evolution predicted by linear theory viz. in dimensional variables $\hat{r}_s(s,t) = \hat{R}_0 + \sum_{i=1}^{40}C_i\left(\hat{t}_0\right)y_i(s)\cos\left[\hat{\lambda}\left(\hat{t}-\hat{t}_0\right)\right]$}
	\label{fig19}
\end{figure}

Fig. \ref{fig19} is to be read column-wise and all results represent axisymmetric cases. The first column presents the simulation discussed earlier in fig. \ref{fig18} for visual reference, starting from dimple formation at $\tilde{t}_0=0.295$ (panel (d) of figure \ref{fig18}, see inset). To demonstrate the subsequent effect of the velocity-field in the liquid outside the bubble on the evolution of the dimple, we have re-run the same simulation by initialising the interface shape as depicted by the red curve in panel (d) of fig. \ref{fig18}, but setting the velocity-field everywhere in the domain, artificially to zero in the simulation initially. The bubble shape evolution in time starting from such an initial condition is shown in the second column of fig. \ref{fig19}. Upon comparison with the first column, it is observed that the presence of the velocity-field is necessary for jet formation and without which the dimple does not evolve into the jet seen in the first column. The third column (first row) of figure \ref{fig19} provides the time evolution of the bubble surface in two ways. The curve in blue corresponds to the bubble shape (also in blue) in panel (d) of fig. \ref{fig18}. This bubble shape is evolved in time in Gerris, once again by setting the velocity field to zero. The bubble shape indicated by magenta plus symbols in the third column of fig. \ref{fig19}, represent the prediction for the shape with time and zero velocity everywhere, had the bubble shape evolved linearly following the linearised prediction $\sum_{i=1}^{40}C_i\left(\hat{t}_0\right)y_i(s)\cos\left[\hat{\lambda}\left(\hat{t}-\hat{t}_0\right)\right]$ with $\tilde{t}_0=0.295$. It is seen that the simulation (blue curve) and the linear theory (magenta symbols) match each other quite closely indicating the modes present at $\tilde{t}_0=0.295$ evolve independent of each other and linearly. However, when one compares the predictions between the second and third column at $\tilde{t}=0.299$ and beyond, differences are seen. These are due to differences at $\tilde{t}=0.295$ between the actual interface shape in the simulation and the summation $\sum_{i=1}^{40}C_i\left(\hat{t}_0\right)y_i(s)$, as seen earlier in panel (d) of figure \ref{fig18}.

\subsection{Local self-similar behaviour}
The generation of a jet due to wave focussing and the large velocities and accelerations associated with this, have been interpreted mathematically \citep{zeff2000singularity,duchemin2002jet,gekle2010generation,lai2018bubble} as being due to a generation of a (near) singularity in interface curvature at the axis of symmetry. Around the time of formation of this singularity, it may be expected that the local flow around the singularity evolves in a self-similar manner following power-law scaling and independent of details such as size of the bubble or the boundary conditions at the CL. Employing cylindrical axisymmetric coordinates $\hat{r},\hat{z}$ (instead of spherical coordinates employed so far), we write the functional dependence for the interface shape $\hat{\eta}$ and disturbance velocity potential $\hat{\phi}$ at any time $\hat{t}$ (hats are used for dimensional variables)
\begin{subequations}\label{eqn4_2}
	\begin{align}
		\hat{\eta} =\hat{f}\left(\hat{r},\;\hat{t},\;\frac{T}{\rho\OUT},\;\hat{R}_0,\;L,\;\hat{a}_0\right) \tag{\theequation b} \\
		\hat{\phi} = \hat{g}\left(\hat{r},\;\hat{z},\;\hat{t},\;\frac{T}{\rho\OUT},\;\hat{R}_0,\;L,\;\hat{a}_0\right) \tag{\theequation b}
	\end{align}
\end{subequations}

\noindent where $L = \hat{R}_0\sin(\alpha)$, see fig. \ref{fig2}. Non-dimensionally eqn. \ref{eqn4_2} a, b may be re-written as
\begin{subequations}\label{eqn4_3}
	\begin{align}
		\frac{\hat{\eta}}{L} = f\left(\frac{\hat{r}}{L},\;\hat{t}\left(\frac{T}{\rho\OUT L^3}\right)^{1/2};\; \epsilon,\;\alpha\right) \tag{\theequation a} \\ \frac{\hat{\phi}}{L^{1/2}\left(\frac{T}{\rho\OUT}\right)^{1/2}} = g\left(\frac{\hat{r}}{L},\;\frac{\hat{z}}{L},\;\hat{t}\left(\frac{T}{\rho\OUT L^3}\right)^{1/2};\; \epsilon,\;\alpha\right) \tag{\theequation b}
	\end{align}
\end{subequations}
where $\epsilon \equiv \hat{a}_0/\hat{R}_0$ and the explicit dependence on $\sin(\alpha)$  has been suppressed in favour of $\alpha$. Sufficiently close to the singularity which occurs at $\hat{t}=\hat{t}_0$, we hypothesize that the time evolution of the interface, locally evolves in a manner which is independent of the variable $L$ in eqns. \ref{eqn4_3}. This is because this variable encodes information about the size of the bubble $\hat{R}_0$ and its contact angle $\alpha$. Sufficiently close to the singularity, we may anticipate that these parameters do not affect the time evolution of the interface in a small region around the singularity with nearly divergent accelerations and velocities therein. If this ansatz is true, then it must be possible to express \ref{eqn4_3} a and b in such a manner that the time evolution of the interface becomes independent of $L$. This implies that the functions $f$ and $g$ in eqn. \ref{eqn4_3} a, b instead of depending explicitly on $\hat{r}/L,\;\hat{z}/L$ and $\hat{t}\left(\frac{T}{\rho\OUT L^3}\right)^{1/2}$ must depend on suitable combinations of these such that both left hand side and right  hand side of \ref{eqn4_3} a and b are independent of $L$.  Dimensional consideration shows that the required scales are unique and must be of the form
\begin{eqnarray}\label{eqn4_4}
		\tilde{\tilde{\eta}} \equiv  \frac{\hat{\eta}-\hat{z}_b}{\left(\hat{t}-\hat{t}_0\right)^{2/3}\left(\frac{T}{\rho\OUT}\right)^{1/3}}  = \tilde{\tilde{f}}\left(\tilde{\tilde{r}}\;;\;\epsilon,\alpha\right), 
\end{eqnarray}
where $\tilde{\tilde{r}} \equiv \frac{\hat{r}}{\left(\hat{t}-\hat{t}_0\right)^{2/3}\left(\frac{T}{\rho\OUT}\right)^{1/3}},\quad \tilde{\tilde{z}} \equiv \frac{\hat{z}}{\left(\hat{t}-\hat{t}_0\right)^{2/3}\left(\frac{T}{\rho\OUT}\right)^{1/3}}$ and $\hat{z}_b$ is the vertical coordinate of the extrema at the base of the dimple.

\begin{figure}
	\centering
	\subfloat[Interface shape]{\includegraphics[scale=0.3]{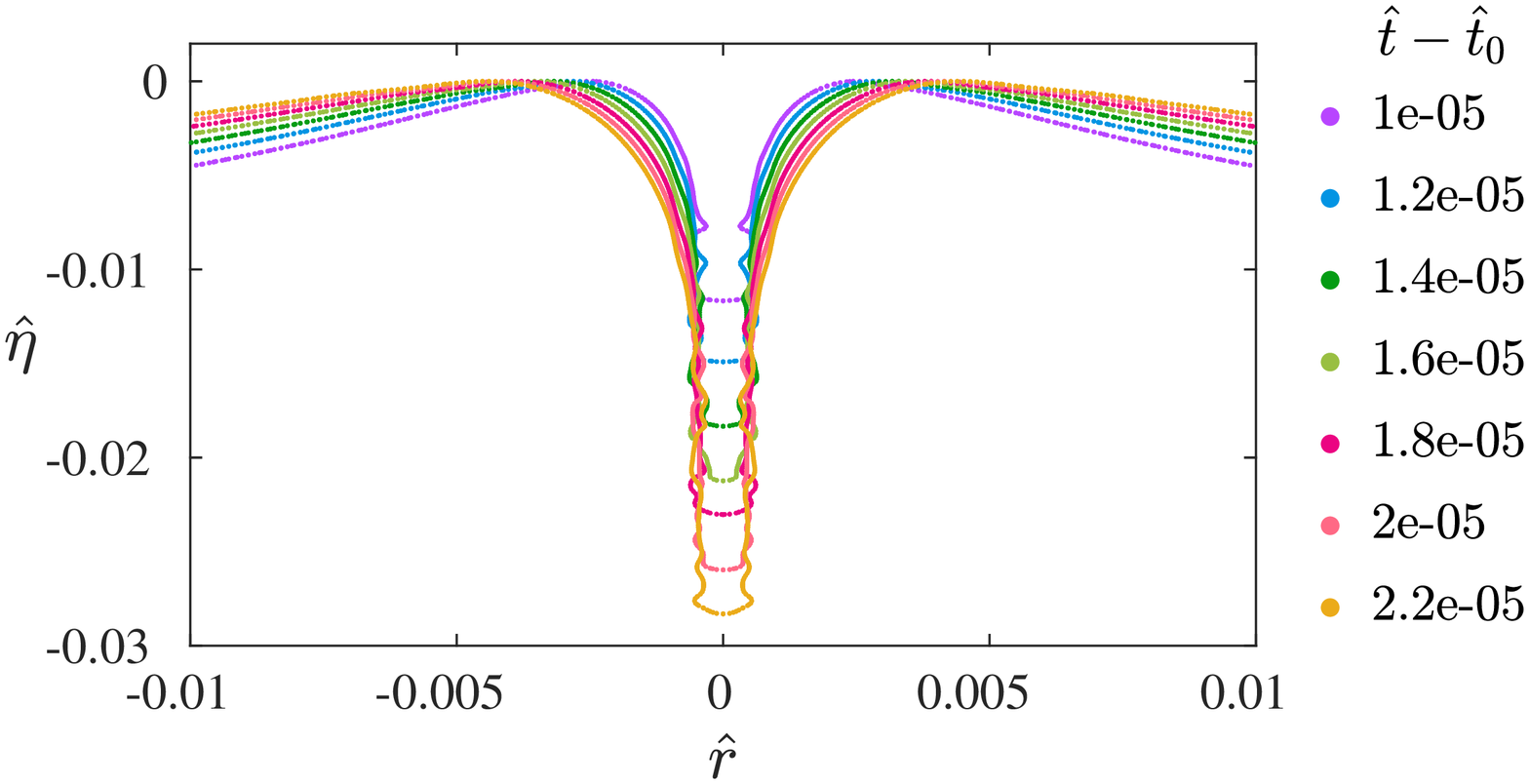}	\label{fig20a}}\\
	\subfloat[Collapse]{\includegraphics[scale=0.3]{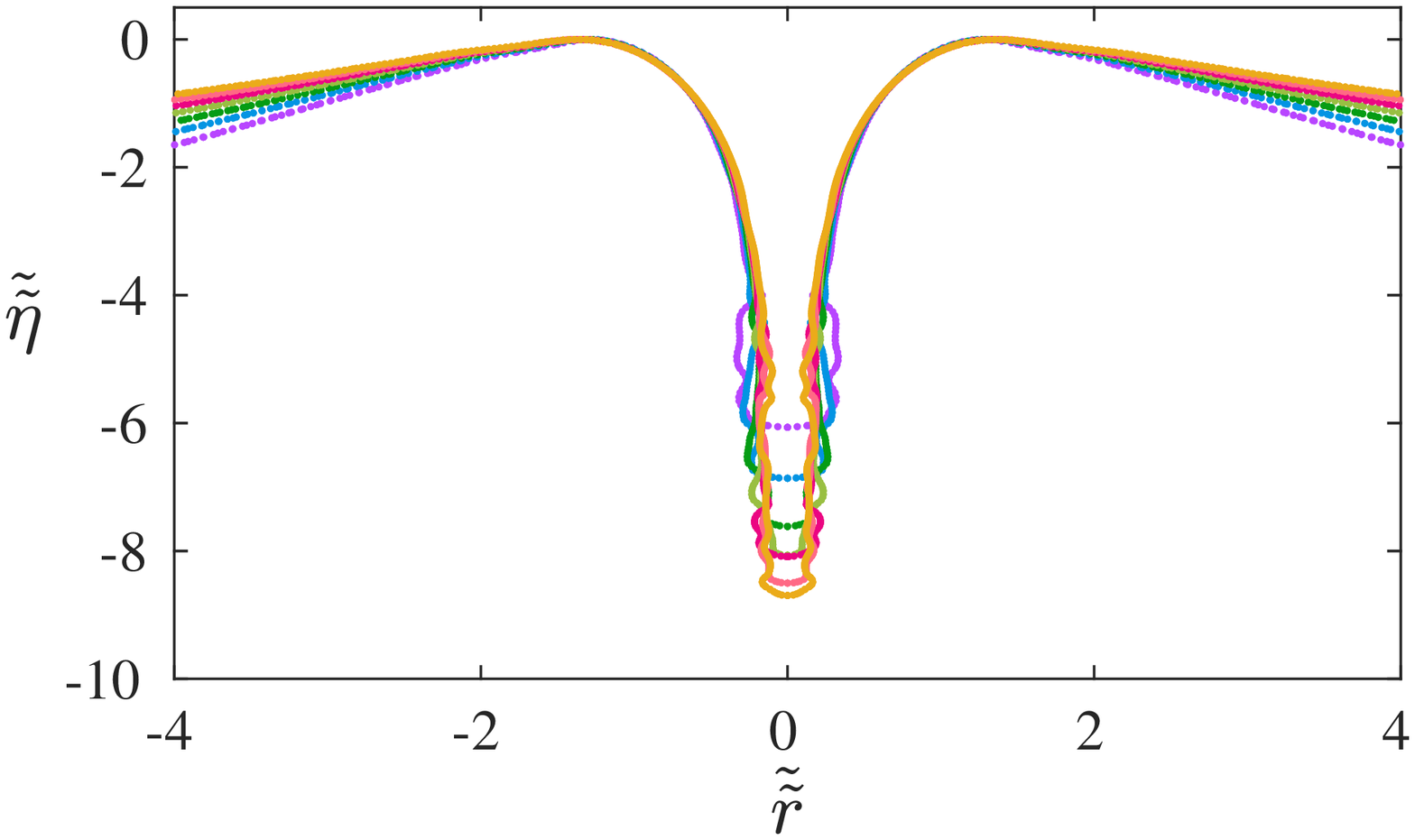}\label{fig20b}}
	\caption{Self-similar evolution of the interface. Panel (a) depicts the interface shape after the emergence of the dimple at various time instants. (b) Collapse of the interfaces at different instants of time on to a single curve using the scales of eqns. \ref{eqn4_4}.}	
	\label{fig20}
\end{figure}

We emphasize that eqn. \ref{eqn4_4} implies that the functional form of $\tilde{\tilde{f}}$ depends on the parameters $\alpha$ and $\epsilon$ but the temporal evolution of the interface happens independent of these implying the interface at different instants of time, can be collapsed onto a single master curve, whose \textit{shape} depends on the parameters $\epsilon$ and $\alpha$. The scales in \ref{eqn4_4} are the well-known similarity scales found by \citet{keller1983surface}. In particular \citet{zeff2000singularity} demonstrated that these exponents in eqn. \ref{eqn4_4} are unique only when surface tension is considered in the fully non-linear governing equations and boundary condition. Without including surface tension, an infinite family of such exponents can be found (see discussion around eqn. $3$ in \citet{duchemin2002jet}). The validity of the scaling in expression  \ref{eqn4_4} for the interface variable $\tilde{\tilde{\eta}}$ is demonstrated in fig. \ref{fig20} where it is seen that after applying the required scaling, the interface at different instants of time around the axis of symmetry collapses into a single master curve (panel (b) of fig. \ref{fig20}). This data in fig. \ref{fig20}
is taken after the emergence of the dimple and is obtained from the simulation shown earlier in column three of fig. \ref{fig14}. It must be remarked here that despite the collapse seen in fig. \ref{fig20}, the collapse extends to less than a decade in time and thus the inertio-capillary balance that has been used to arrive at the self-similar ansatz in eqn. \ref{eqn4_4} needs to be revisited \cite{gordillo2019capillary}. We also note that a purely inertial mechanism of axisymmetric jet formation devoid of gravity or surface tension in the potential flow limit, was proposed by \citet{longuet1983bubbles} and the applicability of this model, to our situation is proposed for future study.  
\subsection{Effect of viscosity and gravity on jet formation}
\begin{figure}
	\centering
	\includegraphics[scale=0.5]{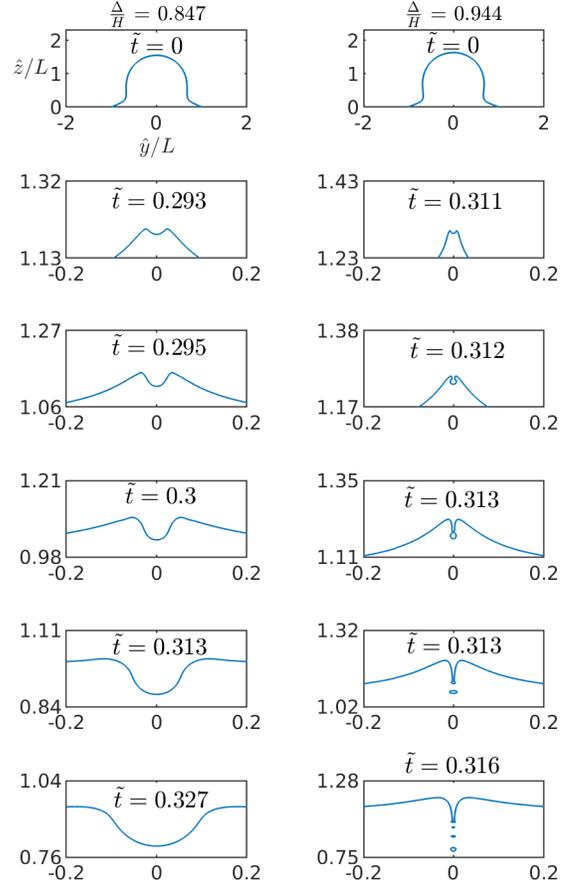}
	\caption{Time evolution of the bubble shape accounting for the outer fluid viscosity value approximated as water i.e. $\mu\OUT = 0.01 \text{gm/cm-s}$. The first column of this figure has the same value of $\Delta/H$ as the third column of fig. \ref{fig14} earlier. Comparing the two figures, we see that taking into account viscosity in the fluid outside, weakens the jet although the production of a dimple-like structure is still evident around $\tilde{t}=0.293$. In the second colum of this figure, we depict a viscous simulation with $\Delta/H=0.944$. It is evident that qualitative behaviour seen in the inviscid simulations in figure \ref{fig14} is recovered even in the viscous case but now for larger values of $\Delta/H$.}
	\label{fig21}
\end{figure}

In this section, we return to a re-evaluation of the effect of viscosity on jet formation. Our inviscid simulations have neglected the effect of numerical viscosity, which can have an effect on the simulations. In the current study, we have ignored this as the effect of numerical viscosity is expected to be substantial only beyond five linear oscillation time-periods (see fig. \ref{fig8}), whereas the jet of current interest, appears within one linear oscillation time-period where the effect of numerical viscosity appears negligble for $(2,0)$ mode, see \ref{fig8}, panel (a). As the linear theory that has been presented earlier is inviscid, we do not have analytical access to the damping rate of the modes. Consequently, our investigation of viscous effects in this section is completely numerical. Studying the effect of viscosity is particularly important because results of the previous section demonstrate that jet formation is accompanied by transfer of energy to a large number of modes in the spectrum, exceeding forty (see last panel of fig. \ref{fig18}). Since viscosity is expected to damp these high frequency modes, we expect that dimple and jet formation at the value of $\Delta/H$ examined so far, to be strongly affected by viscosity. This is demonstrated in figure \ref{fig21}, where the left column of this figure represents the same simulation that was presented earlier in fig. \ref{fig14}, rightmost column but now accounting for the viscosity of the fluid outside. It is seen that while a dimple like structure is still formed, it is significantly wider than the dimple seen in the inviscid case, due to damping of higher modes. The right column of fig. \ref{fig21} establishes that nonlinearity indeed competes with viscosity in generating the dimple and the jet. It is seen from this viscous simulation at a even higher value of $\Delta/H=0.944$, that even for a bubble with viscous liquid outside, sharp dimples and jet like structures can be indeed formed. Interestingly, droplets from the jet tip seem to be ejected earlier in this case compared to the inviscid simulations. Although we have ignored the effect of gravity on our spherical cap bubbles, this will have an effect on larger bubbles for which buoyant forces can distort the bubble shape away from a truncated sphere. For sessile droplets, the effect of gravity on the drop shape and the resultant linearised temporal spectrum has been recently studied in \citet{zhang2023effects}. A similar analysis, also relevant to air bubbles of sizes exceeding the capillary-length where gravity is expected to be important in the base-state is proposed as future work.  

\section{Discussion \& Conclusions}\label{sec:conc}
In this study, we have reformulated the normal mode analysis presented in \citet{ding2022oscillations} as an inviscid IVP, obtaining the corresponding simple harmonic oscillator equation. We have solved the eigenvalue problem and used the eigenmodes to validate the linearised theory for the pinned, shape modes of a sessile bubble via numerical simulations of the incompressible, Euler's equation with surface tension \citep{popinet2009accurate}. Linear theory has been tested for two particular contact angles viz. $\alpha=80^{\circ}$ and $115^{\circ}$. For both cases,  reasonably good agreement is observed between theoretical predictions and simulations for axisymmetric and three-dimensional modes, when the modal amplitude is small. We extend our simulations to the weakly nonlinear regime where reduction in the frequency, is observed consistent with what is known for finite amplitude oscillations of a free bubble. When larger amplitude distortions are considered, a dimple is observed which can develop into a jet, formed via focussing at the symmetry axis. These are reminiscent of similar observations of jets from finite amplitude capillary-gravity \citep{farsoiya2017axisymmetric,basak_farsoiya_dasgupta_2021} or pure capillary waves \cite{kayal2022dimples}.  For such deformations characterised by the non-dimensional parameter $\Delta/H$,  a sharp increase in acceleration at the bubble surface is observed as this parameter is increased by a relatively small amount.  We find dimples ejecting with velocities about twenty times the linear velocity. It is shown that the formation of the dimple coincides with the spread of (surface) energy into a large number of shape modes. Liquid viscosity strongly affects these shape modes and damps out the emergence of the jet; however at sufficiently larger value of $\Delta/H$ even for a viscous simulation, behaviour qualitatively similar to the inviscid behaviour is recovered.

In concluding this study, we return to the approximation discussed the introduction wherein we have ignored energy transfer to the breathing mode, even at large distortion amplitudes. An approximate justification for this was provided in the introduction but requires more careful validation against numerical simulations where the bubble volume is allowed to vary and this is proposed as future work. Our study is closely related to \citet{naude1961mechanism} who solved the weakly nonlinear initial value problem in the potential flow limit for a hemispherical bubble. While their analysis did not allow the  pressure inside the cavity to change and thus corresponds to an incompressible bubble, the monopole term $\frac{\hat{\phi}_0(\hat{t})}{\hat{r}}$ was included in their analysis (see eqn. $6$ in their study), which allowed the cavity volume to shrink in time. At second order in the perturbation parameter, they allowed interactions between the first six axisymmetric shape modes and obtained nonlinear corrections which predicted that a wall directed jet can be generated. We may analogously ask, if it is possible to carry out a weakly nonlinear analytical calculation accounting for interactions between the various shape modes that we see in our modal decomposition to predict the dimple that is seen in fig. \ref{fig14}? Recent studies from our group \cite{basak_farsoiya_dasgupta_2021,kayal2022dimples} have carried out weakly nonlinear calculations starting with a single eigenmode in a cylindrical container. These studies have shown that one can predict the generation of the dimple analytically via weakly nonlinear theory, although the jet remains inaccessible to theory. These interesting possibilities are under investigation and will be reported subsequently.

\section*{Appendix A: Derivation of Green function}
The eigenvalue problem to be solved for determining the temporal spectrum is given by eqn. \ref{eqn2_11}
\begin{eqnarray}
\left(\frac{\partial\Phi_{(j,m)}}{\partial r}\right)_{r=\csc(\alpha)}\hspace{-10mm}\left(x\right) = \lambda_{(j,m)}^2\int_{b}^{1}G(x,y;m)\Phi_{(j,m)}(r=\csc(\alpha),y)dy \nonumber
\end{eqnarray}
The expression for the Green function has been provided by \citet{ding2022oscillations} and a step-by-step derivation for this is presented below for the benefit of the reader. The Green function $G(x,y;m)$ for the operator $\mathbf{K}^{-1}$ respecting pinned boundary condition satisfies
\begin{eqnarray}
&\mathbf{K}\cdot G(x,y;m) = \delta(x-y), \quad x,y \in \left[b,1\right]  \label{eq2} \\
&\text{with}\quad G(b,y;m)=0,\quad G(1,y;m)\rightarrow \text{finite} \nonumber
\end{eqnarray}
where $\mathbf{K} \equiv \left(1-x^2\right)\frac{d^2}{dx^2} - 2x\frac{d}{dx} + \left(2 - \frac{m^2}{1-x^2}\right)$ and $b \equiv \cos(\alpha)$. In order to determine $G(x,y;m)$, we note that the equation for $\mathbf{K}$ is a special case of the associated Legendre differential equation \ref{eq3}, with $q=1$.
\begin{eqnarray}
&&\left[(1-x^2)\frac{d^2}{dx^2} -2x\frac{d}{dx} + \left(q(q+1) - \frac{m^2}{1-x^2}\right)\right]y(x;m)=0 \nonumber\\
&&\label{eq3}
\end{eqnarray}
We thus require two homogenous solutions to eqn. \ref{eq3} with $q=1$. For $m=0,1$ the two linearly independent solutions are \cite{Mathematica2020}.
\begin{eqnarray}
&& y_1(x;0) = \mathbb{P}_1^{(0)}(x)=x \nonumber\\
&& y_2(x;0) = \mathbb{Q}_1^{(0)}(x) = -1 + \frac{x}{2}\ln\left(\frac{1+x}{1-x}\right) \nonumber \\
&& y_1(x;1) = \mathbb{P}_1^{(1)}(x) = -\sqrt{1-x^2} \nonumber\\
&& y_2(x;1) = \frac{1}{2}\mathbb{Q}_1^{(1)}(x) = \frac{1}{2}\left[\frac{-x}{\sqrt{1-x^2}}+\frac{\sqrt{1-x^2}}{2} \ln\left(\frac{1-x}{1+x}\right)\right] \nonumber \\
\label{eq4}
\end{eqnarray}
For $m \geq 2$, the independent solutions to eqn. \ref{eq3} need be chosen differently. We choose these to be proportional to the Ferrers function $\mathbb{P}_1^{(-m)}(x)$ and $\mathbb{P}_1^{(-m)}(-x)$ (see \citet{olver2010nist}, section $14.2(iii)$, page $352$). Thus for $m \geq 2$,
\begin{eqnarray}
&&y_1(x;m\geq 2) = \left(m+x\right)\left(\frac{1-x}{1+x}\right)^{m/2},\nonumber \\
&&y_2(x;m\geq 2) = \frac{1}{2m(m^2-1)}\left(m-x\right)\left(\frac{1+x}{1-x}\right)^{m/2} \label{eq5}
\end{eqnarray}
The factor $\frac{1}{2m(m^2-1)}$ in $y_2(x;m\geq 2)$ in \ref{eq5} is for algebraic convenience as we will see shortly. We have
\begin{equation}
G(x,y;m) = \left\{
\begin{array}{ll}
G_L(x,y;m) \equiv A(y)y_1(x;m) + B(y)y_2(x;m) \\ \hspace{4.3cm} b \leq x < y < 1 \\
G_R(x,y;m) \equiv C(y)y_1(x;m) + D(y)y_2(x;m) \\ \hspace{4.3cm} b < y < x \leq 1 \label{eq6}
\end{array}
\right.
\end{equation}
Note that the expressions for $y_2(x;m)$ in eqns. \ref{eq4} and \ref{eq5} diverge at $x$=$1$ and so we set $D(y)=0$ in expression \ref{eq6}, so that the finiteness condition $G(1,y)\rightarrow \text{finite}$ is satisfied. With $G_L(b,y) = A(y)y_1(b) + B(y)y_2(b) = 0$ implies $B = -A\frac{y_1(b)}{y_2(b)}$.

\begin{eqnarray}\label{eqn7}
G(x,y;m) = \left\{
\begin{array}{l}
G_L(x,y;m) \equiv A(y)\left(y_1(x;m) - \frac{y_1(b)}{y_2(b)}y_2(x;m)\right) \\ \hspace{4.3cm}  b \leq x < y < 1 \\\\
G_R(x,y;m) \equiv C(y)y_1(x;m) \\ \hspace{4.3cm}  b < y < x \leq 1
\end{array}
\right. \nonumber \\
\end{eqnarray}

In addition at $x=y$
\begin{eqnarray}
G_L(y,y) = G_R(y,y), \quad \left(\frac{\partial G_R}{\partial x} - \frac{\partial G_L}{\partial x}\right)_{(y,y)} = \frac{1}{1-y^2} \label{eq8}
\end{eqnarray}
The second relation for the jump discontinuity in \ref{eq8} arises because $\mathbf{K}\cdot G(x,y)= \delta(x-y)$ may be written as
\begin{eqnarray}
\left[\frac{d}{dx}\left\lbrace(1-x^2)\frac{d}{dx}\right\rbrace + \left(2 - \frac{m^2}{1-x^2}\right)\right]G(x,y) = \delta(x-y), \label{eq9}
\end{eqnarray}
integrating which across $x=y$ leads to the required relation. Defining $y_L(x;m) \equiv y_1(x;m) - \frac{y_1(b)}{y_2(b)}y_2(x;m)$ and $y_R(x;m) \equiv y_1(x;m)$, the Green function may be written compactly as
\begin{eqnarray}
G(x,y) = \left\{
\begin{array}{ll}
A(y)y_L(x;m) & \quad b \leq x < y < 1 \\
C(y)y_R(x;m) & \quad b < y < x \leq 1 \label{eq10}
\end{array}
\right.
\end{eqnarray}
The conditions \ref{eq8} lead to two linear equations for $A(y), C(y)$ (prime below indicates differentiation)
\begin{eqnarray}
&& y_L(y;m)A - y_R(y;m)C = 0 \nonumber \\
&& y^{'}_L(y;m)A - y^{'}_R(y;m)C = \frac{1}{1-y^2} \nonumber \\
\label{eq11}
\end{eqnarray}
whose solutions are
\begin{eqnarray}
A(y) &=& \frac{y_R(y;m)}{(1-y^2)\mathscr{W}(y_L(y;m),y_R(y;m))} \nonumber \\ C(y) &=& \frac{y_L(y;m)}{(1-y^2)\mathscr{W}(y_L(y;m),y_R(y;m))} \nonumber \\ \label{eq12}
\end{eqnarray}
where $\mathscr{W}(\cdot,\cdot)$ represents the Wronskian determinant with  $\mathscr{W}(y_L(y;m),y_R(y;m)) = \frac{y_1(b;m)}{y_2(b;m)}\mathscr{W}(y_1(y;m),y_2(y;m))$. Below, we show that the prefactors in eqns. \ref{eq2} and \ref{eq5} are chosen such that the form of the Wronskian determinant remains the same, whatever be the value of $m$.      
\subsection*{$\textbf{Wronskian:}\mathbf{\;\mathscr{W}(y_1(y;m),y_2(y;m)}$ \textbf{for} $\mathbf{m=0}$}
For $m=0$, we note    
\begin{eqnarray}
&& y_1(y;0) = \mathbb{P}_1^{(0)}(y) = y \nonumber \\
&& y_2(y;0) = \mathbb{Q}_1^{(0)}(y) = \frac{y}{2}\ln\left(\frac{1+y}{1-y}\right) -1 \label{eq13} \\
&& y_1^{'}(y;0) = 1 \nonumber \\
&& y_2^{'}(y;0) = \frac{1}{2}\ln\left(\frac{1+y}{1-y}\right) + \frac{y}{1-y^2} \label{eq14}
\end{eqnarray}
Using the formula provided earlier, we obtain
\begin{eqnarray}
\mathscr{W}(y_L(y;0),y_R(y;0)) &=& \frac{y_1(b;0)}{y_2(b;0)}\mathscr{W}(y_1(y;0),y_2(y;0)) \nonumber \\
&=& \frac{y_1(b;0)}{y_2(b;0)}\frac{1}{(1-y^2)} \label{eq15}
\end{eqnarray}
\subsection*{$\textbf{Wronskian:}\mathbf{\;\mathscr{W}(y_1(y;m),y_2(y;m)}$ \textbf{for} $\mathbf{m=1}$}
\begin{eqnarray}
&& y_1(y;1) = \mathbb{P}_1^{(1)}(y) = -\sqrt{1-y^2},\nonumber \\
&& y_2(y;1) = \frac{1}{2}\mathbb{Q}_1^{(1)}(y) = \frac{1}{2}\left[\frac{-y}{\sqrt{1-y^2}}+\frac{\sqrt{1-y^2}}{2} \ln\left(\frac{1-y}{1+y}\right)\right] \nonumber \\ \label{eq16}
\end{eqnarray}
We thus obtain
\begin{eqnarray}
\mathscr{W}(y_L(y;1),y_R(y;1)) &=& \frac{y_1(b;1)}{y_2(b;1)}\mathscr{W}(y_1(y;1),y_2(y;1)) \nonumber \\
&=& \frac{y_1(b;1)}{y_2(b;1)}\frac{1}{1-y^2} \label{eq17}
\end{eqnarray}
\subsection*{$\textbf{Wronskian:}\mathbf{\;\mathscr{W}(y_1(y;m),y_2(y;m)}$ \textbf{for} $\mathbf{m \geq 2}$}
For $m \geq 2$,  
\begin{eqnarray}
&& y_1(y;m) = \left(m+y\right)\left(\frac{1-y}{1+y}\right)^{m/2} \nonumber \\
&& y_2(y;m) = \frac{1}{2m(m^2-1)}\left(m-y\right)\left(\frac{1+y}{1-y}\right)^{m/2} \label{eq18} \\
&& y_1^{'}(y;m) = \left(\frac{1-y}{1+y}\right)^{m/2}\left[1 - m\left(\frac{y+m}{1-y^2}\right)\right] \nonumber \\
&& y_2^{'}(y;m) = \frac{1}{2m(m^2-1)}\left(\frac{1+y}{1-y}\right)^{m/2}\left[1 - m\left(\frac{y-m}{1-y^2}\right)\right] \nonumber \\
&& \label{eq19}
\end{eqnarray}
We thus obtain
\begin{eqnarray}
&& \mathscr{W}(y_L(y;m \geq 2),y_R(y;m \geq 2)) \nonumber \\
&=& \frac{y_1(b;m)}{y_2(b;m)}\mathscr{W}(y_1(y;m),y_2(y;m)) \nonumber \\
&=& \frac{y_1(b;m \geq 2)}{y_2(b;m\geq 2)}\frac{1}{(1-y^2)}  \label{eq20}        
\end{eqnarray}
We note that expressions \ref{eq15}, \ref{eq17} and \ref{eq20} for the Wronskian, have same form independent of the value of $m$; this is the motivation for the prefactors used earlier in certain expressions for $y_2$. Using the form of the Wronskian in \ref{eq10} and \ref{eq12}, we obtain the Green function as
\begin{eqnarray}
G(x,y;m) = \left\{
\begin{array}{ll}
\frac{y_2(b;m)}{y_1(b;m)}y_R(y;m)y_L(x;m) & \quad b \leq x < y < 1 \\\\
\frac{y_2(b;m)}{y_1(b;m)}y_L(y;m)y_R(x;m) & \quad b < y < x \leq 1 \label{eq21}
\end{array}
\right.
\end{eqnarray}
and using the expressions for $y_R$ and $y_L$ obtained earlier, we may write \ref{eq10} using \ref{eq12} as

\begin{eqnarray}\label{eqn22}
G(x,y;m) = \left\{
\begin{array}{r}
y_1(y;m)\left(\frac{y_2(b;m)}{y_1(b;m)}y_1(x;m) - y_2(x;m)\right) \\  b \leq x < y < 1 \\\\
y_1(x;m)\left(\frac{y_2(b;m)}{y_1(b;m)}y_1(y;m) - y_2(y;m)\right) \\ b < y < x \leq 1
\end{array}
\right. \nonumber\\
\end{eqnarray}
\noindent where expressions for $y_1$ and $y_2$ can be read from eqns. \ref{eq4} and \ref{eq5}. This completes the derivation of the Green function.
\subsection*{\textbf{No-penetration at the wall}}
We prove here the constraint needed for satisfying the no-penetration condition at the wall viz.
\begin{eqnarray}
\left(\bm{\nabla}\phi\cdot\mathbf{e}_z\right)_{\theta=\pi/2} = 0 \label{eq37}
\end{eqnarray}
Refer to fig. $3$ in the paper, the components of the unit normal to the substrate may be written as
\begin{eqnarray}
&&\mathbf{e}_z = \cos(\theta)\mathbf{e}_{\rho}  - \sin(\theta)\mathbf{e}_{\theta} \label{eq38}\\
&& \bm{\nabla}\phi = \frac{\partial\phi}{\partial \rho}\mathbf{e}_{\rho} + \frac{1}{\rho}\frac{\partial\phi}{\partial\theta}\mathbf{e}_{\theta} + \frac{1}{\rho \sin(\theta)}\frac{\partial\phi}{\partial\psi}\mathbf{e}_{\psi} \label{eq39}    
\end{eqnarray}
which leads to
\begin{eqnarray}
-\frac{1}{\rho}\left(\frac{\partial\phi}{\partial\theta}\right)_{\theta=\pi/2} = 0 \quad\text{for}\quad \text{at all}\quad \rho,\psi
\end{eqnarray}
Since $\phi \propto \Phi \propto \mathbb{P}_{i}^{(m)}\left(\cos(\theta)\right)$ the above condition implies finding when is the derivative of $\mathbb{P}_i^{(m)}$ with respect to $\theta$ zero at $\theta=\pi/2$. Using some identities for the derivative of $\mathbb{P}_{i}^{(m)}$ we find the required condition viz. $i + m=\texttt{even}$ which is the condition provided in \citet{bostwick2014dynamics}.

\section*{Appendix B}
A simple formula to reproduce results of fig. \ref{fig14} is provided below. The initial shape of the bubble for the third column of fig. \ref{fig14} can be generated from $\hat{r}_s(s,t=0)=\hat{R}_0 + \hat{a}_0\;f(s)$ where $\hat{R}_0=0.1$ cm and $\hat{a}_0=0.07$ cm and $f(s)$ has the expression ($0 < s < \frac{80\pi}{180}$, in radians).

\begin{eqnarray}
f(s)&=& - 2.3981s^{15} + 25.04318s^{14} - 117.73493s^{13} \nonumber \\
&& + 329.3777s^{12} - 610.6851s^{11} + 790.9188s^{10} \nonumber \\
&&
- 734.89142s^9 + 495.3968s^8 - 242.0073s^7 \nonumber \\
&&
+ 84.1644s^6 - 20.8533s^5 + 5.5747s^4 \nonumber \\
&&
- 0.3694s^3 - 2.812s^2 - 0.00071s
+ 1 \nonumber
\end{eqnarray}
In obtaining the above formula from the eigenvalue problem \ref{eqn2_8}, the coefficient of some terms have been rounded off and this leads to a minor effect on the initial shape of the bubble and the subsequent jet that is created in the simulation.
\section*{Appendix C}
\begin{figure}
	\centering
	\includegraphics[scale=0.5]{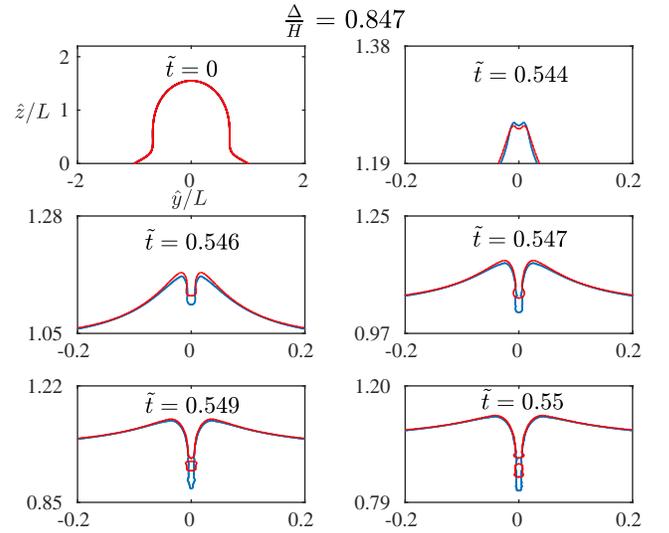}
	\caption{(Red symbols) Adaptive level $11$ at the interface and (blue symbols) is level $12$.}
	\label{fig23}
\end{figure}

Fig. \ref{fig23} presents grid convergence results for $\Delta/H= 0 .847$ of fig. \ref{fig14}.

\section*{Declaration of interests}
The authors report no conflict of interest.

\appendix*
\section*{Acknowledgements}
We thank Mr. Saswata Basak for assistance with the Green function derivation in the early phase of this study. The Ph.D. fellowship for YJD was funded by the Ministry of Education, Govt. of India and IRCC, IIT Bombay and is thankfully acknowledged. Financial support from DST-SERB (Govt. of India) grant CRG/2020/003707 is gratefully acknowledged.

\bibliography{pof}
\end{document}